\documentclass[11pt]{article}
 \pdfoutput=1
\usepackage{amsfonts,amsthm,amsmath,amssymb,slashed}
\usepackage[textwidth = 500 pt, textheight = 630 pt]{geometry}
\usepackage{latexsym,amsfonts,amsmath,amssymb,theorem,dsfont,wasysym}
\usepackage{amssymb}
\usepackage{amsmath}
\usepackage{amstext}
\usepackage{graphicx,epsfig}
\usepackage{epsfig}
\usepackage{verbatim} 
\usepackage{fancyhdr}
\usepackage{fancybox}
\usepackage{color}
\usepackage{ulem,bbold}
\usepackage{enumitem}
\usepackage{subfigure}
\usepackage{bbm}
\usepackage{parskip}
\usepackage[numbers,sort&compress]{natbib}

\newcommand{\Comment}[1]{{}}
\definecolor{MyDarkBlue}{rgb}{0.15,0.15,0.45}
\usepackage[linktocpage=true]{hyperref}
\hypersetup{
colorlinks=true,
citecolor=MyDarkBlue,
linkcolor=MyDarkBlue,
urlcolor=MyDarkBlue,
}

\usepackage[numbers,sort&compress]{natbib}
\usepackage{hypernat}

\newcommand{\la}{\langle}
\newcommand{\ra}{\rangle}
\newcommand{\half}{\frac{1}{2}}
\newcommand{\fourth}{\frac{1}{4}}

\newcommand\ignore[1]{}
\def\one{{\,\hbox{1\kern-.8mm l}}}

\def\ra{\rangle}\def\la{\langle}

\def\dag{\dagger}

\def\({\left(}
\def\){\right)}

\newcommand{\Cset}{{\,\,{{{^{_{\pmb{\mid}}}}\kern-.45em{\mathrm C}}}}}

\newcommand{\be}{\begin{equation}}
\newcommand{\bea}{\begin{eqnarray}}

\newcommand{\ee}{\end{equation}}
\newcommand{\eea}{\end{eqnarray}}

\newcommand{\nn}{\nonumber}

\begin{document}

\renewcommand{\thefootnote}{\fnsymbol{footnote}}

\makeatletter
\@addtoreset{equation}{section}
\makeatother
\renewcommand{\theequation}{\thesection.\arabic{equation}}

\rightline{}
\rightline{}
   \vspace{1.8truecm}


\vspace{10pt}


\begin{center}
{\Large \bf{An Infinite Set of Ward Identities for Adiabatic Modes in Cosmology}}
\end{center} 
 \vspace{1truecm}
\thispagestyle{empty} \centerline{
{\large  {Kurt Hinterbichler${}^{a}$, Lam Hui${}^{b}$ and Justin Khoury${}^{c}$}}
}

\vspace{1cm}

\centerline{{\it ${}^a$ 
Perimeter Institute for Theoretical Physics,}}
 \centerline{{\it 31 Caroline St. N, Waterloo, Ontario, Canada, N2L 2Y5}} 
 
 \vspace{1cm}

 \centerline{{\it ${}^b$ 
Physics Department and Institute for Strings, Cosmology and Astroparticle Physics,}}
\centerline{{\it Columbia University, New York, NY 10027, USA}}

 \vspace{1cm}

\centerline{{\it ${}^c$ 
 Center for Particle Cosmology, Department of Physics \& Astronomy, University of Pennsylvania,}}
 \centerline{{\it  209 South 33rd Street, Philadelphia, PA 19104}}
 
 \vspace{1cm}

\begin{abstract}
We show that the correlation functions of any single-field cosmological model with constant growing-modes are constrained
by an infinite number of novel consistency relations, which relate $N+1$-point correlation functions with a soft-momentum scalar or tensor mode
to a symmetry transformation on $N$-point correlation functions of hard-momentum modes. We derive these consistency relations from
Ward identities for an infinite tower of non-linearly realized global
symmetries governing scalar and tensor perturbations. These symmetries
can be labeled by an integer $n$.
At each order $n$, the consistency relations constrain --- completely for $n=0,1$, and partially for $n\geq 2$ --- the $q^n$ behavior of the soft
limits.  The identities at $n=0$ recover Maldacena's original consistency relations for a soft scalar and tensor mode, $n=1$ gives
the recently-discovered conformal consistency relations, and the identities for $n\geq 2$ are new.  As a check,
we verify directly that the $n=2$ identity is satisfied by known correlation functions in slow-roll inflation.
\end{abstract}

\newpage
\setcounter{page}{1}

\renewcommand{\thefootnote}{\arabic{footnote}}
\setcounter{footnote}{0}

\linespread{1.1}
\parskip 4pt


\section{Introduction}
\label{intro}

The consistency relation is one of the most powerful probes of early universe physics.
It states that the 3-point function of the curvature perturbation $\zeta$~\cite{Bardeen:1983qw,Salopek:1990jq},
in the squeezed or local limit where one of the modes becomes soft,
is determined by the scale transformation of the 2-point function\footnote{Here $P_\zeta$ denotes the power spectrum,
and $\la \ldots \ra'$ are correlators without the momentum-conserving $\delta$-function.}~\cite{Maldacena:2002vr,Creminelli:2004yq,Cheung:2007sv}:
\be
\lim_{\vec{q}\rightarrow 0} \frac{1}{P_\zeta(q)} \la \zeta_{\vec{q}}\zeta_{\vec{k}_1}\zeta_{\vec{k}_2}\ra'  = - \vec{k}_1\cdot \frac{\partial}{\partial \vec{k}_1} \la \zeta_{\vec{k}_1}\zeta_{\vec{k}_2} \ra'\,.
\label{cons1}
\ee
The consistency relation holds under very general conditions~\cite{Cheung:2007sv}: any early universe scenario involving a single
scalar degree of freedom (or single `clock') whose perturbations grow to a constant at late times, must satisfy~(\ref{cons1}).
This encompasses {\it all} single-field inflationary
models~\cite{Starobinsky:1979ty,Guth:1980zm,Albrecht:1982wi,Linde:1981mu,Mukhanov:1981xt},
including hybrid models~\cite{Linde:1993cn}, as well as some
non-inflationary
scenarios~\cite{Khoury:2009my,Khoury:2011ii,Joyce:2011ta,ArmendarizPicon:2003ht,ArmendarizPicon:2006if,Piao:2006ja,Magueijo:2008pm,Magueijo:2008sx,Piao:2008ip,Bessada:2009ns,Khoury:2010gw}. Conversely,
to observe a significant 3-point function in this limit requires
violating one of the assumptions above, for instance through
additional scalar
fields~\cite{Lyth:2001nq,Dvali:2003em,Kofman:2003nx,Lehners:2007ac,Buchbinder:2007ad,Creminelli:2007aq}
or an unstable
background~\cite{Wands:1998yp,Finelli:2001sr,Cai:2009fn,Khoury:2008wj}\footnote{One
  exception, for instance, is the Solid Inflation model proposed
  recently~\cite{Endlich:2012pz}.}. The consistency
relation~(\ref{cons1}) has been established through explicit
calculations~\cite{Maldacena:2002vr,Cheung:2007sv} and can be
understood using ``background-wave" arguments\footnote{The
  background-wave argument basically goes as follows. The long
  ($\vec{q}\rightarrow 0$) mode freezes out much earlier than the
  short ($\vec{k}_{1,2}$) modes, and hence acts as a classical
  background for the generation of these modes. Since the growing mode
  solution for $\zeta$ is a constant, the background 3-metric
  (ignoring tensors) experienced by the short modes is $h_{ij} =
  e^{2\zeta_{\rm L}} a^2(t)\delta_{ij}$, where $\zeta_{\rm L}$ is
  approximately constant. This constant background is an adiabatic
  mode~\cite{Weinberg:2008zzc,Weinberg:2003sw}, which can be removed
  by a local rescaling of the short
  modes.}~\cite{Maldacena:2002vr,Creminelli:2004yq}. It holds whether
the short-wavelength modes are inside or outside the
horizon~\cite{Senatore:2012wy}. Recent papers~\cite{Tanaka:2011aj,Pajer:2013ana} show that
the consistency relation can be rewritten as the vanishing
of a certain repackaged 3-point function in the squeezed limit.
Nonetheless, whether this repackaged 3-point function vanishes
or not --- as in the case of muliple fields --- has observational
consequences and is in principle testable.

In this paper we will show that inflationary correlation functions --- and more generally those of any single-field cosmology with constant growing mode solutions for scalars and tensors --- are constrained by {\it an infinite number of consistency relations}. Similarly to~(\ref{cons1}), they relate the soft limit $\vec{q}\rightarrow 0$ of a scalar or tensor mode in an $N+1$-point correlation function to a symmetry transformation on an $N$-point function. We will show how these arise as Ward identities for an infinite number of non-linearly realized symmetries, akin to the soft-pion theorems of chiral perturbation theory~\cite{Adler:1964um,Weinberg:1966kf}. At each order $n$, the consistency relations constrain --- completely for $n=0,1$, and partially for $n\geq 2$ --- the $q^n$ behavior of the soft limits of correlation functions. 

As the simplest example, we will recover the original consistency relation~(\ref{cons1}) from the Ward identity for spontaneously broken spatial dilations. 
This was already pointed out in~\cite{Maldacena:2002vr} and recently shown in~\cite{Assassi:2012zq}. The operator approach followed here in translating the Ward identity to
a consistency relation has some overlap with~\cite{Assassi:2012zq}.
The approach establishes the non-perturbative nature of 
the consistency relations.
In a parallel investigation, Goldberger, Hui and Nicolis~\cite{goldberger} have used the path integral approach
to derive~(\ref{cons1}) from a Ward identity. See~\cite{Schalm:2012pi,Bzowski:2012ih} for recent derivations of~(\ref{cons1}) using holographic arguments. As another example, we will recover the tensor consistency relation~\cite{Maldacena:2002vr}
\be
\lim_{\vec{q}\rightarrow 0} \frac{1}{P_\gamma(q)} \la \gamma^{s}(\vec{q})\zeta_{\vec{k}_1}\zeta_{\vec{k}_2} \ra'  = -\frac{1}{2}  \epsilon_{i\ell_0}^{s}(q) k^{i}_1\frac{\partial}{\partial k^{\ell_0}_1} \la   \zeta_{\vec{k}_1}\zeta_{\vec{k}_2} \ra'    \,,
\label{tens1}
\ee
where $\epsilon_{i\ell_0}^s$ is the polarization tensor, from the Ward identity for spontaneously broken anisotropic spatial rescaling. 

Recently, it has been pointed out~\cite{Creminelli:2012ed,Hinterbichler:2012nm} that scalar perturbations in spatially flat single-field cosmology are governed by
the symmetry breaking pattern
\be
SO(4,1)\rightarrow {\rm spatial} \; {\rm rotations} + {\rm translations}\,,
\label{symbreak}
\ee
with $\zeta$ playing the role of the Goldstone boson (or dilaton) for the $SO(4,1)$ conformal symmetries on $\mathds{R}^3$. The origin of conformal symmetry is
most easily seen in comoving gauge, where the scalar field is unperturbed, and the spatial metric (ignoring tensors) is $h_{ij} = a^2(t) e^{2\zeta(\vec{x},t)}\delta_{ij}$.
Since this 3-metric is conformally flat, the 10 conformal transformations on $\mathds{R}^3$ preserve this form and hence are symmetries of the scalar sector.\footnote{This does not contradict the usual notion that ``comoving gauge completely fixes the gauge," since this statement assumes that gauge transformations fall off at infinity --- conformal transformations clearly do not.  As such, conformal transformations are residual diffeomorphisms mapping field configurations which fall off at infinity into those which do not. Nevertheless, as shown in~\cite{Creminelli:2012ed,Hinterbichler:2012nm} and reviewed in Sec.~\ref{adiab}, they can be extended to transformations which do fall off at infinity, and hence generate new physical solutions or adiabatic modes~\cite{Weinberg:2008zzc,Weinberg:2003sw}.} The dilation and special conformal transformations (SCTs) are non-linearly realized on $\zeta$, whereas spatial translations and rotations are linearly realized. The conformal symmetries are restored (and linearly realized on the fields) in the limit that the background becomes exact de Sitter space~\cite{Antoniadis:1996dj,Antoniadis:2011ib,Maldacena:2011nz,Creminelli:2011mw,Kehagias:2012pd,Kehagias:2012td,Mata:2012bx,Parthasarathy:2012gp}.

As another special case of our general Ward identities, we will show that the conformal consistency relation, recently derived using background-wave arguments~\cite{Creminelli:2012ed},
is a consequence of the Ward identity for non-linearly realized SCTs. The conformal consistency relation relates the order $q$ behavior of correlation functions with a soft-$\zeta$ mode
to an SCT acting on the correlation functions of the hard modes. Since SCTs result in a departure from the transverse, traceless conditions for tensor modes, they are strictly speaking a good symmetry of the scalar sector only~\cite{Creminelli:2012ed,Hinterbichler:2012nm}. However, we will show how the SCTs can be corrected, order by order in the tensors, to become full-fledged symmetries of the theory of scalar and tensor perturbations. These result in corrections to the conformal consistency relation to all orders in tensors. (In contrast, the spatial dilation symmetry is exact, including tensors, and~(\ref{cons1}) receives no correction.) There is also a tensor consistency relation constraining the order $q$ behavior of the correlation functions with a soft tensor mode~\cite{Creminelli:2012ed}, which we will also recover from our Ward identities. 

More generally, we will show in Sec.~\ref{setup} that the theory of scalar and tensor perturbations in comoving gauge,
\be
h_{ij}=a^2(t)e^{2\zeta(\vec{x},t)}\(e^{\gamma(\vec{x},t)} \)_{ij}\,,
\label{comovgauge}
\ee
where\footnote{Unless otherwise stated, spatial indices are everywhere raised/lowered using $\delta_{ij}$.} $\gamma^i_{\ i}=0$, $\partial_i\gamma^i_{\ j}=0$, is constrained by infinitely-many non-linearly realized symmetries. These are residual diffeomorphisms, which do not fall off at infinity and which
leave the form of the 3-metric~(\ref{comovgauge}) intact. The symmetries are defined perturbatively in the tensors, and the corresponding field
transformations include terms to all orders in $\gamma$. No quasi-de Sitter or slow-roll approximation will be assumed --- the symmetries uncovered
here hold on any spatially-flat cosmological background. 

Since the residual diffeomorphisms diverge at infinity, they map field configurations which fall off at infinity into those which do not.
Nevertheless, in Sec.~\ref{adiab} we will generalize Weinberg's argument~\cite{Weinberg:2003sw}
to show that certain linear combinations of these transformations can be smoothly extended to physical configurations which fall off at infinity. In other words,
they correspond to {\it adiabatic modes}. Such transformations can be thought of as the $q\rightarrow 0$ limit of transformations which do fall off at infinity,
and therefore generate new physical solutions.

In Sec.~\ref{wardgen} we will show how the Ward identities associated with the non-linearly symmetries lead to consistency relations for
the soft limits of correlation functions. The derivation, while somewhat technical, is very general and allows us to derive at once
all consistency relations. (In contrast, the background-wave method, while straightforward for the dilation consistency relation~\cite{Creminelli:2004yq},
is already considerably more intricate for the conformal consistency relation~\cite{Creminelli:2012ed}.) Another upshot of the operator approach is that
the regime of applicability of the consistency relations is sharply defined --- Ward identities are non-perturbative statements, and in particular
do not rely on semi-classical approximations.

The consistency relations we obtain are of the following
schematic form (with $n\geq 0$)
\bea
\lim_{\vec{q}\rightarrow 0} {\partial^n \over \partial q^n}
\left( {1\over P_\zeta (q)} \langle \zeta(\vec q) {\cal O}\rangle'_c + 
{1\over P_\gamma (q)} \langle \gamma(\vec q) {\cal O}\rangle'_c\right)
\sim - {\partial^n \over \partial k^n} \left( \langle {\cal O}
  \rangle'_c
+ \langle \tilde {\cal O} \rangle'_c \right) \, ,
\label{schematic}
\eea
where ${\cal O}$
represents a product of $N$ scalar $\zeta$ and tensor
$\gamma$ perturbations with momenta labeled
by $\vec k_1, ..., \vec k_N$;
$\tilde{\cal O}$ represents
the same set of fields with one of the fields
($\zeta$ or $\gamma$) replaced by another $\gamma$;
$\langle \,\, \rangle'_c$ represents the connected
momentum space correlation functions with the
overall delta function removed.
On the left hand side, we have $N+1$-point functions
with a soft (small $q$) $\zeta$ or $\gamma$ leg
(some relations have only soft $\zeta$, some only soft $\gamma$,
and some have both);
on the right hand side, we have $N$-point functions
with the soft leg removed. As such, these consistency
relations resemble the well-known soft-pion theorems, and indeed
they follow from Ward identities applied to nonlinearly realized
symmetries just like the soft-pion theorems do.
There is an infinite set of consistency relations, labeled by 
the integer $n \geq 0$, each controlling
the $q^n$ behavior of the relevant $N+1$-point function.
The other momenta $\vec k_1, ..., \vec k_N$ are hard compared
to $\vec q$, but there is no assumption on whether they
are inside or outside the horizon \cite{Senatore:2012wy}. 
There are 3 independent relations for $n=0$
(one involving a soft scalar and two involving
a soft tensor), 7 relations for $n=1$
(three involving a soft scalar and four involving
a soft tensor),
and 6 for each $n\geq 2$
(four involving a soft tensor and two involving
mixtures of soft scalar and tensor). 
The $n=0$ and $n=1$ relations are known.
The $n\geq 2$ relations are new.

In Sec.~\ref{knownegs} we will show that the $n=0$ identities reproduce the dilation and anisotropic scaling consistency relations, of which~(\ref{cons1}) and~(\ref{tens1})
are the simplest renditions. Moreover, we will recover from the $n=1$ identity the conformal consistency relation and the linear-gradient tensor analogue relation.
While the $n=0,1$ consistency relations completely fix the $q^0$ and $q$ behavior of the soft limits, the $n\geq 2$ identities only partially fix the $q^n$
behavior of the correlation functions. Of the $6$ algebraically-independent consistency relations at $n\geq 2$, 4 involve soft-$\gamma$ correlation functions only,
while the remaining 2 involve a linear combination of soft-$\zeta$ and soft-$\gamma$ correlators. (This is a key difference compared to the lower-order identities --- there are no
``pure scalar" consistency relations for $n\geq 2$.) As an example of a novel consistency relation, we study in Sec.~\ref{newexamples} the $n=2$ tensor Ward identity relating the $q^2$
behavior of $\la \gamma_{ij}(\vec{q})\zeta_{\vec{k}_1}\zeta_{\vec{k}_2}\ra$ in the limit $\vec{q}\rightarrow 0$ to derivatives of the scalar 2-point function
$\la \zeta_{\vec{k}_1}\zeta_{\vec{k}_2}\ra$. As a check of our result, we verify that this new consistency relation is satisfied by the 3-point function computed by Maldacena in slow-roll inflation~\cite{Maldacena:2002vr}. 
As another non-trivial check on our identities, in Sec.~\ref{dotzeta3check} we verify that the contribution to $\langle \zeta\zeta\zeta\rangle$ from a particular operator is annihilated by our derivative operators to all orders in $n$.
We conclude in Sec.~\ref{conclu} with a discussion of possible future investigations. Many of the technical details of our derivation are described in a series of Appendices. 

The general Ward identities derived here constitute the complete list of consistency relations that any single-field
cosmological scenario with constant growing modes must satisfy. The Planck data, which has found no evidence of non-Gaussianity in
the squeezed limit~\cite{Ade:2013ydc}, is so far consistent with the lowest-order consistency relation~(\ref{cons1}). The higher-order relations involving scalars and tensors, while unlikely to be tested in the near future, offer the complete checklist with which current and future-generation observations will put single-field inflation to test.
 
\section{Symmetries of Cosmological Perturbations}
\label{setup}

Consider perturbations on a spatially-flat, Friedmann-Robertson-Walker (FRW) background driven by a scalar field $\phi = \phi(t)$. Following~\cite{Maldacena:2002vr}, we work in comoving (or uniform-density) gauge\footnote{The form of $h_{ij}$ in this gauge is closely related to the conformal decomposition introduced by York~\cite{York:1971hw} and Lichnerowicz~\cite{lich}.}, where the scalar field is unperturbed, 
\be 
\phi=\phi(t)\,,\ \ \ h_{ij}=a^2(t)e^{2\zeta(\vec{x},t)}\(e^{\gamma(\vec{x},t)} \)_{ij}\,,\ \ \ \gamma^i_{\ i}=0\,,\ \  \partial_i\gamma^i_{\ j}=0\,.
\label{comovgauge2}
\ee
In this gauge, scalar inhomogeneities are captured by $\zeta$, while tensor modes are encoded in $\gamma_{ij}$. 

If all the fields and gauge parameters are assumed to die off sufficiently fast at spatial infinity, then this gauge choice completely fixes the gauge,
leaving no residual symmetries. Nevertheless, there is still room for diffeomorphisms whose gauge parameters do not die off at infinity. In the absence
of tensors, for instance, it has been argued recently that conformal transformations on spatial slices are residual symmetries of the
scalar sector~\cite{Hinterbichler:2012nm,Creminelli:2012ed}. Our goal in this Section is to generalize these results to include tensor perturbations to all orders. Although we will eventually be interested in deriving Ward identities for inflationary correlation functions, we should stress that no slow-roll
or quasi-de Sitter approximations will be necessary to identify these residual symmetries. They hold on any (spatially-flat) FRW background.

Consider a (possibly time-dependent) spatial diffeomorphism, $\xi^i(\vec x,t)$. 
Since this diffeomorphism is purely spatial, it does not disturb the gauge choice $\phi=\phi(t)$. 
If we can identify field transformations $\delta\zeta$ and $\delta\gamma_{ij}$ such that 
\be 
\delta \(e^{2\zeta}\(e^\gamma\)_{ij}\)={\cal L}_\xi \(e^{2\zeta}\(e^\gamma\)_{ij}\)\,,
\label{fulleq}
\ee
then these transformations will act like diffeomorphisms which leave
the gauge fixed action invariant, and hence constitute a symmetry.  
The Lie derivative ${\cal L}_\xi (g_{ij})$ is defined as
$\xi^k \partial_k g_{ij} + \partial_i \xi^k  g_{kj} +
\partial_j \xi^k  g_{ik}$. 
We will solve~(\ref{fulleq})
order by order in powers of $\gamma$, expanding the field variations and diffeomorphism parameter as follows:
\bea 
\nonumber
\delta \gamma_{ij}&=& \delta \gamma_{ij}^{(\gamma^0)}+ \delta \gamma_{ij}^{(\gamma^1)}+\ldots\\
\nonumber
\delta\zeta&=&\delta\zeta^{(\gamma^0)}+\delta\zeta^{(\gamma^1)}+\ldots\\
\xi_i&=&\xi_i^{(\gamma^0)}+\xi_i^{(\gamma^1)}+\ldots
\label{tensexpand}
\eea
At each order we will impose that $\delta\gamma_{ij}$ be transverse and traceless.

\subsection{Zeroth-order in tensors}

We first focus on the zeroth-order diffeomorphisms, $\xi_i^{(\gamma^0)}$, leaving the derivation of the first-order correction $\xi_i^{(\gamma^1)}$ to Sec.~\ref{higherorder} below.

At zeroth-order in $\gamma$, expanding~(\ref{fulleq}) gives
\be 
2\delta \zeta^{(\gamma^0)} \delta_{ij}+\delta\gamma_{ij}^{(\gamma^0)}=2\xi_{k}^{(\gamma^0)}\partial^k\zeta\delta_{ij}+\partial_i\xi_j^{(\gamma^0)} +\partial_j\xi_i^{(\gamma^0)} \,.
\label{eq1sta}
\ee
Since $\delta\gamma_{ij}$ must be traceless, taking the trace allows us to solve for $\delta\zeta^{(\gamma^0)}$,
\be 
\delta\zeta ^{(\gamma^0)}={1\over 3}\partial^i\xi_i^{(\gamma^0)} + \xi_i^{(\gamma^0)} \partial^i\zeta\,.
\label{zerot1}
\ee
Plugging back into~\eqref{eq1sta}, we can solve for $\delta\gamma_{ij}^{(\gamma^0)}$,
\be 
\delta\gamma_{ij}^{(\gamma^0)} =\partial_i\xi_j^{(\gamma^0)} +\partial_j\xi_i^{(\gamma^0)}-{2\over 3}\partial^k\xi_k^{(\gamma^0)} \delta_{ij}\,.
\label{zerot2}
\ee
Taking the divergence of this expression yields an equation for $\xi_i^{(\gamma^0)}$,
\be 
\vec{\nabla}^2 \xi_i ^{(\gamma^0)}+{1\over 3}\partial_i\partial^j\xi_j^{(\gamma^0)}=0\,.
\label{xieqn0}
\ee
Note that~(\ref{xieqn0}) admits no non-trivial solutions which vanish at infinity.  We see this by taking its divergence, $\vec{\nabla}^2 (\partial^i\xi_i^{(\gamma^0)})=0$, which for fields that vanish
at infinity implies $\partial^i\xi_i^{(\gamma^0)}=0$. Plugging this back into~(\ref{xieqn0}) gives $\vec{\nabla}^2\xi_i^{(\gamma^0)}=0$, which then implies $\xi_i^{(\gamma^0)}=0$. Hence $\xi_i^{(\gamma^0)}$ cannot vanish at infinity. As a corollary, even though we allow for $\xi_i^{(\gamma^0)}$ to depend on $\zeta$, the argument above implies that it is in fact independent of $\zeta$. To see this, imagine expanding $\xi_i^{(\gamma^0)}$ in powers of $\zeta$, where each term in this expansion must separately satisfy~(\ref{xieqn0}). Since higher-order terms are proportional to powers of $\zeta$, and thus
vanish at infinity (since $\zeta$ does), they must be trivial.

Any (possibly time-dependent) $\xi_i^{(\gamma^0)}$ satisfying~\eqref{xieqn0} preserves comoving gauge, to zeroth order in tensors, with field
transformations~\eqref{zerot1} and~\eqref{zerot2}.  The condition \eqref{xieqn0} is the divergence of the conformal Killing equation on $\mathbb{R}^3$, $\partial_i \xi_j+\partial_j\xi_i={2\over 3}\partial_k\xi^k\delta_{ij}$, so any of the transformations of the conformal group on $\mathbb{R}^3$ (with possibly time dependent parameters) will satisfy it.  These include, in addition to (time-dependent) spatial rotations and translations, the spatial dilation
and 3 special conformal transformations (SCTs)~\cite{Creminelli:2012ed,Hinterbichler:2012nm} 
\bea
\nonumber
\xi_i^{\rm dilation} &=& \lambda(t) x_i \\
\xi_i^{\rm SCT} &=& 2b^j(t)x_j x_i-\vec x^2b_i(t)\,.
\label{xidilSCT}
\eea

\subsection{Higher order in tensors}
\label{higherorder}

It is worth emphasizing that while
SCTs were initially believed to be good symmetries of the scalar sector only~\cite{Creminelli:2012ed,Hinterbichler:2012nm}, since
they result in a departure from the transverse, traceless conditions for tensor modes, here we show that $\xi_i^{\rm SCT}$
can be corrected order by order in $\gamma$ to preserve the transverse, traceless conditions, and thus is promoted to a
full-fledged gauge-preserving transformation. In contrast, we also show that spatial dilation is exact to all orders in tensors.

At first order in $\gamma$,~(\ref{fulleq}) gives\footnote{We thank Lasha Berezhiani and Junpu Wang for correcting important typos in the original ${\cal O}(\gamma^1)$ expressions.}
\bea
\nonumber 
& & 2\delta \zeta^{(\gamma^1)} \delta_{ij}+2\delta \zeta^{(\gamma^0)} \gamma_{ij}+\delta\gamma_{ij}^{(\gamma^1)} + \frac{1}{2}\left(\delta\gamma_{ik}^{(\gamma^0)}\gamma^k_{~j} +\gamma_i^{~k}  \delta\gamma_{kj}^{(\gamma^0)}\right) \\
& & =2\xi_k^{(\gamma^1)}\partial^k\zeta\delta_{ij}+2\xi_k^{(\gamma^0)}\partial^k\zeta\gamma_{ij}+\partial_i\xi_j^{(\gamma^1)}+\partial_j\xi_i^{(\gamma^1)}+{\cal L}_{\xi^{(\gamma^0)}}\gamma_{ij}\,.
\label{eq2st} 
\eea
The tensor variation $\delta\gamma_{ij}^{(\gamma^1)}$ must be transverse and traceless, so as before, taking the trace allows us to solve for the scalar variation,
\be 
\delta\zeta^{(\gamma^1)}=\xi_i^{(\gamma^1)}\partial^i\zeta+{1\over 3}\partial^i\xi_i^{(\gamma^1)}\,.
\label{dzeta1st}
\ee
Plugging back into~\eqref{eq2st}, we can solve for the tensor variation,
\bea 
\delta\gamma_{ij}^{(\gamma^1)}=&&\partial_i\xi_j^{(\gamma^1)}+\partial_j\xi_i^{(\gamma^1)}-{2\over 3}\partial^k\xi_k^{(\gamma^1)}\delta_{ij}\nn\\
&& +  \frac{1}{2} \left(2\xi_k^{(\gamma^0)} \partial^k\gamma_{ij} +\partial_i \xi_k^{(\gamma^0)}\gamma^k_{~j} + \partial_j \xi_k^{(\gamma^0)}\gamma^k_{~i} - \partial_k \xi_i^{(\gamma^0)}\gamma^k_{~j} - \partial_k \xi_j^{(\gamma^0)}\gamma^k_{~i} \right)\,.
\label{dgamma1st}
\eea
Taking a divergence, we find the following equation for $\xi_i^{(1)}$,
\be 
\nabla^2 \xi_i^{(\gamma^1)}+{1\over 3}\partial_i\(\partial^j\xi_j^{(\gamma^1)}\)=-\frac{\partial^j}{2}\(2\xi_k^{(\gamma^0)} \partial^k\gamma_{ij} +\partial_i \xi_k^{(\gamma^0)}\gamma^k_{~j} + \partial_j \xi_k^{(\gamma^0)}\gamma^k_{~i} - \partial_k \xi_i^{(\gamma^0)}\gamma^k_{~j} - \partial_k \xi_j^{(\gamma^0)}\gamma^k_{~i} \)\,.
\label{xi1eq}
\ee
The right-hand side contains known zeroth-order quantities only and thus sources $\xi_i^{(\gamma^1)}$.  We are instructed to solve this equation subject to the boundary condition that $\xi_i^{(\gamma^1)}$ vanish at infinity. (Since $\xi_i^{(\gamma^1)}$ is first-order in $\gamma$, which itself goes to zero at infinity, there is no room for a homogeneous solution that would not be proportional to $\gamma$.)  By taking a divergence of \eqref{xi1eq} it is easy to see that the solution is unique, albeit spatially non-local. Explicitly, the solution for $\xi_i^{(\gamma^1)}$ with these boundary conditions is
\be
\xi_i^{(\gamma^1)} = \frac{1}{2} \frac{\partial^k}{\nabla^2}\left(\delta_i^{\;\ell} - \frac{1}{4}\frac{\partial_i\partial^\ell}{\nabla^2}\right) \left(-{\cal L}_{\xi^{(\gamma^0)}}\gamma_{k\ell}  + \partial^m\xi^{(\gamma^0)}_\ell \gamma_{mk} + \partial^m\xi^{(\gamma^0)}_m\gamma_{k\ell} \right)  \,.
\label{xigamm1}
\ee

There is no obstruction to extending this to any order in $\gamma$.  At $m$-th order in the tensors,~\eqref{fulleq} reads
\be 
2\delta \zeta^{(\gamma^m)} \delta_{ij}+\delta\gamma_{ij}^{(\gamma^m)}=2\xi_k^{(\gamma^m)}\partial^k\zeta\delta_{ij}+\partial_i\xi_j^{(\gamma^m)}+\partial_j\xi_i^{(\gamma^m)}+\(\rm known\ lower\ order\ pieces\)\,.
\label{eq1st}
\ee
As before, taking the trace allows us to solve for $\delta\zeta^{(\gamma^m)}$, and then plugging back we can solve for $\delta\gamma_{ij}^{(\gamma^m)}$.  Taking the divergence then yields an equation for $\xi_i^{(\gamma^m)}$ which reads
\be 
\nabla^2 \xi_i^{(\gamma^m)}+{1\over 3}\partial_i\(\partial_j\xi^{j(\gamma^m)}\)=\(\rm known\ lower\ order\ pieces\)\,.
\label{xineq}
\ee
The same line of argument as before leads us to conclude that~(\ref{xineq}) has a unique solution.

Note that the dilation transformation, corresponding to $\xi^{(\gamma^0),\;{\rm dil.}}_i = \lambda x_i$, is exceptional in that it does not get corrected at any order in the tensors. Indeed, from~(\ref{xigamm1}) it is straightforward to show that $\xi^{(\gamma^1),\;{\rm dil.}}_i  = 0$ in this case. This is true to all orders: $\xi^{\rm dil.}_i = \lambda x_i$ is an {\it exact} symmetry, with 
\bea
\nonumber
\delta^{\rm dilation}\zeta(\vec{x}) &=& \lambda \left(1 + x^i \partial_i \zeta(\vec{x})\right)\,,\\
\delta^{\rm dilation}\gamma_{ij}(\vec{x}) &=& \lambda x^k\partial_k \gamma_{ij}(\vec{x})\,.
\label{exactdilationtransfns}
\eea

\subsection{Adiabatic modes}
\label{adiab}

By the argument above, the residual transformations $\xi_i$ have a field-independent part which does not fall off at spatial infinity, and
therefore map field configurations which fall off at infinity into those which do not. Nevertheless, by generalizing Weinberg's adiabatic mode
argument~\cite{Weinberg:2003sw}, we will see --- at least to linear order in perturbations --- that a subset of these
transformations can be extended to physical configurations with suitable fall-off behavior at infinity. In other words,
this subset of transformations can be thought of as the $q\rightarrow 0$ limit of transformations which do fall off at infinity,
and therefore generate new physical solutions. The results are adiabatic modes: physical solutions which locally approximate pure-gauge profiles.

To be extendible to a physical mode, a configuration must satisfy the equations of motion away from zero momenta, that is, it cannot ``accidentally'' solve the equations simply because it is being hit by spatial derivatives. The only equations for which this may happen are the constraint equations of General Relativity, since the evolution equations have terms with second time derivatives and hence no spatial derivatives (since all the equations are second order). At linear order, the momentum and Hamiltonian constraints are
\bea
\nonumber
& & 2\partial_i\left(H  N_1-\dot\zeta\right)-{1\over 2}\vec\nabla^2 N_i+{1\over 2}\partial_i\left(\partial_jN^j\right)=0\,; \\
& & \partial_i \left(\partial^i\zeta + HN^i\right) + \frac{a^2\dot{H}}{c_s^2}N_1 + 3a^2H\left(HN_1-\dot{\zeta} \right) = 0 \,.
\label{momHcons}
\eea
Here, $N$ and $N_i$ represent the lapse and shift of the metric, $N_1$ is the perturbation of $N$, $H$ is the Hubble parameter, and $c_s$ is the sound speed.
We focus on physical configurations, with suitable fall-off behavior at spatial infinity. In this case, we can solve these equations for $N_1$ and $N^i$ by inverting
$\vec{\nabla}^2$ in the usual way~\cite{Maldacena:2002vr}:
\be
N_1 = \frac{\dot{\zeta}}{H} \,; \qquad N_i = - \frac{\partial_i\zeta}{H}  -  \frac{a^2\dot{H}}{H\,c_s^2}\frac{\partial_i}{\vec{\nabla}^2}N_1 \,.
\label{momHconssoln}
\ee

Next we consider a profile which locally looks pure gauge. To linear order we only need the non-linear part of the transformation laws,
\be\label{lineartrans}
\delta \zeta \simeq {1\over 3}\partial_i\xi^i, \ \ \ \delta N^i\simeq \dot{\xi}^i,\ \ \ \delta N_1\simeq 0\,,
\ee
where ``$\simeq$'' indicates that this is only true locally. Since this represents a diffeomorphism from the unperturbed solution, and the equations of General Relativity are diffeomorphism invariant, these transformations automatically preserve the constraints~(\ref{momHcons}). To be extendible to a physical field configuration with suitable fall-off behavior, however, it must also be consistent with the solution~(\ref{momHconssoln}). To start with, since $\delta N_1 = 0$, the first of~(\ref{momHconssoln}) implies
\be
\partial_i\dot{\xi}^i = 0\,.
\label{cond1}
\ee
Note that this criterion is local, and can be understood as a constraint on a locally-pure-gauge profile to
be extendible to a physical solution. Substituting $N_1 = 0$ into the second of~\eqref{momHconssoln}, the solution for $N_i$ also
becomes local: $N_i = - \partial_i\zeta/H$. Consistency with~(\ref{lineartrans}) then requires
\be
\dot{\xi}^i = - \frac{1}{3H} \partial_i\partial_j\xi^j \,.
\label{mainrel}
\ee
Any solution to~(\ref{cond1}),~(\ref{mainrel}) and the gauge-preserving condition~(\ref{xieqn0})
represents a gauge transformation of the unperturbed solution which can be extended to a physical
field configuration, {\it i.e.}, an {\it adiabatic mode}.

Let us decompose the diffeomorphism as
\be
\xi^i = \bar{\xi}^i + \xi^i_{\rm T}\,,
\ee
where $\partial^i \xi_i^{\rm T} = 0$. While not unique (since $\xi^i $ does not fall off at infinity), such decomposition can always be done.
From~(\ref{cond1}), we conclude that 
\be
\partial_i\dot{\bar{\xi}}^i = 0\,.
\ee
Any time-dependent contribution to $\bar{\xi}^i$ must be divergence-free, and hence can be absorbed into a redefinition of $\xi^i_{\rm T}$.
We can therefore assume $\bar{\xi}^i$ is time-independent without loss of generality. Equation~(\ref{mainrel}) then reduces to
\be
\dot{\xi}^i_{\rm T} =   - \frac{1}{3H} \partial_i\partial_j\bar{\xi}^j \,,
\ee
with solution
\be
\xi^i_{\rm T} =  - \frac{1}{3} \int^t \frac{{\rm d} t'}{H(t')}  \partial_i\partial_j\bar{\xi}^j \,.
\ee
Finally, to satisfy~(\ref{xieqn0}) at all times, $\bar{\xi}^i$ must itself be a solution, {\it i.e.},
\be
\vec{\nabla}^2 \bar{\xi}_i + \frac{1}{3} \partial_i\partial^j\bar{\xi}_j = 0\,.
\label{cond0bar}
\ee
The physically allowed diffeomorphisms are therefore given by
\be
{\xi^i  = \left(1 + \int^t \frac{{\rm d}t'}{H(t')}\vec{\nabla}^2\right) \bar{\xi}^i \,.}
\label{xigen}
\ee
where $\bar{\xi}_i$ is any time-independent, gauge-preserving transformation, that is, any time-independent transformation satisfying~(\ref{cond0bar}).
For example, the physical extension of a time-independent SCT $\bar{\xi}_i^{\rm SCT} = 2b^j x_j x_i-\vec x^2b_i$, which as mentioned earlier satisfies~(\ref{cond0bar}),
is
\be
\xi_i =2b^j x_j x_i-\vec x^2b_i - 2b_i \int^t \frac{{\rm d}t'}{H(t')}\,,
\ee
where the correction is recognized as a time-dependent translation, in agreement with~\cite{Creminelli:2012ed,Hinterbichler:2012nm}. A time-independent
spatial dilation, $\bar{\xi}_i^{\rm dilation} = \lambda x^i$, on the other hand, does not get corrected and therefore corresponds to
an adiabatic mode~\cite{Weinberg:2003sw}. 

More generally, note that $\zeta$ always transforms linearly under the time-dependent correction $\xi^i_{\rm T}$, since $\partial_i \xi^i_{\rm T} = 0$ by definition. Specifically,
the field transformations (to zeroth order in tensors) are
\be
\delta\zeta = \frac{1}{3} \partial_i\bar{\xi}^i +  \xi_i \partial^i\zeta \,;\qquad \delta\gamma_{ij} = \bar{\delta}\gamma_{ij} + \vec{\nabla}^2 \left(\partial_i\bar{\xi}_j + \partial_j\bar{\xi}_i\right) \int^t \frac{{\rm d}t'}{H(t')}\,,
\ee
where, according to~(\ref{zerot2}), $\bar{\delta}\gamma_{ij}
= \partial_i
\bar\xi_j +\partial_j\bar\xi_i-2 \partial^k\bar\xi_k
\delta_{ij}/3$. 

If $\bar{\xi}^i$ is itself transverse such
that the two terms of~(\ref{cond0bar}) vanish separately, then the physical symmetry receives no time-dependent correction:
\be
\xi^i = \bar{\xi}^i_{\rm T}\,,
\label{xitensorsym}
\ee
with $\partial_i \bar\xi^i_{\rm T} = \vec{\nabla}^2 \bar\xi^i_{\rm T}
= 0$. We will refer to such symmetries as ``tensor symmetries", in the
sense that they effect nonlinear shifts in the tensor but not the
scalar perturbations.

The constraints that we derive in this Section,
that our residual diffeomorphism in fact solves the constraints at
finite momenta, are necessary only for the $O(\gamma^0)$ part of
the diffeomorphism (we adopt the implicit notation that $\xi_i$ without
superscript ``$\gamma^0$'' is the $O(\gamma^0)$ part).
For the $O(\gamma^1)$ and higher parts of the diffeomorphism,
since $\gamma$ falls off at infinity, there is no need to check that
the constraints are satisfied, and we simply extend the $O(\gamma^1)$ part of the physical transformation to higher order in the tensors using the procedure of Sec.~\ref{higherorder}.

\subsection{Taylor expansion}
\label{Taylorexpand}

As shown in~(\ref{xigen}), the physical symmetries are expressed in terms of time-independent diffeomorphisms 
$\bar{\xi}^i$ satisfying the gauge preserving condition~(\ref{cond0bar}). In general, this can be solved in a power series:
\be 
\bar{\xi}_i \equiv \sum_{n=0}^\infty \bar{\xi}_i^{(n)} = \sum_{n=0}^\infty \frac{1}{(n+1)!}M_{i\ell_0 \ldots \ell_n}  x^{\ell_0}\cdots x^{\ell_n}\,,
\label{xiexpansion}
\ee
where the array $M_{i\ell_0\cdots \ell_n}$ is constant and symmetric in its last $n+1$ indices.\footnote{We will ignore the constant term
in the Taylor series expansion, $\bar{\xi}_i = M_i$, since this represents an unbroken (time-independent) spatial translation.} Equation~(\ref{xieqn0})
translates to the condition:
\be
M_{i \ell\ell \ell_2\ldots \ell_n}  = - \frac{1}{3} M_{\ell i \ell \ell_2\ldots \ell_n}\qquad (n \geq 1) \,.
\label{Mcond}
\ee
Each diffeomorphism $\bar{\xi}_i^{(n)}$ in the series generates its own field transformations~(\ref{zerot1}) and~(\ref{zerot2}). 
To derive the Ward identities consistently at lowest non-trivial orders in the tensors, we will need to work to linear order 
in $\gamma$ for the field transformations. Using~(\ref{xigamm1}), the
transformations in momentum space for each $n$ are given
by\footnote{See Appendix~\ref{convs} for Fourier transform
  conventions. Our symmetrization convention is $T_{(ab)} \equiv
  \frac{1}{2}\left(T_{ab} + T_{ba}\right)$.
Note that the transformations on $\zeta$ and $\gamma$
expressed here are affected by the time-independent
diffeomorphism $\bar\xi^i$, not its 
time-dependent deformation (into an adiabatic mode).
}
\bea
\nonumber
\delta \zeta^{(n)}(\vec{k}) &=& \frac{(-i)^n}{3n!} M_{ii\ell_1 \ldots \ell_n}\frac{\partial^{n}}{\partial k_{\ell_1}\cdots \partial k_{\ell_n}}\left((2\pi)^3\delta^3(\vec{k})\right) \\
\nonumber
& -& \frac{(-i)^n}{n!} M_{i \ell_0 \ldots \ell_n} \Bigg( \delta^{i\ell_0} \frac{\partial^{n}}{\partial k_{\ell_1}\cdots \partial k_{\ell_n}} + \frac{k^{i}}{n+1}  \frac{\partial^{n+1}}{\partial k_{\ell_0} \cdots \partial k_{\ell_n}}\Bigg)\zeta(\vec{k}) \\
\nonumber
&+& \frac{(-i)^n}{n!} M_{\ell \ell_0\ldots \ell_n}\Upsilon^{\ell\ell_0ij}(\hat{k}) \frac{\partial^{n}}{\partial k_{\ell_1}\cdots \partial k_{\ell_n}}  \gamma_{ij}(\vec{k}) + \ldots \\
\nonumber
\delta \gamma_{ij}^{(n)}(\vec{k}) &=& \frac{(-i)^n}{n!} \left( M_{ij\ell_1 \ldots \ell_n} + M_{j i\ell_1 \ldots \ell_n} - \frac{2}{3}\delta_{ij} M_{\ell\ell\ell_1 \ldots \ell_n}\right) \frac{\partial^{n}}{\partial k_{\ell_1}\cdots \partial k_{\ell_n}}\left((2\pi)^3\delta^3(\vec{k})\right) \\
\nonumber
&-&  \frac{(-i)^n}{n!} M_{\ell \ell_0\ldots \ell_n}\Bigg(\delta^{\ell\ell_0} \frac{\partial^{n}}{\partial k_{\ell_1}\cdots \partial k_{\ell_n}} + \frac{k^{\ell}}{n+1}\frac{\partial^{n+1}}{\partial k_{\ell_0} \cdots \partial k_{\ell_n}}\Bigg) \gamma_{ij}(\vec{k})\\
&+&  \frac{(-i)^n}{n!} M_{ab \ell_1 \ldots \ell_n}  \Gamma^{ab}_{\;\;\;ij k\ell}(\hat{k}) \frac{\partial^{n}}{\partial k_{\ell_1}\cdots \partial k_{\ell_n}}  \gamma^{k\ell}(\vec{k}) + \ldots
\label{delnmom}
\eea
where\footnote{We are most grateful to Lasha Berezhiani and Junpu Wang for correcting some typos in the original expressions for $\Upsilon$ and $\Gamma$.}
\bea
\nonumber
\Upsilon_{abcd}(\hat{k})  &\equiv & \frac{1}{4}\delta_{ab}\hat{k}_c\hat{k}_d -\frac{1}{8}\delta_{ac}\hat{k}_b\hat{k}_d -\frac{1}{8}\delta_{ad}\hat{k}_b\hat{k}_c\,;\\
\Gamma_{a b i j k \ell}(\hat k)&=& -\frac{1}{2} \left(\delta_{ij}+{\hat k}_i{\hat k}_j\right) \left(\delta_{ab}{\hat k}_k{\hat k}_\ell -\frac{1}{2}\delta_{a k}{\hat k}_\ell{\hat k}_b -\frac{1}{2}\delta_{a \ell}{\hat k}_k{\hat k}_b \right)
+\delta_{b(i}\delta_{j)(k}\delta_{\ell)a} -\delta_{a(i}\delta_{j)(k}\delta_{\ell)b} \nonumber \\
&-& \delta_{b(i}{\hat k}_{j)}\delta_{a (k } {\hat k}_{\ell)} +\delta_{a(i}{\hat k}_{j)}\delta_{b (k } {\hat k}_{\ell)} -\delta_{a (k }\delta_{\ell) (i}{\hat k}_{j)}{\hat k}_b -\delta_{b (k }\delta_{\ell) (i}{\hat k}_{j)}{\hat k}_a
+2\delta_{a b} {\hat k}_{(i}\delta_{j)(k}{\hat k}_{\ell)}\,,
\label{Gamdef}
\eea
with $\hat{k}^i \equiv k^i/k$.  The ellipses in~(\ref{delnmom}) indicate higher-order corrections in the fields. Specifically, in $\delta \zeta$ the ellipses include terms of order $\zeta\gamma$, $\gamma^2$, {\it etc.}, while in $\delta\gamma$ they include terms $\sim \gamma^2$ and higher-order in $\gamma$.

One further ``adiabatic" condition must be imposed for these transformations to correspond to physically-realized symmetries. 
As before, we want to consider only those configurations which can be smoothly extended to ({\it i.e.}, thought of as the long-wavelength limit of) a {\it physical} mode with suitable fall-off behavior at spatial infinity. The additional condition arises by demanding that the non-linear shift in $\gamma_{i\ell_0}$ in~(\ref{delnmom}) remains transverse
when extended to a physical mode. To see the constraint, we imagine
smoothing out the momentum profile around $\vec{q} = 0$. To ensure
that transversality is preserved in Fourier space at finite momentum,
$\hat{q}^i {\delta} \gamma_{i\ell_0}(\vec{q}) = 0$, we let the
$M_{i\ell_0\cdots \ell_n}$ coefficients become $\hat{q}$-dependent
such that\footnote{Here, we need only check that
the nonlinear (field-independent) part of $\delta\gamma_{i\ell0} (\vec
q)$ is transverse. The field-dependent parts are guaranteed to be
transverse by construction i.e. $\partial^i
\delta\gamma_{i\ell0}^{(\gamma^1)}
= 0$ (Section \ref{higherorder}).}
\be
\hat{q}^i  \left( M_{i\ell_0\ell_1 \ldots \ell_n}(\hat{q}) + M_{\ell_0 i\ell_1 \ldots \ell_n}(\hat{q}) - \frac{2}{3}\delta_{i\ell_0} M_{\ell\ell\ell_1 \ldots \ell_n}(\hat{q})\right) = 0\,.
\label{Mtrans}
\ee
To determine the space of possible $M$'s, we proceed as follows: we first determine the $M$'s for some reference direction in momentum space, say $\hat q=\hat z=(0,0,1)$.  These are all arrays which satisfy \eqref{Mtrans} for $\hat q=\hat z$ as well as \eqref{Mcond}, and which are symmetric in the last $n+1$ indices.  We then get the remaining $q$ dependence by applying to all the indices of $M$ some standard rotation which takes $\hat z$ into $\hat q$.

To illustrate the construction, we discuss the lowest-order transformations in some detail:

\begin{itemize}

\item $n=0$:  In this simplest case, the symmetry transformation is linear, $\bar{\xi}_i^{(n=0)} =M_{i\ell_0}x^{\ell_0}$.
There are no constraints coming from symmetry in the last indices or from \eqref{Mcond}.  A general array $M_{i\ell_0}$ can be
decomposed into a trace part, a symmetric traceless part and an anti-symmetric part: 
\be
M_{i\ell_0} = \lambda \delta_{i\ell_0} + S_{i\ell_0} + \omega_{i\ell_0} \,,
\label{M1gendecomp}
\ee
where $S_{i\ell_0} = S_{\ell_0 i}$, $\omega_{i\ell_0} = - \omega_{\ell_0 i}$, and $S_{ii} = 0$. The anti-symmetric part, parametrized by $\omega_{i\ell_0}$,
corresponds to spatial rotations, which are linearly realized and hence will not concern us. The trace part, parametrized by $\lambda$, corresponds to a dilation, $M_{i\ell_0}^{\rm dilation} = \lambda \delta_{i\ell_0}$. Finally, the symmetric traceless part, parametrized by $S_{i\ell_0}$, describes a volume-preserving, anisotropic
rescaling of coordinates, $M_{i\ell_0}^{\rm anise} = S_{i\ell_0}$, under which $\zeta$ transforms linearly and $\gamma_{i\ell_0}$ shifts by a constant.
The constraint~\eqref{Mtrans} tells us that $S_{i\ell_0}(\vec{q})$ is transverse:
\be
\hat{q}^iS_{i\ell_0}(\vec{q})  = 0\,.
\label{Strans}
\ee
The array $S_{i\ell_0}(\vec{q})$ is therefore symmetric, transverse and traceless, and thus describes 2 physical tensor symmetries.
Therefore, for $n=0$ we have in total 3 physical symmetries (1 dilation and 2 anisotropic rescalings).

\item $n=1$: In this next simplest case,~\eqref{Mcond} requires
\be
M_{i \ell\ell}  = - \frac{1}{3} M_{\ell i \ell}\,.
\label{n=2cond}
\ee
This gives 3 conditions on the 18 algebraically independent components of $M_{i\ell_0\ell_1}$.
These include the 3 SCTs~(\ref{xidilSCT}) of the scalar sector, with constant $b^i$:
\be
M_{i\ell_0\ell_1}^{\rm SCT} = b_{\ell_1} \delta_{i\ell_0} + b_{\ell_0}\delta_{i\ell_1} - b_i\delta_{\ell_0\ell_1}\,.
\label{MSCT}
\ee
Under the SCTs, $\zeta$ transforms non-linearly while $\gamma_{i\ell_0}$ transforms linearly. In particular,~(\ref{Mtrans}) is
automatically satisfied, for $M_{i\ell_0\ell_1}^{\rm SCT} + M_{\ell_0 i \ell_1}^{\rm SCT} - \frac{2}{3}\delta_{i\ell_0} M_{\ell\ell\ell_1}^{\rm SCT} = 0$.
The non-linear shift in $\zeta$ describes in real space the generation of a linear-gradient profile for the curvature perturbation.

The remaining symmetries are ``tensor" symmetries, 
\be
M_{i \ell\ell} = M_{\ell i \ell} = 0\qquad (n=1~{\rm tensor}~{\rm symmetries}) \,,
\label{n=1Mtensor}
\ee
under which $\zeta$ transforms linearly while $\gamma$ transforms non-linearly. The shift in $\gamma$ describes the
generation of a linear-gradient tensor mode. The transversality condition~\eqref{Mtrans} requires
\be
\hat{q}^i\left(M_{i\ell_0\ell_1}(\vec{q}) + M_{\ell_0 i\ell_1}(\vec{q}) \right) = 0 \,.
\ee
This imposes 8 conditions on the 12 algebraically independent components satisfying~(\ref{n=1Mtensor}), leaving us with 4 tensor symmetries.
Therefore, for $n=1$ we have in total 7 physical symmetries (3 SCTs and 4 tensor linear-gradient transformations).

\end{itemize}

For $n\geq 2$, the counting of symmetries works as follows. The array $M_{i\ell_0\cdots \ell_n}$, being symmetric in its last $n+1$ indices, starts out with
$\frac{3}{2}(n+3)(n+2)$ algebraically independent components. The trace constraint~(\ref{Mcond}) gives
$\frac{3}{2}(n+1)n$ conditions, while the transversality condition~(\ref{Mtrans}) imposes $3(2n+1)$ relations on the coefficients.
In total, we therefore have
\be
\frac{3}{2} (n+3)(n+2) - \frac{3}{2}(n+1)n - 3(2n+1) = 6~{\rm symmetries}
\label{counting}
\ee
at each order $n \geq 2$. These include 4 tensor symmetries, under which only $\gamma$ transforms non-linearly, plus
2 ``mixed" symmetries, under which both $\zeta$ and $\gamma$ transform non-linearly. 

\subsection{Physical interpretation}

To shed light on the physical origin of these symmetries, consider expanding the metric about the origin
\be
h_{ij} = \sum_{n=0}^\infty \frac{1}{n!}H_{ij\ell_1\cdots\ell_n} x^{\ell_1}\cdots x^{\ell_n}\,,
\ee
where the array $H_{ij\ell_1\cdots\ell_n}$ is constant, and is symmetric both in its first 2 indices and its last $n$ indices. 
At linear order, the metric is given by $h_{ij} = (1+2\zeta) \delta_{ij} + \gamma_{ij}$. In particular, $\gamma_{ij} =  h_{ij} - \frac{1}{3}\delta_{ij} h^\ell_{\;\ell}$ is transverse,
which implies the comoving gauge condition
\be
\partial^j h_{ij} = \frac{1}{3}\partial_i h^\ell_{\; \ell} \,.
\label{gp}
\ee
This translates into a trace condition on the coefficients
\be
H_{i\ell\ell \ell_2\cdots \ell_n} = \frac{1}{3} H_{\ell\ell i \ell_2\cdots\ell_n}\,.
\label{Hcond}
\ee
(As a check, for the pure gauge configuration $h_{ij} = \partial_i\bar{\xi}_j  + \partial_j\bar{\xi}_i = \sum_{n=0}^\infty \frac{1}{n!} (M_{ij \ell_1\cdots\ell_n} + M_{ji \ell_1\cdots\ell_n})x^{\ell_1}\cdots x^{\ell_n}$, this trace condition reproduces~(\ref{Mcond}).) 

As before, we restrict our attention to metric configurations which can be smoothly extended to a physical profile with suitable fall-off behavior at spatial infinity.
In momentum space, the tensor profile is 
\be
\gamma_{ij}(\vec{q}) = \sum_{n=0}^\infty \frac{(-i)^n}{n!}\left( H_{ij\ell_1\cdots\ell_n}  - \frac{1}{3}\delta_{ij}  H_{\ell\ell\ell_1\cdots\ell_n}\right) \frac{\partial^n}{\partial q_{\ell_1}\cdots \partial q_{\ell_n}} \bigg((2\pi)^3\delta^3(\vec{q})\bigg)\,.
\ee
To enforce the finite-momentum transversality condition $\hat{q}^i \gamma_{ij}(\vec{q}) = 0$, the $H_{ij\ell_1\cdots\ell_n}$ must become $\hat{q}$-dependent such that
\be
\hat{q}^i \left(H_{ij\ell_1\cdots\ell_n}(\hat{q})  - \frac{1}{3}\delta_{ij}  H_{\ell\ell\ell_1\cdots\ell_n}(\hat{q}) \right) = 0\,.
\label{Htrans}
\ee
This is the generalization of~(\ref{Mtrans}).

The counting works as follows. The array $H_{ij\ell_1\cdots\ell_n}$, being symmetric in its first 2 and last $n$ indices, starts out with $3(n+1)(n+2)$ algebraically independent components.
The trace condition~(\ref{Hcond}) gives $\frac{3}{2}(n+1)n$ conditions, while the transversality condition~(\ref{Htrans}) imposes additional relations on the coefficients.
At $n=0$ and $n=1$, we find that the number of $H$ coefficients matches the number of $M$'s, respectively 3 and 7. This reflects the fact that $h_{ij}$ and its first-derivatives can be made trivial at a point, which defines Riemann normal coordinates. At $n=2$, $H$ has 12 components, while recall that $M$ only has 6. This leaves us with 6 physical components, which matches the number of algebraically independent components of the Riemann tensor $R_{ikj\ell}$ in 3 dimensions.\footnote{For $n\geq 3$, we find fewer components than the most general coefficients in the Riemann normal expansion, which are given by derivatives of Riemann. For instance, at $n=3$, $H$ has 18 components, while $M$ again has 6, leaving us with 12 physical components. This is 3 fewer than the number of independent $\partial_m R_{ik j\ell}$ coefficients in the Riemann normal expansion. Although we have not checked this in detail, this must be because the metric profile we are considering is not the most general, but is restricted by the adiabatic transversality condition~(\ref{Htrans}). This must translate in Riemann normal coordinates as adiabatic transversality constraints on derivatives of the Riemann tensor.}

\subsection{Noether charges}
\label{noethercharge}

The Noether charge associated with the symmetry transformation
$(\delta\zeta,  \delta\gamma_{ij})$ is:
\be
Q = \frac{1}{2} \int {\rm d}^3x \bigg( \{\Pi_\zeta(x), \delta \zeta(x) \} + \{\Pi_\gamma^{ij}(x), \delta \gamma_{ij}(x) \}\bigg)\,,
\label{Qdef1}
\ee
where $\Pi_\zeta \equiv \delta{\cal L}/\delta \dot{\zeta}$, $\Pi_\gamma^{ij}\equiv \delta{\cal L}/\delta \dot{\gamma}_{ij} $ are the canonical momenta,
and where the anti-commutator, denoted by $\{\ ,\ \}$, makes $Q$
Hermitian in the quantum theory. The integration over an infinite
volume may be divergent and the charge undefined, just as in theories
with massless Goldstone bosons, but the commutator of $Q$ with local fields is well-defined. 
In particular, it is straightforward to check with the above definition that the charge generates the symmetry transformation, $[Q,\zeta] = -i\delta\zeta$ and $[Q,\gamma_{ij}] =
-i\delta\gamma_{ij}$, as it should.
In practice, we will regularize the charge by
turning the integral over volume into a Fourier space
quantity with the associated momentum understood
to be taken to zero eventually.

For the symmetry~(\ref{xigen}), the Noether charge $Q$ takes the form
\be
Q = \bar{Q} + \Delta\bar{Q} \int^t \frac{{\rm d}t'}{H(t')}\,, 
\label{Qmaindef}
\ee
where $\bar{Q}$ and $\Delta\bar{Q}$ have no explicit time dependence. 
$Q$ has explicit time dependence, but since it is a symmetry its total time derivative
vanishes. We will be particularly interested in
the part of these operators that generate the non-linear field transformations
\bea
\nonumber
\bar{Q}_0 &=& \int {\rm d}^3 x  \, \partial_i \bar{\xi}_j(\vec{x})
\left( \frac{1}{3}\delta^{ij} \Pi_\zeta(\vec{x}) + 2
  \Pi_\gamma^{ij}(\vec{x})\right) = i  \int {{\rm d}^3 q  \over
  (2\pi)^3} \, q_i \bar{\xi}_j(-\vec{q})  \left( \frac{1}{3}\delta^{ij} \Pi_\zeta(\vec{q}) + 2 \Pi_\gamma^{ij}(\vec{q})\right)\,;\\
\Delta\bar{Q}_{0} &=& \int {\rm d}^3 x  \,
\left(\vec{\nabla}^2 \partial_i \bar{\xi}_j(\vec{x})\right)
2\Pi_\gamma^{ij}(\vec{x})  =   - i  \int {{\rm d}^3 q \over (2\pi)^3} \, q^2 q_i \bar{\xi}_j(-\vec{q})  2\Pi_\gamma^{ij}(\vec{q}) \,.
\label{Q0reg}
\eea
For the diffeomorphisms $\bar{\xi}_i^{(n)}$ in the Taylor expansion~(\ref{xiexpansion}), the generators $\bar{Q}_0^{(n)}$ are explicitly given by:
\be
\bar{Q}^{(n)}_0 =  \lim_{\vec{q}\rightarrow 0} \frac{(-i)^n}{n!} M_{i\ell_0 \ldots \ell_n} \frac{\partial^{n}}{\partial q_{\ell_1}\cdots \partial q_{\ell_n}} \left( \frac{1}{3} \delta^{i\ell_0} \Pi_\zeta(\vec{q}) + 2\Pi_\gamma^{i\ell_0}(\vec{q}) \right)\,,
\label{Q0reg2}
\ee
where we have used $\Pi^{\,i}_{\gamma\; i} = 0$. 

It is worth emphasizing that $\Delta \bar Q$
is associated with the diffeomorphism
$\xi^i = \vec \nabla^2 \bar\xi^i$, with
$\bar\xi$ satisfying $\vec\nabla^2\bar\xi_i + \partial_i\partial^j
\bar\xi_j/3 = 0$ (see (\ref{cond0bar}) \& (\ref{xigen})), implying
the diffeomorphism $\xi^i = \vec \nabla^2 \bar\xi^i$ 
is divergence free and is thus a tensor symmetry.
As such, by the argument at the end of Sec. \ref{adiab},
the diffeomorphism receives no time-dependent correction
and $\Delta \bar Q$ by itself is a good conserved charge.

\section{Inflationary Consistency Relations as Ward Identities}
\label{wardgen}

In this Section we derive the Ward identities associated with the non-linearly realized symmetries identified above.
Analogously to the low-energy theorems for pions~\cite{Adler:1964um,Weinberg:1966kf}, the Ward identities constrain the soft limit of various inflationary
correlation functions. As particular cases of these, we will recover in Sec.~\ref{knownegs} the standard consistency
relations~\cite{Maldacena:2002vr,Creminelli:2004yq,Cheung:2007sv}, which determine the $q^0$ and $q$ behavior of the soft limits
in terms of lower-order correlation functions. Beyond these, the Ward identities yield an infinite network of further
consistency relations, which at each order partially constrain the $q^n$ behavior of correlation functions.

The Ward identities are obtained by taking the in-in vacuum expectation value of the action of the charges
\be
\langle \Omega \vert [Q, {\cal O}]  \vert\Omega \rangle =  -i  \langle \Omega \vert  \delta {\cal O} \vert\Omega \rangle  \,,
\label{wardbegin}
\ee
where $|\Omega\rangle$ is the (Heisenberg picture) in-vacuum of the interacting theory, and ${\cal O}(\vec{k}_1,\ldots,\vec{k}_N)$ denotes an equal-time product of $N$ scalar and tensor fields:\footnote{Since this is an equal-time product, the fields all commute with each other, hence their ordering is irrelevant.}
\be
{\cal O}(\vec{k}_1, \ldots,\vec{k}_N)  = {\cal O}^\zeta(\vec{k}_1,\ldots,\vec{k}_M) \cdot {\cal O}^\gamma_{i_{M+1} j_{M+1},\ldots,i_Nj_N} (\vec{k}_{M+1},\ldots,\vec{k}_N) \,,
\label{Aproddef}
\ee
where $0 \leq M \leq N$, with ${\cal O}^\zeta \equiv \prod_{a=1}^{M}\zeta(\vec{k}_a,t)$ and ${\cal O}^\gamma_{i_{M+1} j_{M+1},\ldots,i_Nj_N}  \equiv \prod_{b=M+ 1}^{N}\gamma_{i_bj_b}(\vec{k}_b,t)$.
Note that the tensor indices are arbitrary --- a subset of these can be contracted among themselves or not. To avoid cluttering the notation, 
we will refrain from explicitly writing these indices unless
necessary.
We work within the Heisenberg picture unless stated otherwise.

\subsection{Time-independent identity}

Substituting the general expression~(\ref{Qmaindef}) for $Q$, the Ward identity becomes
\be
\langle \Omega \vert [\bar{Q}, {\cal O}]  \vert\Omega \rangle +  \langle \Omega \vert [\Delta\bar{Q}, {\cal O}]  \vert\Omega \rangle  \int^t \frac{{\rm d}t'}{H(t')} =  -i  \langle \Omega \vert  \bar{\delta} {\cal O} \vert\Omega \rangle   -i  \langle \Omega \vert  \bar{\delta}_{\Delta} {\cal O} \vert\Omega \rangle \int^t \frac{{\rm d}t'}{H(t')}   \,,
\label{wardbegin2}
\ee
Although this explicitly depends on time, we now argue that the identity holds for the time-independent components $\bar{Q}$ and $\Delta\bar{Q}$ separately.
The argument is straightforward. First, consider a time-independent tensor symmetry $\bar{\xi}_{\rm T}$, {\it i.e.}, one under which $\gamma_{ij}$ shifts non-linearly
while $\zeta$ transforms linearly. By definition, this satisfies $\partial_i \xi^i_{\rm T} = \vec{\nabla}^2 \xi^i_{\rm T} = 0$.  As discussed at the end of Sec.~\ref{adiab}, such a diffeomorphism receives no time-dependent correction and represents an honest-to-goodness symmetry. The corresponding charge $Q = \bar{Q}_{\rm T}$ therefore satisfies
\be
\langle \Omega \vert [\bar{Q}_{\rm T}, {\cal O}]  \vert\Omega \rangle =  -i  \langle \Omega \vert  \bar{\delta}_{\rm T} {\cal O} \vert\Omega \rangle  \,.
\label{wardtensor}
\ee
In particular, since the generator $\Delta\bar{Q}$ appearing in~(\ref{wardbegin2}) is itself a tensor symmetry, as argued in Sec.~\ref{adiab}, we have
\be
\langle \Omega \vert [\Delta\bar{Q}, {\cal O}]  \vert\Omega \rangle =  -i  \langle \Omega \vert  \bar{\delta}_{\Delta} {\cal O} \vert\Omega \rangle  \,.
\label{wardDeltensor}
\ee
The time-dependent pieces in~\eqref{wardbegin2} therefore cancel out, leaving us with
\be
\langle \Omega \vert [\bar{Q}, {\cal O}]  \vert\Omega \rangle =  -i  \langle \Omega \vert  \bar{\delta} {\cal O} \vert\Omega \rangle \,.
\label{wardbar}
\ee
Thus the Ward identity holds for any time-independent spatial diffeomorphism $\bar{\xi}^i$, subject to the gauge-preserving condition~\eqref{xieqn0}.
In particular, it holds for each $\bar{\xi}^{(n)}$ in the Taylor expansion~(\ref{xiexpansion}).

\subsection{The left-hand side}
\label{LHSsec}

We begin with the left-hand side
of (\ref{wardbar}):
\begin{eqnarray}
\label{lhsWard}
\langle \Omega \vert [\bar{Q}, {\cal O}]  \vert\Omega \rangle =
- 2 i {\,\rm Im} \langle \Omega \vert {\cal O} \bar{Q} \vert\Omega
\rangle \, ,
\end{eqnarray}
where we have used the fact that $\bar Q$ and ${\cal O}$ are hermitian. 
The operators are assumed to be evaluated at some (late) time $t$.
For the most part, we suppress this $t$ dependence,
but its presence will be important in some of our arguments below.

The in-vacuum $\vert \Omega \rangle$
of the full interacting theory ({\it i.e.}, the Bunch-Davies vacuum)
is related to the free vacuum $\vert 0 \rangle$ by:
\begin{eqnarray}
\label{fullfreeVac}
\vert \Omega\rangle = \Omega(-\infty) \vert 0 \rangle \,,
\end{eqnarray}
where
\be
\Omega(t_i) \equiv U^\dagger (t_i,0) U_0(t_i,0)\,,
\ee
with $U$ and $U_0$ denoting respectively the full and free
time evolution operators. This kind of statement should strictly speaking be understood
in the context of a wave-packet (see \cite{weinbergQFT}).
In a similar way:
\begin{eqnarray}
\label{QUU0}
  \Omega^\dagger(-\infty) 
\, \bar Q  \,  \Omega(-\infty) 
= \bar Q_0\,,
\end{eqnarray}
where $\bar Q_0$ is the free part of $\bar Q$,
and it generates transformations in $\zeta$ and $\gamma$
that are independent of the fields i.e. only the nonlinear transformations
(\ref{Q0reg}), since these are the symmetries of the action of quadratic fluctuations.
This statement, that a Heisenberg operator sandwiched between 
$ \Omega(-\infty)$ and its inverse equals
its free part, is strictly true for a general operator only if it is
evaluated in the far past, for appropriate asymptotic states \cite{weinbergQFT}. 
However, if the operator happens to be independent of time,
the time of evaluation becomes inconsequential.
For our argument here, we treat $\bar Q$ as time independent,
and apply (\ref{QUU0}) with $\bar Q$ evaluated at some time $t$.
In reality this is not quite correct, and a full justification for our procedure
is provided in Appendix \ref{chargeAppendix}.

Thus, we have
\begin{eqnarray}
\label{QUU1}
\bar Q |\Omega \rangle 
=  \Omega(-\infty)  \bar Q_0 | 0 \rangle \, .
\end{eqnarray}
The action of $\bar Q_0$ on the free vacuum $| 0 \rangle$ can be worked out by
using wave-functionals
\cite{Guth:1985ya,Guven:1987bx}. 
Let us use $\vert \zeta_0, \gamma_0 \rangle$ to
denote eigenstates of the {\it free} Heisenberg operators $\zeta_0$
and $\gamma_0$ (suppressing the indices on $\gamma_0$ to avoid clutter).
Inserting a complete set of eigenstates, and using (\ref{Q0reg}), we
find:\footnote{
Focusing on scalar perturbations for simplicity,
the general relation in real space for the momentum
acting on an arbitrary state $|s \rangle$ is
$\Pi_\zeta (\vec x) |s \rangle = \int D\zeta |\zeta \rangle [-i
\delta/\delta\zeta(\vec x)] \langle \zeta | s \rangle$.
Fourier transforming into $\Pi_\zeta (\vec q) 
= \int {\rm d}^3 x \Pi_\zeta (\vec x) e^{i \vec q \cdot \vec x}$ thus
gives $\Pi_\zeta (\vec q) = \int D\zeta | \zeta \rangle
[-i \delta/\delta\zeta(-\vec q)] \langle \zeta | s \rangle$. 
Similarly, 
$\langle s | \Pi_\zeta (\vec x) = \int D\zeta [i
\delta/\delta\zeta(\vec x) ] \langle s | \zeta \rangle \langle \zeta |
$ implies
$\langle s | \Pi_\zeta (\vec q) = \int D\zeta [i
\delta /\delta\zeta(-\vec q)] \langle s | \zeta \rangle \langle \zeta
|$.
}
\begin{eqnarray}
\label{wavefunc0}
\bar Q_0 |0 \rangle &=& \int D\zeta_0 D\gamma_0 \, \bar Q_0 \vert
\zeta_0, \gamma_0 \rangle \langle \zeta_0, \gamma_0 \vert 0 \rangle
\nonumber \\
&=& i \int {{\rm d}^3 q \over (2\pi)^3} q_i \bar\xi_j (-\vec q) \int D\zeta_0 D\gamma_0 \,
\left( {1\over 3} \delta^{ij} \Pi_{\zeta_0}(\vec q) + 2 \Pi^{ij}_{\gamma_0}
  (\vec q) \right) \vert
\zeta_0, \gamma_0 \rangle \langle \zeta_0, \gamma_0 \vert 0 \rangle
\nonumber \\
&=& i \int {{\rm d}^3 q \over (2\pi)^3} q_i \bar\xi_j (-\vec q) \int D\zeta_0 D\gamma_0 \,
\vert \zeta_0, \gamma_0 \rangle
\left( - {1\over 3} \delta^{ij} i {\delta \over \delta \zeta_0 (-\vec q)}
- 2 i {\delta \over \delta \gamma_0 {}_{ij} (-\vec q)}
\right) \langle \zeta_0, \gamma_0 \vert 0 \rangle \, .
\end{eqnarray}
The free vacuum wavefunctional $\langle \zeta_0, \gamma_0 | 0 \rangle$
takes the Gaussian form:
\begin{eqnarray}
\label{wavefunc1}
\langle \zeta_0, \gamma_0 | 0 \rangle = {\cal N}
\exp\left[ - \int \frac{{\rm d}^3k}{(2\pi)^3} \,\left(\frac{1}{2}
    \zeta_0(\vec{k})D_\zeta(k)\zeta_0(-\vec{k}) +
    \frac{1}{4}\gamma_0 {}_{ij} (\vec{k}) D_\gamma(k) \gamma_0 {}^{ij}(-\vec{k})\right) \right]\,,
\label{Psi0}
\end{eqnarray}
where ${\cal N}$ is an irrelevant normalization.
The real part of the kernels $D$ is fixed by the power spectra\footnote{See Appendix~\ref{convs} for details on our conventions for the power spectra.}
\be
\label{DzetaDgamma}
{\rm Re} D_\zeta (k) = \frac{1}{2P_\zeta(k)}\,;\qquad {\rm Re} D_\gamma(k) = \frac{1}{2P_\gamma(k)}\,.
\ee
Meanwhile, the imaginary part is related to the real part by the
Schr\"odinger's equation on the wavefunctional~\cite{Guven:1987bx}, though we do not need its explicit form.
Taking functional derivatives of the free vacuum 
wavefunctional we have
\begin{eqnarray}
\label{Qbar00}
\bar Q_0 | 0 \rangle = - \int {{\rm d}^3 q \over (2\pi)^3} q_i \bar \xi_j
(-\vec q) \left( {1\over 3} D_{\zeta} (q) \delta^{ij} \zeta_0 (\vec q)
+ D_{\gamma} (q) \gamma_0 {}^{ij} (\vec q) \right) \vert 0 \rangle \, .
\end{eqnarray}

At this point we would like to repeat (in reverse) the same argument as in~(\ref{QUU1}), so that we can
pull the {\it free} $\zeta_0 (\vec q)$ and $\gamma_0 (\vec q)$ through 
$\Omega(-\infty)$ to obtain the {\it full} $\zeta (\vec q)$ and
$\gamma (\vec q)$. Recall that this argument strictly works only for
(Heisenberg) operators in the far past and between appropriate asymptotic states, {\it i.e.}, focusing on $\zeta$ for the moment:
\begin{eqnarray}
\label{tiPast}
\lim_{t_i \rightarrow -\infty} \Omega^\dagger (t_i) \zeta
(\vec q, t_i) \Omega(t_i) = \lim_{t_i \rightarrow -\infty} \zeta_0
(\vec q, t_i) \, ,
\end{eqnarray}
whereas what interests us are $\zeta$ 
and $\zeta_0$ in the far future
(which we
denote by $\zeta(\vec q)$ and $\zeta_0(\vec q)$ without any time argument).
What saves us is that, ultimately, we are interested in the 
small-$q$ limit, because the diffeomorphism in momentum space
$\bar\xi_j$ (in \ref{Qbar00}) contains derivatives of the
delta function $\delta^3 (\vec q)$. 
For any given $t_i$, the following holds:
\begin{eqnarray}
\label{zetaconstant}
\zeta(\vec q, t_i) = \zeta(\vec q) \left[ 1 + \Delta \right]
\quad , \quad
\zeta_0(\vec q, t_i) = \zeta_0(\vec q) \left[ 1 + \Delta_0 \right] \quad {\,\rm where} \quad
\lim_{q|\tau_i| \rightarrow 0} \Delta, \Delta_0 = 0 \, ,
\end{eqnarray}
{\it i.e.}, $\zeta$ and $\zeta_0$ approach a constant in the long-wavelength/super-horizon limit.
Here, $\tau_i$ is the conformal time, to be distinguished from the
proper time $t_i$; the combination $q |\tau_i|$ serves to compare
the wavelength with the horizon.\footnote{Note that 
$t \rightarrow -\infty$ corresponds to $\tau \rightarrow
-\infty$, while $t \rightarrow \infty$ corresponds to $\tau
\rightarrow 0$ in de Sitter space.}
The constancy of $\zeta$ in the low momentum limit has been
shown quantum-mechanically in~\cite{Assassi:2012et,Senatore:2012ya}.
Plugging (\ref{zetaconstant}) into (\ref{tiPast}), we conclude that
\begin{eqnarray}
\label{trickPrecise}
\lim_{t_i \rightarrow -\infty,~q|\tau_i| \rightarrow 0}
\Omega^\dagger (t_i) \zeta
(\vec q) \Omega(t_i) = \zeta_0 (\vec q) \, .
\end{eqnarray}
An analogous expression holds for the tensor mode $\gamma$ as well.\footnote{Note the somewhat intrincate limit implicit in
  (\ref{trickPrecise}): $q$ should be sent to zero before $t_i$ is
  sent to $-\infty$.}
Thus, combining (\ref{QUU1}) and (\ref{Qbar00}), we find
\begin{eqnarray}
\label{Qbarfull}
\bar Q \vert \Omega \rangle = - \int {{\rm d}^3 q \over (2\pi)^3} q_i \bar \xi_j
(-\vec q) \left( {1\over 3} D_{\zeta} (q) \delta^{ij} \zeta (\vec q)
+ D_{\gamma} (q) \gamma {}^{ij} (\vec q) \right) \vert \Omega \rangle
\, , 
\end{eqnarray}
where we have used $\Omega(-\infty) \zeta_0 (\vec q) =
\zeta (\vec q) \Omega(-\infty)$, 
whose precise meaning should be
understood
as (\ref{trickPrecise}) (likewise for $\gamma$).\footnote{Recall that $\xi_j(\vec{q})$, being the Fourier transform of~(\ref{xiexpansion}), is given by $q$-derivatives of $\delta^3(\vec{q})$, hence the
integrand in~(\ref{Qbarfull}) only has support at $\vec{q}=0$.}
One issue deserves more discussion: is (\ref{trickPrecise})
really adequate, given that we ultimately will
be taking derivatives of $\zeta(\vec q)$ (hidden in $\xi_j(-\vec q)$)
{\it before} sending $q$ to zero?
Let us postpone the discussion to the end of this subsection.

Substituting (\ref{Qbarfull}) into (\ref{lhsWard}), and recognizing that
$\bar\xi_j^* (\vec q) = \bar\xi_j (-\vec q)$, 
$\zeta^\dagger (\vec q) = \zeta (-\vec q)$,
$\gamma^\dagger (\vec q) = \gamma (-\vec q)$, and
$D_\zeta$,
$D_\gamma$ depend on only the magnitude of $\vec q$,
we arrive at
\be
\langle \Omega \vert [\bar{Q}, {\cal O}]  \vert\Omega \rangle = \int
 {{\rm d}^3 q  \over (2\pi)^3} \, q_i \bar{\xi}_j(-\vec{q})
 \Bigg(\frac{\delta^{ij}}{3P_\zeta(q)} \la \Omega| \zeta(\vec{q})
 {\cal O} |\Omega\ra  + \frac{1}{P_\gamma(q)} \la \Omega |
 \gamma^{ij}(\vec{q}){\cal O}|\Omega\ra  \Bigg)\, .
\label{lhsalmostthere}
\ee
We have used the fact that ${\cal O}$ consists
of a bunch of $\zeta$'s and $\gamma$'s at the same
time as $\zeta(\vec q)$ and $\gamma(\vec q)$, and thus
they commute.
For each term in the Taylor expansion~(\ref{xiexpansion}), this gives
\be
\langle \Omega \vert [\bar{Q}^{(n)}, {\cal O}]  \vert\Omega \rangle = \lim_{\vec{q}\rightarrow 0}\frac{(-i)^{n+1}}{n!} M_{i\ell_0 \ldots \ell_n} \frac{\partial^{n}}{\partial q_{\ell_1}\cdots \partial q_{\ell_n}} \Bigg(\frac{1}{P_\gamma(q)} \la \gamma^{i\ell_0}(\vec{q}){\cal O}\ra + \frac{\delta^{i\ell_0}}{3P_\zeta(q)} \la \zeta(\vec{q}) {\cal O} \ra  \Bigg)\,. 
\label{lhsmostgeneral}
\ee
Hence each Ward identity will constrain the $q^n$ behavior of correlation functions in the soft limit.
This is the main result of this subsection.
Let us close by returning to an issue raised earlier.
From (\ref{lhsmostgeneral}), it is evident that what we need is
something stronger than (\ref{trickPrecise}) {\it i.e.}, we need instead:
\begin{eqnarray}
\label{trickPrecise2}
\lim_{t_i \rightarrow -\infty,~q|\tau_i| \rightarrow 0}
\Omega^\dagger (t_i) {\partial_q} {}^n \left[ f (q) \zeta
(\vec q) \right] \Omega(t_i) = {\partial_q} {}^n\left[ f (q) 
\zeta_0 (\vec q) \right] \, ,
\end{eqnarray}
where $f(q)$ is some function of $q$, and $\partial_q {}^n$ represents
some general $n$ derivatives with respect to $q$ (and likewise for
$\gamma$). To justify
this, we start from what we know to be true:
\begin{eqnarray}
\label{tiPast2}
\lim_{t_i \rightarrow -\infty} \Omega^\dagger (t_i) 
\partial_q {}^n [ f(q) \zeta
(\vec q, t_i) ]\Omega(t_i) = \lim_{t_i \rightarrow -\infty} 
\partial_q {}^n [ f(q)\zeta_0
(\vec q, t_i) ]\, ,
\end{eqnarray}
which follows from (\ref{tiPast}).
Substituting (\ref{zetaconstant}) into
$\partial_q {}^n [f(q) \zeta(\vec q, t_i)]$, we have
\begin{eqnarray}
&& \partial_q {}^n [f(q) \zeta(\vec q, t_i)] \sim
\partial_q {}^n [f(q) \zeta(\vec q)] + \partial_q {}^n [f(q)
\zeta(\vec q)] \Delta + \partial_q {}^{n-1} [f(q)
\zeta(\vec q)] \partial_q \Delta \nonumber \\ 
&& \quad \quad + \partial_q {}^{n-2} [f(q)
\zeta(\vec q)] \partial_q {}^2 \Delta + ... \, ,
\end{eqnarray}
where we have been cavalier about numerical coefficients.
The important point is that 
$\partial_q {}^{n-m} [f(q) \zeta(\vec q)] \sim q^{-m} 
\partial_q {}^n [f(q) \zeta(\vec q)]$, and thus as long
as 
\begin{eqnarray}
\label{dDelta}
\lim_{q|\tau_i| \rightarrow 0} q^m \partial_q {}^m \Delta = 0 \quad
{\rm for} \quad m \ge 0
\end{eqnarray}
(and likewise for $\Delta_0$), all the corrections from $\Delta,
\Delta_0$ and their derivatives drop out in the long-wavelength limit, 
and (\ref{trickPrecise2}) is established.
Therefore, the precise long-wavelength-constancy 
requirement on $\zeta$ is (\ref{zetaconstant})
supplemented by (\ref{dDelta}).
It is worth noting that (\ref{dDelta}) is obeyed by many different
possible $\Delta$'s. For instance,
$1 + \Delta = (1 + i q\tau_i) e^{-iq\tau_i}$ ({\it i.e.}, Hankel function)
gives $\Delta \sim (q\tau_i)^2 /2 + ...$ in the long-wavelength limit,
while $1 + \Delta = e^{iq\tau_i}$ gives $\Delta \sim i q \tau_i +
...$. Both satisfy (\ref{dDelta}). 
The former describes the behavior in standard inflationary models,
while the latter arise in more unusual scenarios including
ones with no expansion \cite{goldberger}. 
It is important to emphasize that whether $\Delta$ goes like $q$ or
$q^2$ in the soft limit has no bearing on which
of the consistency relations ({\it i.e.}, which $n$ in
(\ref{lhsmostgeneral})) is satisfied. 
As long as (\ref{dDelta}) holds, the consistency relations are
expected to hold for {\it all} $n$. 

\subsection{The right-hand side}
\label{RHSsec}

At each order in $n$, the variation of ${\cal O}$ can be split into a part that includes the non-linear shifts in the fields and a part that includes the linear transformations:
\be
\bar{\delta}^{(n)} {\cal O}  = \bar{\delta}_{\rm non-lin.}^{(n)} {\cal O} + \bar{\delta}_{\rm lin.}^{(n)}{\cal O}\,.
\ee
As shown in Appendix~\ref{connWard}, the non-linear part contributes to disconnected diagrams in the Ward identities
and therefore drops out when considering connected correlation functions. For instance, from~(\ref{delnmom}) the non-linear variation
of $\zeta$ schematically gives contributions of the form
\bea
\nonumber
\la  \bar{\delta}^{(n)}_{\rm non-lin.} {\cal O}  \ra &\sim & 
\lim_{\vec{q}\rightarrow 0} 
\frac{\partial^{n}}{\partial k^{n}_a} \delta^3(\vec q + \vec{k}_a) \la
\zeta(\vec{k}_1) \cdots \zeta(\vec{k}_{a-1})\zeta(\vec{k}_{a+1})\cdots
\zeta(\vec{k}_M) {\cal O}^\gamma (\vec{k}_{M+1},\ldots,\vec{k}_N)\ra  
\eea
describing a 2-point function 
(or more precisely, $\langle \zeta(\vec q) \zeta (\vec k_a) \rangle
/P_\zeta (q)$)
disconnected from an $N-1$ point function,
which is precisely canceled by a disconnected term from the left-hand
side
of the Ward identities.
Focusing on {\it connected} correlators, substitution of the field transformations~(\ref{delnmom})
therefore gives
\be
\begin{split}\label{rhsfinal}
 \langle  \bar{\delta}^{(n)} {\cal O} \rangle_c {=} & -  \frac{(-i)^{n}}{n!} M_{i\ell_0 \ldots \ell_n}  \Bigg\{ \sum_{a=1}^N\Bigg( \delta^{i\ell_0} \frac{\partial^{n-1}}{\partial k_{\ell_1}^a\cdots \partial k_{\ell_n}^a} + \frac{k^{i}_a}{n+1}   \frac{\partial^{n+1}}{\partial k_{\ell_0}^a \cdots \partial k_{\ell_n}^a}\Bigg)\la {\cal O}(\vec{k}_1,\ldots,\vec{k}_N) \ra_c \\
& -\sum_{a=1}^M \Upsilon^{i\ell_0i_aj_a}(\hat{k}_a)  \frac{\partial^{n}}{\partial k_{\ell_1}^a\cdots \partial k_{\ell_n}^a} \la {\cal O}^\zeta(\vec{k}_1,\ldots,\vec{k}_{a-1},\vec{k}_{a+1},\ldots \vec{k}_M)\gamma_{i_aj_a}(\vec{k}_a) {\cal O}^\gamma (\vec{k}_{M+1},\ldots,\vec{k}_N)\ra_c \\
& - \sum_{b=M+1}^N  \Gamma^{i\ell_0\;\;\;\;\;k_b\ell_b}_{\;\;\;\;i_bj_b}(\hat{k}_b)\frac{\partial^{n}}{\partial k_{\ell_1}^b\cdots \partial k_{\ell_n}^b} \la {\cal O}^\zeta(\vec{k}_1,\ldots,\vec{k}_M) {\cal O}^\gamma_{i_{M+1} j_{M+1},\ldots,k_b\ell_b,\ldots i_Nj_N}(\vec{k}_{M+1},\ldots,\vec{k}_N) \ra_c \Bigg\} + \ldots
\end{split}
 \ee
Combining with~(\ref{lhsmostgeneral}), we obtain the connected version of the Ward identities:
\be
\begin{split}\label{wardalmostthere}
& \lim_{\vec{q}\rightarrow 0} M_{i\ell_0 \ldots \ell_n} \frac{\partial^{n}}{\partial q_{\ell_1}\cdots \partial q_{\ell_n}} \Bigg(\frac{1}{P_\gamma(q)} \la \gamma^{i\ell_0}(\vec{q}){\cal O}(\vec{k}_1,\ldots,\vec{k}_N) \ra_c + \frac{\delta^{i\ell_0}}{3P_\zeta(q)} \la  \zeta(\vec{q}) {\cal O}(\vec{k}_1,\ldots,\vec{k}_N) \ra_c  \Bigg) \\
& \quad ~~=  -  M_{i\ell_0 \ldots \ell_n} \Bigg\{ \sum_{a=1}^N \Bigg( \delta^{i\ell_0} \frac{\partial^{n}}{\partial k_{\ell_1}^a\cdots \partial k_{\ell_n}^a} + \frac{k^{i}_a}{n+1}  \frac{\partial^{n+1}}{\partial k_{\ell_0}^a \cdots \partial k_{\ell_n}^a}\Bigg) \la  {\cal O}(\vec{k}_1,\ldots,\vec{k}_N) \ra_c  \\
& \quad~~-\sum_{a=1}^M \Upsilon^{i\ell_0i_aj_a}(\hat{k}_a)  \frac{\partial^{n}}{\partial k_{\ell_1}^a\cdots \partial k_{\ell_n}^a} \la {\cal O}^\zeta(\vec{k}_1,\ldots,\vec{k}_{a-1},\vec{k}_{a+1},\ldots \vec{k}_M)\gamma_{i_aj_a}(\vec{k}_a) {\cal O}^\gamma (\vec{k}_{M+1},\ldots,\vec{k}_N)\ra_c \\& \quad ~~-\sum_{b=M+1}^N  \Gamma^{i\ell_0\;\;\;\;\;k_b\ell_b}_{\;\;\;\;i_bj_b}(\hat{k}_b )\frac{\partial^{n}}{\partial k_{\ell_1}^b\cdots \partial k_{\ell_n}^b} \la {\cal O}^\zeta(\vec{k}_1,\ldots,\vec{k}_M) {\cal O}^\gamma_{i_{M+1} j_{M+1},\ldots,k_b\ell_b,\ldots i_Nj_N}(\vec{k}_{M+1},\ldots,\vec{k}_N) \ra_c\Bigg\} + \ldots 
\end{split}
 \ee
The ellipses indicate terms on the right-hand side that are higher-order in the fields. One further simplification is possible. As shown in Appendix~\ref{deltaremove}, the momentum-conserving delta functions implicit on both sides of~(\ref{wardalmostthere}) can be canceled, leaving us with an identity for ``primed" correlation functions, defined by removing the delta function~\cite{Maldacena:2011nz}
\be
\la {\cal O}(\vec{q}, \vec{k}_1, \ldots ,\vec{k}_N)\ra  = (2\pi)^3\delta^3(\vec{P}) \la {\cal O}(\vec{q}, \vec{k}_1, \ldots ,\vec{k}_N)\ra' \,,
\label{primecor}
\ee
where $\vec{P} \equiv \vec{q} +\vec{k}_1 + \ldots + \vec{k}_N$. The primed correlator is thus defined on shell. It is a function of only $N$ momenta, which can be chosen to be $\vec{k}_1,\ldots, \vec{k}_{N}$. As shown in Appendix~\ref{deltaremove}, the Ward identities in terms of primed correlators are given by
\bea
\nonumber
 & & \lim_{\vec{q}\rightarrow 0} M_{i\ell_0 \ldots \ell_n} \frac{\partial^{n}}{\partial q_{\ell_1}\cdots \partial q_{\ell_n}} \Bigg(\frac{1}{P_\gamma(q)} \la \gamma^{i\ell_0}(\vec{q}){\cal O}(\vec{k}_1,\ldots,\vec{k}_N) \ra_c' + \frac{\delta^{i\ell_0}}{3P_\zeta(q)} \la  \zeta(\vec{q}) {\cal O}(\vec{k}_1,\ldots,\vec{k}_N) \ra_c'  \Bigg) \\
\nonumber
& & ~~ =  - M_{i\ell_0 \ldots \ell_n} \Bigg\{ \sum_{a=1}^N \Bigg( \delta^{i\ell_0} \frac{\partial^{n}}{\partial k_{\ell_1}^a\cdots \partial k_{\ell_n}^a} - \frac{\delta_{n0}}{N}\delta^{i\ell_0}
+ \frac{k^{i}_a}{n+1}  \frac{\partial^{n+1}}{\partial k_{\ell_0}^a \cdots \partial k_{\ell_n}^a}\Bigg) \la  {\cal O}(\vec{k}_1,\ldots,\vec{k}_N) \ra_c'  \\
\nonumber
& & \;\;\;\;~~-\sum_{a=1}^M \Upsilon^{i\ell_0i_aj_a}(\hat{k}_a)  \frac{\partial^{n}}{\partial k_{\ell_1}^a\cdots \partial k_{\ell_n}^a} \la {\cal O}^\zeta(\vec{k}_1,\ldots,\vec{k}_{a-1},\vec{k}_{a+1},\ldots \vec{k}_M)\gamma_{i_aj_a}(\vec{k}_a) {\cal O}^\gamma (\vec{k}_{M+1},\ldots,\vec{k}_N)\ra_c' \\
\nonumber
& & \;\;\;\; ~~ -\sum_{b=M+1}^N  \Gamma^{i\ell_0\;\;\;\;\;k_b\ell_b}_{\;\;\;\;i_bj_b}(\hat{k}_b )\frac{\partial^{n}}{\partial k_{\ell_1}^b\cdots \partial k_{\ell_n}^b}\la {\cal O}^\zeta(\vec{k}_1,\ldots,\vec{k}_M) {\cal O}^\gamma_{i_{M+1} j_{M+1},\ldots,k_b\ell_b,\ldots i_Nj_N}(\vec{k}_{M+1},\ldots,\vec{k}_N) \ra_c'\Bigg\}  \\
& &  \;\;\;\; ~~+ \ldots 
\label{wardalmosttheren>2}
\eea
Apart from the $\delta_{n0}$ term on the second line, this takes the same form as the unprimed identity~(\ref{wardalmostthere}).
The general Ward identities~(\ref{wardalmosttheren>2}) relate $N+1$-point correlation functions with a $\vec{q} = 0$ scalar or tensor insertion to the symmetry transformations of $N$-point correlation functions.

\subsection{Component Ward identities}
\label{physward}

The Ward identities~(\ref{wardalmosttheren>2}) involve the coefficient arrays $M_{i\ell_0 \ldots \ell_n}(\hat{q})$.
 In order for these to arise as the $\vec{q} \rightarrow 0$ limit of correlation functions involving {\it physical} modes, the $M_{i\ell_0 \ldots \ell_n}$ parameters of the transformations must acquire $\vec{q}$-dependence, as discussed in Sec.~\ref{Taylorexpand}. Specifically, the regulated array $M_{i\ell_0 \ldots \ell_n}(\hat{q})$ must satisfy the transversality condition~(\ref{Mtrans}).  For each choice of $M$'s satisfying these conditions as well as the symmetry property \eqref{Mcond}, we can plug into (\ref{wardalmosttheren>2}) and obtain a Ward identity.

Alternatively, we can remove the $M_{i\ell_0 \ldots \ell_n}(\hat{q})$ coefficients from~(\ref{wardalmosttheren>2}) by projecting the indices contracted with the $M$'s onto an appropriate subspace.   We must not only ensure that the resulting identities are symmetric in $(\ell_0,\ldots, \ell_n)$ and consistent with the trace condition~(\ref{Mcond}), but that they are also transverse in the sense of~(\ref{Mtrans}). This can be achieved by introducing operators $P_{i\ell_0\ldots \ell_n j m_0\ldots m_n}(\hat{q)}$ with the following properties:

\begin{enumerate}

\item $P_{i\ell_0\ldots \ell_n j m_0\ldots m_n}$ is symmetric in the $(\ell_0,\ldots,\ell_n)$ indices and in the $(m_0 \ldots m_n)$ indices.

\item $P_{i\ell_0\ldots \ell_n j m_0\ldots m_n}$ is symmetric under the interchange of sets of indices: $P_{i\ell_0\ldots \ell_n  j m_0 \ldots m_n} = P_{j m_0 \ldots m_n i \ell_0\ldots \ell_n}$.

\item For $n\geq 1$, $P_{i\ell_0\ldots \ell_n j m_0\ldots m_n}$ obeys the trace condition~(\ref{Mcond}):
\be
P_{i \ell\ell \ell_2\ldots \ell_n j m_0\ldots m_n}  = - \frac{1}{3} P_{\ell i \ell \ell_2\ldots \ell_n j m_0\ldots m_n}\qquad (n \geq 1) \,.
\label{Ptracecond}
\ee

\item $P_{i\ell_0\ldots \ell_n j m_0\ldots m_n}$ satisfies the transverse condition~(\ref{Mtrans}):
\be
\hat{q}^i  \left( P_{i\ell_0\ell_1 \ldots \ell_n jm_0\ldots m_n}(\hat{q}) + P_{\ell_0 i\ell_1 \ldots \ell_n jm_0\ldots m_n}(\hat{q}) - \frac{2}{3}\delta_{i\ell_0} P_{\ell\ell\ell_1 \ldots \ell_n jm_0\ldots m_n}(\hat{q})\right) = 0\,.
\label{Ptrans}
\ee
\end{enumerate}
In other words, $P$ has the same properties as $M$ under either sets of indices and is symmetric under the interchange of sets of indices.
In Appendix~\ref{projectors}, we will explain how to systematically construct these physical operators and give explicit expressions for
the first few values of $n$.  (As noted in the Appendix, the various projectors obtained using this method will in general form an over-complete set.  This means the identities below will not all be independent, though none will be missing.)

In terms of these operators, the Ward identities~(\ref{wardalmosttheren>2}) become
\be
\boxed{
\begin{split}\label{wardfinaln>2}
 &  \lim_{\vec{q}\rightarrow 0} P_{i\ell_0 \ldots \ell_n j m_0\ldots m_n}(\hat{q}) \frac{\partial^{n}}{\partial q_{m_1}\cdots \partial q_{m_n}} \Bigg(\frac{1}{P_\gamma(q)} \la \gamma^{jm_0}(\vec{q}){\cal O}(\vec{k}_1,\ldots,\vec{k}_N) \ra_c' + \frac{\delta^{jm_0}}{3P_\zeta(q)} \la  \zeta(\vec{q}) {\cal O}(\vec{k}_1,\ldots,\vec{k}_N)\ra_c'  \Bigg) \\
& =  - P_{i\ell_0 \ldots \ell_n j m_0\ldots m_n}(\hat{q}) \Bigg\{ \sum_{a=1}^N \Bigg( \delta^{jm_0} \frac{\partial^{n}}{\partial k_{m_1}^a\cdots \partial k_{m_n}^a} - \frac{\delta_{n0}}{N}\delta^{jm_0} + \frac{k^{j}_a}{n+1}  \frac{\partial^{n+1}}{\partial k_{m_0}^a \cdots \partial k_{m_n}^a}\Bigg) \la  {\cal O}(\vec{k}_1, \ldots, \vec{k}_N)\ra_c'  \\
& ~-\sum_{a=1}^M\Upsilon^{jm_0i_aj_a}(\hat{k}_a)  \frac{\partial^{n}}{\partial k_{m_1}^a\cdots \partial k_{m_n}^a} \la {\cal O}^\zeta(\vec{k}_1,\ldots,\vec{k}_{a-1},\vec{k}_{a+1},\ldots \vec{k}_M)\gamma_{i_aj_a}(\vec{k}_a) {\cal O}^\gamma (\vec{k}_{M+1},\ldots,\vec{k}_N)\ra_c' \\
&  ~-\sum_{b=M+1}^N \Gamma^{jm_0\;\;\;k_b\ell_b}_{\;\;\;\;\;i_bj_b}(\hat{k}_b )\frac{\partial^{n}}{\partial k_{m_1}^b\cdots \partial k_{m_n}^b}\la {\cal O}^\zeta(\vec{k}_1,\ldots,\vec{k}_M) {\cal O}^\gamma_{i_{M+1} j_{M+1},\ldots,k_b\ell_b,\ldots i_Nj_N} (\vec{k}_{M+1},\ldots,\vec{k}_N)\ra_c'\Bigg\} + \ldots \\
\end{split}}
\ee
These component Ward identities are the main result of this paper. A few remarks are in order:

\begin{itemize}

\item At each $n$, the identities~(\ref{wardfinaln>2}) constrains the $q^n$ behavior of the soft limits. At lowest order, $n=0,1$, the $q^0$ and $q$ behavior is completely fixed~\cite{Maldacena:2002vr,Creminelli:2012ed}. At higher order, $n\geq 2$, the Ward identities only constrain part of the $N+1$ correlator in the soft limit.

\item The primed correlation functions are defined in~(\ref{primecor}) as on-shell correlators. Correlators on the left-hand side, such as $\la  \zeta(\vec{q}) {\cal O}(\vec{k}_1,\ldots,\vec{k}_N)\ra'$, are functions of $N$ momenta. Without loss of generality, these can be chosen to be $\vec{q},\vec{k}_1,\ldots, \vec{k}_{N-1}$, in which case one should make the replacement $\vec{k}_N = -\vec{q} -\vec{k}_1-\ldots-\vec{k}_{N-1}$ {\it before} differentiating with respect to $q$. Similarly, the primed correlator $\la {\cal O}(\vec{k}_1,\ldots,\vec{k}_N)\ra'$ on the right-hand side is a function of $N-1$ momenta. Choosing these to be $\vec{k}_1,\ldots, \vec{k}_{N-1}$, one should make the replacement $\vec{k}_N = -\vec{k}_1-\ldots-\vec{k}_{N-1}$ before differentiating with respect to $k_a$, where $a = 1,\ldots,N-1$.

\item The ellipses denote corrections that are higher-order in the fields, as discussed below~(\ref{delnmom}). (As argued in Sec.~\ref{higherorder}, however, the dilation symmetry is exceptional, and the field transformations~(\ref{exactdilationtransfns}) are exact in this case.)

\item The Ward identities~(\ref{wardfinaln>2}) hold independently of whether the hard modes with momenta $\vec{k}_1,\ldots,\vec{k}_N$ are inside or outside the horizon. In particular,
following~\cite{Senatore:2012wy} it would be interesting to generalize the analysis to include correlation functions with time or spatial derivatives acting on the
short modes, as this can be useful for loop
calculations~\cite{Weinberg:2005vy,Weinberg:2006ac,Senatore:2009cf,Senatore:2012nq,Pimentel:2012tw}.

\item Note that even if the operator ${\cal O}$ contains
no $\gamma$, the right hand side does involve $\gamma$
through the third line of (\ref{wardfinaln>2}), which replaces successively
each $\zeta$ by $\gamma$. However, the fourth line would vanish
in such a case.

\item The Ward identities also have implications for correlation functions with a soft internal line, where the sum $\vec{q}\equiv -(\vec{k}_1 + \ldots \vec{k}_M)$ of $M$ of external momenta approaches zero. In this limit, the correlation function is dominated by the exchange of a soft scalar or tensor mode, and the answer factorizes into the product of $N-M + 1$- and $M+ 1$-point correlators 
in the soft limit~\cite{Seery:2008ax,Leblond:2010yq}. For instance, for a product of scalar modes~\cite{Creminelli:2012ed},
\bea
\nonumber
\lim_{\vec{q}\rightarrow 0}  \la {\cal O}^\zeta(\vec{k}_1,\ldots,\vec{k}_N)\ra_c' & = & \lim_{\vec{q}\rightarrow 0}\Bigg( \la \zeta(\vec{q}) {\cal O}^\zeta(\vec{k}_1,\ldots \vec{k}_M)\ra_c' \frac{1}{P_\zeta(q)} \la \zeta(-\vec{q}) {\cal O}^\zeta(\vec{k}_{M+1},\ldots \vec{k}_N)\ra_c' \\
& + & \sum_s \la \gamma^s(\vec{q}) {\cal O}^\zeta(\vec{k}_1,\ldots \vec{k}_M) \ra_c' \frac{1}{P_\gamma(q)} \la \gamma^s(-\vec{q}) {\cal O}^\zeta(\vec{k}_{M+1},\ldots \vec{k}_N) \ra_c'  \Bigg)\,.
\eea
The identities~(\ref{wardfinaln>2}) can be applied to the soft limits of the individual correlators in the products, and thus constrain the form of correlation functions
with soft internal lines.

\end{itemize}

\noindent In the following Sections, we will study special cases of these identities. Specifically, we will show in Sec.~\ref{knownegs} that the $n=0$ identities reproduce Maldacena's original consistency relations for scalars and tensors~\cite{Maldacena:2002vr}. We will also show that the $n=1$ identities reproduce the conformal consistency relation and the linear-gradient tensor consistency relation derived recently in~\cite{Creminelli:2012ed}. In Sec.~\ref{newexamples}, we will show in detail that the $n=2$ identities lead to novel consistency relations, and check their validity with the 3-point graviton correlation function in slow-roll inflation.

\section{Recovering Known Consistency Relations}
\label{knownegs}

In this Section we will recover, as particular cases of our general Ward identities~(\ref{wardfinaln>2}), the known consistency relations for single-field inflation.
Specifically, for $n=0$ we will recover Maldacena's original dilation consistency relation and anisotropic scaling relation~\cite{Maldacena:2002vr}. For $n=1$,
we will reproduce the linear-gradient consistency relations derived recently in~\cite{Creminelli:2012ed}.

\subsection{Dilation consistency relation}

For a dilation, the transverse condition~(\ref{Mtrans}) is trivially satisfied since tensors transform linearly. We can simply go back to~(\ref{wardalmosttheren>2}) and substitute $M_{i\ell_0} = \lambda\delta_{i\ell_0}$ to obtain\footnote{Alternatively, one can work directly with~(\ref{wardfinaln>2}) by substituting $P^{\rm dil.}_{i\ell_0 j m_0} = \frac{1}{3}\delta_{i\ell_0}\delta_{j m_0}$, which has all the desired properties. Tracing the resulting identity over $(i,\ell_0)$ yields~(\ref{warddilation}).} 
\be 
\lim_{\vec{q}\rightarrow 0} \frac{1}{P_\zeta(q)} \la  \zeta(\vec{q}) {\cal O}(\vec{k}_1,\ldots,\vec{k}_N)\ra_c'   =  
- \Bigg( 3(N-1)  +   \sum_{a=1}^N  \vec{k}_a\cdot \frac{\partial}{\partial \vec{k}_a}\Bigg)\la   {\cal O}(\vec{k}_1, \ldots,\vec{k}_N)\ra_c'   \,,
\label{warddilation}
\ee
which is the well-known consistency relation of inflation~\cite{Maldacena:2002vr,Creminelli:2004yq,Cheung:2007sv}. Note that this identity receives no higher-order corrections in the fields, since the dilation field transformations~(\ref{exactdilationtransfns}) are exact. 

\subsection{Anisotropic scaling consistency relation}

For the anisotropic scaling, we substitute in~(\ref{wardfinaln>2}) $M_{i\ell_0} = S_{i\ell_0}$ with $S_{ii} = 0$ and $\hat{q}^iS_{i\ell_0}(\hat{q}) = 0$. For concreteness, let us
specialize to the case where the hard modes are scalars: ${\cal O} = {\cal O}^\zeta= \prod_{a=1}^N\zeta(\vec{k}_a)$. In this case, the Ward identity reduces to
\bea
\nonumber
 \lim_{\vec{q}\rightarrow 0} P_{i\ell_0 jm_0}^{\rm T}(\hat{q}) \frac{1}{P_\gamma(q)} \la \gamma^{j m_0}(\vec{q}){\cal O}^\zeta(\vec{k}_1,\ldots,\vec{k}_N)\ra_c'   = &-& P_{i\ell_0 jm_0}^{\rm T}(\hat{q}) \sum_{a=1}^N \Bigg\{k^{j}_a\frac{\partial}{\partial k_{m_0}^a} \la   {\cal O}^\zeta(\vec{k}_1, \ldots,\vec{k}_N)\ra_c'  \\
 \nonumber
& -&\frac{1}{2}  \la   {\cal O}^\zeta(\vec{k}_1, \ldots,\vec{k}_{a-1},\vec{k}_{a+1},\ldots,\vec{k}_N)\gamma_{jm_0}(\vec{k}_a) \ra_c' \Bigg\} \\ 
&+& \ldots
\label{wardaniso}
\eea
where we have used $\Upsilon^{jm_0i_aj_a}(\hat{k}_a)\gamma_{i_aj_a}(\vec{k}_a) = \frac{1}{2}\gamma_{jm_0}(\vec{k}_a)$.
The required operator $P_{i\ell_0 jm_0}^{\rm T}$ has the same permutation properties as $M$ for each pair of indices, that is, it is 
symmetric and traceless in $(i,\ell_0)$ and $(j,m_0)$. Moreover, it is symmetric under the exchange of pairs of
indices $(i,\ell_0) \leftrightarrow (j,m_0)$, as well as transverse $\hat{q}^i P_{i \ell_0 j m_0}^{\rm T}  = 0$. The operator with these properties and suitably normalized
can be readily inferred:
\be
P_{i\ell_0 jm_0}^{\rm T} = P_{ij}P_{\ell_0 m_0} + P_{i m_0}P_{j\ell_0} - P_{i\ell_0}P_{j m_0}\,,
\label{PTT}
\ee
where $P_{i\ell_0}= \delta_{i\ell_0} - \hat{q}_i\hat{q}_{\ell_0}$ is the transverse projector. (The operator $P_{i\ell_0 jm_0}^{\rm T}$ appears in the completeness relation for the polarization tensors --- see~(\ref{compeps}).) Since $\gamma_{j m_0}$ is itself transverse and traceless, we have $P_{i\ell_0 jm_0}^{\rm T}(\hat{q}) \gamma^{j m_0}(\vec{q}) = 2\gamma_{i\ell_0}(\vec{q})$, and~(\ref{wardaniso})
simplifies to
\bea
\nonumber
\lim_{\vec{q}\rightarrow 0} \frac{1}{P_\gamma(q)} \la \gamma_{i \ell_0}(\vec{q}){\cal O}^\zeta(\vec{k}_1,\ldots,\vec{k}_N)\ra_c'  = &-& \frac{1}{2} P_{i\ell_0 jm_0}^{\rm T}(\hat{q}) \sum_{a=1}^N \Bigg\{k^{j}_a\frac{\partial}{\partial k_{m_0}^a} \la   {\cal O}^\zeta(\vec{k}_1, \ldots,\vec{k}_N)\ra_c'  \\
& -&\frac{1}{2}  \la   {\cal O}^\zeta(\vec{k}_1, \ldots,\vec{k}_{a-1},\vec{k}_{a+1},\ldots,\vec{k}_N)\gamma_{jm_0}(\vec{k}_a) \ra_c'  \Bigg\}+ \ldots
\label{wardaniso2}
\eea
Since this relation is usually expressed in the helicity basis, reviewed in Appendix~\ref{convs}, we can project both sides with the polarization tensor $\epsilon^{s}_{i\ell_0}(q)$.
Using the orthonormality condition~(\ref{ortho}), we obtain
\bea
\nonumber
\lim_{\vec{q}\rightarrow 0} \frac{1}{P_\gamma(q)} \la \gamma^{s}(\vec{q}){\cal O}^\zeta(\vec{k}_1,\ldots,\vec{k}_N)\ra_c'  = &-&\frac{1}{2}  \epsilon_{i\ell_0}^{s}(\hat{q}) \sum_{a=1}^N \Bigg\{k^{i}_a\frac{\partial}{\partial k^{\ell_0}_a} \la   {\cal O}^\zeta(\vec{k}_1, \ldots,\vec{k}_N)\ra_c'  \\
&-& \frac{1}{2}  \la   {\cal O}^\zeta(\vec{k}_1, \ldots,\vec{k}_{a-1},\vec{k}_{a+1},\ldots,\vec{k}_N)\gamma_{i\ell_0}(\vec{k}_a) \ra_c' \Bigg\}+\ldots 
\eea
For $N=2$, the last line vanishes (since $\la \zeta\gamma\ra = 0$), and the result agrees with~\cite{Maldacena:2002vr}.
For general $N$, the above agrees with~\cite{Creminelli:2012ed} to lowest order in the tensors, that is, as long as the last line is neglected.

\subsection{Linear-gradient consistency relations}

It has been argued recently that the order $q$ behavior of soft correlators is also fixed by consistency relations~\cite{Creminelli:2012ed,Creminelli:2011rh}. 
We will show that these relations follow from the $n=1$ Ward identities. To make contact with~\cite{Creminelli:2012ed}, we again specialize to a product
of scalars for the hard modes: ${\cal O}= {\cal O}^\zeta$. 

Recall that at $n=1$ we have a total of 7 symmetries, consisting of 3 SCTs and 4 tensor symmetries. Starting with the SCTs,
we substitute $M_{i\ell_0\ell_1}^{\rm SCT} =  b_{\ell_1} \delta_{i\ell_0} + b_{\ell_0}\delta_{i\ell_1} - b_i\delta_{\ell_0\ell_1}$ in~(\ref{wardalmosttheren>2})
to obtain
\bea
\nonumber
\lim_{\vec{q}\rightarrow 0}  \frac{\partial}{\partial q^i}\Bigg(\frac{1}{P_\zeta(q)} \la  \zeta(\vec{q}) {\cal O}^\zeta(\vec{k}_1,\ldots,\vec{k}_N)\ra_c'  \Bigg) &=& -\frac{1}{2}\sum_{a=1}^{N}\left( 6\frac{\partial}{\partial k^i_a} - k_a^i\frac{\partial^2}{\partial k_a^j\partial k_a^j} + 2 k_a^j\frac{\partial^2}{\partial k_a^j\partial k_a^i}\right)\la {\cal O}^\zeta(\vec{k}_1,\ldots,\vec{k}_N)\ra_c' \\
&+& \ldots \,.
\label{SCTconsistency}
\eea
(Note that the $\Upsilon$ terms in~(\ref{wardalmosttheren>2}) vanish identically in this case.) This agrees with the conformal consistency relation --- see Eq.~(54) of~\cite{Creminelli:2012ed}. Although it was originally believed that the SCTs were restricted to the scalar sector only~\cite{Creminelli:2012ed,Hinterbichler:2012nm}, we have now generalized these transformations to include tensors. In particular, the tensor corrections encoded in the ellipses in~(\ref{SCTconsistency}) can be computed explicitly using the expressions given in Sec.~\ref{higherorder}.

For the tensor symmetries, described by fully traceless $M_{i\ell_0\ell_1}$, the Ward identity~(\ref{wardfinaln>2}) gives
\bea
\nonumber
& & \lim_{\vec{q}\rightarrow 0}P_{i\ell_0\ell_1 jm_0m_1}^{\rm T} (\hat{q}) \frac{\partial}{\partial q^{m_1}}\Bigg(\frac{1}{P_\gamma(q)} \la \gamma^{jm_0}(\vec{q}){\cal O}^\zeta(\vec{k}_1,\ldots,\vec{k}_N) \ra_c' \Bigg) \\
\nonumber 
& & \;\;\;\;\;\;\;\;\;\;\;\;\;\;\;\;\;\; =  -   P_{i\ell_0\ell_1 jm_0m_1}^{\rm T} (\hat{q}) \sum_{a=1}^N \Bigg\{ \frac{1}{2} k_a^{j}\frac{\partial^2}{\partial k^a_{m_0}\partial k^a_{m_1}}  \la {\cal O}^\zeta(\vec{k}_1, \ldots,\vec{k}_N)\ra_c'  \\
& & \;\;\;\;\;\;\;\;\;\;\;\;\;\;\;\;\;\;\;\;\;\;  - \Upsilon^{jm_0i_aj_a}(\hat{k}_a) \frac{\partial}{\partial k_{m_1}^a}   \la   {\cal O}^\zeta(\vec{k}_1, \ldots,\vec{k}_{a-1},\vec{k}_{a+1},\ldots,\vec{k}_N)\gamma_{i\ell_0}(\vec{k}_a) \ra_c' \Bigg\} +\ldots \,,
\label{tensorlinv1}
\eea
where $P_{i\ell_0\ell_1 jm_0m_1}^{\rm T}$ is symmetric in $(\ell_0,\ell_1)$ and $(m_0,m_1)$, fully traceless in $(i,\ell_0,\ell_1)$ and $(j,m_0,m_1)$,
symmetric under the interchange $(i,\ell_0,\ell_1)\leftrightarrow (j,m_0,m_1)$, and transverse:
\be
\hat{q}^i\bigg( P_{i\ell_0\ell_1 jm_0m_1}^{\rm T}(\hat{q}) + P_{\ell_0 i \ell_1 jm_0m_1}^{\rm T}(\hat{q})\bigg) = 0\,.
\ee
As described in Appendix~\ref{projectors}, we may construct three linearly independent operators satisfying these conditions:
\bea
\nonumber
P_{i\ell_0\ell_1 jm_0m_1}^{{\rm T}~(1)}(\hat{q})  &=&  P_{ikm_0m_1}^{\rm T}P_{jk\ell_0\ell_1}^{\rm T} \\
\nonumber
P_{i\ell_0\ell_1 jm_0m_1}^{{\rm T}~(2)}(\hat{q})  &=& P_{ikjm_1}^{\rm T}P_{\ell_0\ell_1m_0k}^{\rm T} + P_{ikm_0j}^{\rm T}P_{\ell_0\ell_1m_1k}^{\rm T} - \frac{2}{3} \left(P_{\ell_0km_0m_1}^{\rm T}P_{i\ell_1 jk}^{\rm T} + P_{\ell_1km_0m_1}^{\rm T}P_{\ell_0ijk}^{\rm T}\right)   \\
\nonumber
P_{i\ell_0\ell_1 jm_0m_1}^{{\rm T}~(3)}(\hat{q})  &=& -\hat{q}^j \left(\hat{q}^{\ell_1} P_{i\ell_0 m_0m_1}^{\rm T} + \hat{q}^{\ell_0}P_{i\ell_1 m_0 m_1 }^{\rm T} - \hat{q}^i P_{\ell_0\ell_1 m_0 m_1}^{\rm T} \right) + \hat{q}^{m_0}  \left(\hat{q}^{\ell_1} P_{i\ell_0 j m_1}^{\rm T}+ \hat{q}^{\ell_0}P_{i\ell_1 j m_1}^{\rm T} - \hat{q}^i P_{\ell_0\ell_1 jm_1 }^{\rm T} \right) \\
& + & \hat{q}^{m_1}  \left(\hat{q}^{\ell_1} P_{i\ell_0 j m_0}^{\rm T} + \hat{q}^{\ell_0}P_{i\ell_1 j m_0}^{\rm T} - \hat{q}^i P_{\ell_0\ell_1 jm_0 }^{\rm T} \right)\,,
\eea
where $P_{i\ell_0 jm_0}^{\rm T}$ is defined in~(\ref{PTT}). To cast~(\ref{tensorlinv1}) in the helicity basis, we focus on $P_{i\ell_0\ell_1 jm_0m_1}^{{\rm T}~(3)}$ and contract both sides with $q_{\ell_1} \epsilon_{i\ell_0}^s(\vec{q})$. Using the orthonormality~(\ref{ortho}) and completeness~(\ref{compeps}) relations of the polarization tensors, the result is
\bea
\nonumber
 \lim_{\vec{q}\rightarrow 0} q^{\ell_1}\frac{\partial}{\partial q^{\ell_1}} \Bigg(\frac{1}{P_\gamma(q)} \la \gamma^{s}(\vec{q}){\cal O}^\zeta(\vec{k}_1,\ldots,\vec{k}_N)\ra_c'\Bigg) = &-& \frac{1}{2} q^{\ell_1}\epsilon^{s}_{i\ell_0}(\vec{q})\sum_{a=1}^N \Bigg\{\left(k_a^i \frac{\partial}{\partial k_a^{\ell_1}} - \frac{k_a^{\ell_1}}{2} \frac{\partial}{\partial k^a_{i}} \right) \frac{\partial}{\partial k^a_{\ell_0}} \la {\cal O}^\zeta(\vec{k}_1,\ldots,\vec{k}_N) \ra_c'  \\
 \nonumber
 & - &  \left(2\Upsilon^{i\ell_0i_aj_a}(\hat{k}_a) \frac{\partial}{\partial k_{\ell_1}^a} -\Upsilon^{\ell_1 i i_aj_a}(\hat{k}_a) \frac{\partial}{\partial k_{\ell_0}^a} \right)  \\
&& ~~~\times  \la   {\cal O}^\zeta(\vec{k}_1, \ldots,\vec{k}_{a-1},\vec{k}_{a+1},\ldots,\vec{k}_N)\gamma_{i\ell_0}(\vec{k}_a) \ra_c' \Bigg\} + \ldots
\eea
This is consistent with~\cite{Creminelli:2012ed} to leading order in the tensors.

\section{Example of Novel Consistency Relation}
\label{newexamples}

The first novel consistency relation arises for $n=2$, which constrains the $q^2$ behavior of correlation functions in the soft limit.
Unlike the $n=0$ and $n=1$ consistency relations of Sec.~\ref{knownegs}, which fully constrain the $q^0$ and $q^1$ behavior
of the correlators, the $n=2$ Ward identities only {\it partially} constrain the $q^2$ behavior of the correlators in the soft limit.

To illustrate this, let us focus on the $n=2$ ``tensor" symmetries, for which $M_{i\ell_0\ell_1 \ell_2}$ is fully
traceless. The corresponding $P_{i\ell_0 \ell_1\ell_2 j m_0m_1 m_2}^{\rm T}$ is symmetric in the $(\ell_0,\ell_1,\ell_2)$ indices and in the  $(m_0,m_1,m_2)$ indices, traceless in the $(i,\ell_0,\ell_1,\ell_2)$ indices and in the $(j,m_0,m_1,m_2)$ indices, and 
symmetric under the exchange $(i,\ell_0,\ell_1,\ell_2) \leftrightarrow (j,m_0,m_1,m_2)$, and transverse in the sense of~(\ref{Ptrans}). See Appendix~\ref{projectors} for a description of how to arrive at operators with these properties. Furthermore, to allow comparison with existing computations, we specialize to the case of $N=2$ hard scalar modes, ${\cal O}(\vec{k}_1,\vec{k}_2) = \zeta_{\vec{k}_1}\zeta_{\vec{k}_2}$. In this case, the $\Upsilon$ terms are absent since $\la \zeta\gamma\ra = 0$.

With these assumptions, the Ward identity becomes
\bea
\nonumber
\lim_{\vec{q}\rightarrow 0} P_{i\ell_0 \ell_1\ell_2 j m_0m_1 m_2}^{\rm T}(\hat{q}) \frac{\partial^{2}}{\partial q_{m_1}\partial q_{m_2}} \Bigg(\frac{1}{P_\gamma(q)} \la \gamma^{jm_0}(\vec{q}) \zeta_{\vec{k}_1}\zeta_{\vec{k}_2}\ra' \Bigg) &=&  - P_{i\ell_0 \ell_1\ell_2 j m_0m_1 m_2}^{\rm T}(\hat{q}) \sum_{a=1}^2 \frac{k^{j}_a}{3}  \frac{\partial^{3}\la  \zeta_{\vec{k}_1}\zeta_{\vec{k}_2}\ra'}{\partial k_{m_0}^a \partial k_{m_1}^a\partial k_{m_2}^a}\,. \\
\label{wardfinalnewn=2}
\eea
Note that we have neglected the ``..." corrections which are higher-order in the fields. In slow-roll inflation, these corrections are higher-order in the slow-roll parameters and hence can be consistently neglected to leading-order in slow-roll. We will check this identity using the three-point function computed by Maldacena~\cite{Maldacena:2002vr}\footnote{To translate from Maldacena's expression in the helicity basis, we multiplied both sides by $\epsilon_{jm_0}^s(\vec{q})$, summed over helicities and used the completeness relation~(\ref{compeps}).}
\be
\frac{1}{P_\gamma(q)} \la \gamma_{i\ell_0}(\vec{q}) \zeta_{\vec{k}_1}\zeta_{\vec{k}_2}\ra'  =P^{\rm T}_{i\ell_0 jm_0}(\hat{q}) \frac{H^2}{4\epsilon k_1^3k_2^3}  k_1^jk_2^{m_0} \left(-K + \frac{(k_1 + k_2)q + k_1k_2}{K} + \frac{qk_1k_2}{K^2}\right) \,,
\label{maldacena3pt}
\ee
where $K \equiv q + k_1 + k_2$, and $P^{\rm T}_{i\ell_0 jm_0}(\hat{q})$ is defined in \eqref{PTT}.   Meanwhile, the scalar two-point function is
\be
\langle  \zeta_{\vec{k}_1}\zeta_{\vec{k}_2}\rangle'  = \frac{H^2}{4\epsilon k_1^3}\,.
\label{maldacena2pt}
\ee
Both~(\ref{maldacena3pt}) and~(\ref{maldacena2pt}) are valid to leading order in slow-roll parameters. Before verifying~(\ref{wardfinalnewn=2}), we quickly check that the lowest-order consistency relations are satisfied.

\begin{itemize}

\item Anisotropic scaling consistency relation ($n=0$):  In the limit $\vec{q}\rightarrow 0$, we have $\vec{k}_2\rightarrow -\vec{k}_1$,
and the left-hand side of~(\ref{wardaniso2}) gives
\be
\lim_{\vec{q}\rightarrow 0} \frac{1}{P_\gamma(q)} \langle  \gamma_{i\ell_0}(\vec{q})  \zeta_{\vec{k}_1}\zeta_{\vec{k}_2}\rangle'  = P_{i\ell_0jm_0}^{\rm T}(\hat{q}) \frac{H^2}{4\epsilon k_1^3} \frac{3}{2} \hat{k}_1^j\hat{k}_1^{m_0}\,.
\label{n=0lhs}
\ee
Differentiating~(\ref{maldacena2pt}), the right-hand side of~(\ref{wardaniso2}) becomes
\be
-  \frac{1}{2} P_{i\ell_0jm_0}^{\rm T}(\hat{q}) \sum_{a=1}^2 k^{j}_a \frac{\partial}{\partial k_{m_0}^a} \langle \zeta_{\vec{k}_1}\zeta_{\vec{k}_2}\rangle'  = P_{i\ell_0jm_0}^{\rm T}(\hat{q}) \frac{H^2}{4\epsilon k_1^3} \frac{3}{2} \hat{k}_1^j\hat{k}_1^{m_0}\,,
\label{n=0rhs}
\ee
which agrees with~(\ref{n=0lhs}). Thus the $n=0$ Ward identity~(\ref{wardaniso2}) is satisfied. 

\item Linear-gradient consistency relation ($n=1$): To compute the $q$-derivative of the three-point function, we must let $\vec{k}_2 = -\vec{k}_1 - \vec{q}$ and work consistently to linear order in $q$: 
\bea
\nonumber
\frac{\partial}{\partial q^{m_1}} \left( \frac{1}{P_\gamma(q)} \langle  \gamma_{jm_0}(\vec{q})  \zeta_{\vec{k}_1}\zeta_{\vec{k}_2}\rangle' \right) &=& \frac{H^2}{4\epsilon k_1^3} \left(\frac{\partial}{\partial q^{m_1}} P_{jm_0 k\ell}^{\rm T}(\hat{q})\right) \frac{3}{2} \hat{k}_1^k\hat{k}_1^\ell \\
\nonumber
&+& \frac{H^2}{4\epsilon k_1^4}  P_{jm_0 k\ell}^{\rm T}(\hat{q}) 3\hat{k}_1^\ell \left(\delta^k_{\;m_1} - \frac{5}{2} \hat{k}_1^k \hat{k}^1_{m_1} \right) \\
&+& ({\rm higher}-{\rm order~in}~q)\,.
\label{n=1lhsv0}
\eea
Next we contract this with $P_{i\ell_0\ell_1 jm_0 m_1}^{\rm T}$. In doing so, we will use the following identities, which straightforwardly follow from the properties of  $P_{i\ell_0\ell_1 jm_0 m_1}^{\rm T}$ and the explicit form~(\ref{PTT}) for $P_{jm_0 k\ell}^{\rm T}$:
\be
P_{i\ell_0\ell_1}^{{\rm T}\;\;\;\; jm_0 m_1} \frac{\partial}{\partial q^{m_1}}P_{jm_0 k\ell}^{\rm T}    = 0\;;\qquad
P_{i\ell_0\ell_1 \;\;\;\; m_1}^{{\rm T}\;\;\; jm_0}  P_{jm_0 k\ell}^{\rm T} = P_{i\ell_0\ell_1k\ell m_1}^{\rm T} + P_{i\ell_0\ell_1\ell k m_1 }^{\rm T}  \,.
\ee
Applying this operator, the left-hand side of~(\ref{tensorlinv1}) becomes
\be
\lim_{\vec{q}\rightarrow 0} P_{i\ell_0\ell_1jm_0m_1}^{\rm T}(\hat{q}) \frac{\partial}{\partial q_{m_1}} \left(\frac{1}{P_\gamma(q)}\langle \gamma^{jm_0 }(\vec{q}) \zeta_{\vec{k}_1}\cdots\zeta_{\vec{k}_N})\rangle'_c\right)
= - \frac{15}{2} \cdot \frac{H^2}{4\epsilon k_1^4} \cdot \hat{k}_1^j \hat{k}_1^{m_0} \hat{k}_1^{m_1} P_{i\ell_0\ell_1jm_0m_1}^{\rm T}(\hat{q}) \,.
\label{n=1lhsv2}
\ee

Meanwhile, on the right-hand side of the identity~(\ref{tensorlinv1}), we have
\be
-\frac{1}{2}P_{i\ell_0\ell_1 jm_0m_1}^{\rm T}(\hat{q}) \sum_{a=1}^2 k_a^j\frac{\partial^2}{\partial k^a_{m_0}\partial k^a_{m_1}} \langle \zeta_{\vec{k}_1}\zeta_{\vec{k}_2}\rangle'   = -\frac{15}{2}\cdot \frac{H^2}{4\epsilon k_1^4} \cdot \hat{k}_1^j \hat{k}_1^{m_0} \hat{k}_1^{m_1} P_{i\ell_0\ell_1jm_0m_1}^{\rm T}(\hat{q}) \,,
\ee
where we have used the fact that $P_{i\ell_0\ell_1 jm_0 m_1}^{\rm T}$ is traceless. This agrees with~(\ref{n=1lhsv2}), which verifies the $n=1$ tensor Ward identity~(\ref{tensorlinv1}). 

\item Novel tensor consistency relation ($n=2$): We now verify that Maldecena's 3-point function~(\ref{maldacena3pt}) satisfies the $n=2$ Ward identity~(\ref{wardfinalnewn=2}). On the left-hand side, we find
\bea
\nonumber
 \frac{\partial^{2}}{\partial q^{m_1}\partial q^{m_2}} \Bigg(\frac{1}{P_\gamma(q)} \la \gamma^{jm_0}(\vec{q}) \zeta_{\vec{k}_1}\zeta_{\vec{k}_2} \ra_c' \Bigg) &=& \frac{H^2}{4\epsilon k_1^3}\Bigg\{ \frac{3}{2}\hat{k}_1^k\hat{k}_1^\ell \frac{\partial^2P_{jm_0k\ell}^{\rm T}(\hat{q}) }{\partial q^{m_1}\partial q^{m_2}}  \\
 \nonumber
& + & \frac{3\hat{k}_1^\ell}{2k_1}\left(\frac{\partial P_{jm_0\ell m_2}^{\rm T}(\hat{q}) }{\partial q^{m_1}} + \frac{\partial P_{jm_0\ell m_1}^{\rm T}(\hat{q}) }{\partial q^{m_2}}\right)\\
\nonumber
 &+ & \frac{15}{4} \frac{\hat{k}_1^k\hat{k}_1^\ell}{k_1}\left(\hat{k}_1^{m_2}\frac{\partial P_{jm_0k\ell}^{\rm T}(\hat{q}) }{\partial q^{m_1}} + \hat{k}_1^{m_1}\frac{\partial P_{jm_0k\ell}^{\rm T}(\hat{q}) }{\partial q^{m_2}} \right) \\
 \nonumber
 &-& \frac{15}{4}\frac{\hat{k}_1^\ell}{k_1^2}\left(\hat{k}_1^{m_2}P_{jm_0\ell m_1}^{\rm T}(\hat{q})  +\hat{k}_1^{m_1}P_{jm_0\ell m_2}^{\rm T}(\hat{q}) \right) \\
 \nonumber
 &-& \frac{5}{2} \frac{\hat{k}_1^k\hat{k}_1^\ell}{k_1^2} P_{jm_0k\ell}^{\rm T}(\hat{q}) \delta_{m_1m_2}  + \frac{35}{2k_1^2} \hat{k}_1^k  \hat{k}_1^\ell \hat{k}_1^{m_1} \hat{k}_1^{m_2} P_{jm_0k\ell}^{\rm T}(\hat{q})   \Bigg\} \\
&+&  ({\rm higher-order~in}~q)\,.
\label{lhsn=2step1}
\eea
When contracting this with $P_{i\ell_0 \ell_1\ell_2 j m_0m_1 m_2}^{\rm T}$, we will use the identities
\bea
\nonumber
P_{i\ell_0\ell_1\ell_2 \;\;\;\; m_1m_2}^{{\rm T}\;\;\;\;\;\;\; jm_0} \frac{\partial^2P_{jm_0k\ell}^{\rm T}}{\partial q^{m_1}\partial q^{m_2}} &=& 0 \;; \qquad  P_{i\ell_0\ell_1\ell_2 \;\;\;\;m_1m_2}^{{\rm T}\;\;\;\;\;\;\; jm_0} \frac{\partial P_{jm_0k\ell}^{\rm T}}{\partial q^{m_1}} = 0\;; \\
P_{i\ell_0\ell_1\ell_2\;\;\;\; m_1m_2}^{{\rm T}\;\;\;\;\;\;\; jm_0}P_{jm_0k\ell}^{\rm T} &=&  P_{i\ell_0\ell_1\ell_2 k\ell m_1m_2}^{\rm T} + P_{i\ell_0\ell_1\ell_2\ell k m_1m_2}^{\rm T}\,,
\eea
which follow from the properties of $P_{i\ell_0 \ell_1\ell_2 j m_0m_1 m_2}^{\rm T}$ and the explicit expression~(\ref{PTT}) for $P_{jm_0 k\ell}^{\rm T}$.
It turns out that only the last term in~(\ref{lhsn=2step1}) gives a non-vanishing contribution:
\be
\lim_{\vec{q}\rightarrow 0} P_{i\ell_0\ell_1\ell_2  j m_0 m_1 m_2}^{\rm T}(\hat{q})  \frac{\partial^{2}}{\partial q_{m_1}\partial q_{m_2}} \Bigg(\frac{1}{P_\gamma(q)} \la \gamma^{jm_0}(\vec{q}) \zeta_{\vec{k}_1}\zeta_{\vec{k}_2} \ra_c' \Bigg)
= \frac{H^2}{4\epsilon k_1^3} \frac{35}{k_1^2}  \hat{k}_1^j  \hat{k}_1^{m_0} \hat{k}_1^{m_1} \hat{k}_1^{m_2} P_{i\ell_0\ell_1\ell_2 jm_0m_1m_2}^{\rm T}(\hat{q})  \,.
\label{n=2lhsfinal}
\ee
Meanwhile, the right-hand side of the Ward identity~(\ref{wardfinalnewn=2}) becomes
\be
- \sum_{a=1}^2 \frac{k^{j}_a}{3}  \frac{\partial^{3}\langle \zeta_{\vec{k}_1} \zeta_{\vec{k}_2} \rangle_c' }{\partial k_{m_0}^a \partial k_{m_1}^a  \partial k_{m_2}^a} 
= \frac{H^2}{4\epsilon k_1^3} \frac{\hat{k}_1^j}{k_1^2} \Bigg[35 \hat{k}_1^{m_0}\hat{k}_1^{m_1}\hat{k}_1^{m_2} -5\left(\hat{k}_1^{m_0}\delta_{m_1m_2} + \hat{k}_1^{m_1}\delta_{m_0m_2} + \hat{k}^{m_2} \delta_{m_0m_1}\right)\Bigg]\,.
\ee
Contracting with $P_{i\ell_0\ell_1\ell_2 km_0m_1m_2}^{\rm T}$, and using the fact that this operator is traceless, we obtain
\be
-P_{i\ell_0\ell_1\ell_2  j m_0 m_1 m_2}^{\rm T}(\hat{q}) \sum_{a=1}^2 \frac{k^{j}_a}{3}  \frac{\partial^{3}\langle \zeta_{\vec{k}_1} \zeta_{\vec{k}_2} \rangle_c'}{\partial k^{m_0}_a \partial k^{m_1}_a  \partial k^{m_2}_a}   =  \frac{H^2}{4\epsilon k_1^3}\frac{35}{k_1^2}  \hat{k}_1^j  \hat{k}_1^{m_0} \hat{k}_1^{m_1} \hat{k}_1^{m_2} P_{i\ell_0\ell_1\ell_2 jm_0m_1m_2}^{\rm T}(\hat{q})  \,.
\ee
This matches~(\ref{n=2lhsfinal}), which verifies the $n=2$ tensor Ward identity.

\end{itemize}

Although we have focused for concreteness on the tensor symmetries in this check, there are of course 2 additional ``mixed" symmetries at $n=2$, under which both scalar and tensor modes shift non-linearly. It will be very interesting to study this identity in detail, particularly in models with reduced sound speed.

\section{Another Non-Trivial Check}
\label{dotzeta3check}

As another check on our identities, consider the contribution to the 3-point amplitude $\langle \zeta\zeta\zeta\rangle$ coming from the $\dot{\zeta}^3$ vertex in the cubic action.\footnote{We thank Daniel Green for suggesting this check.}
This operator is invariant under our symmetries, hence its contribution to the 3-point function should by itself satisfy {\it all} of the consistency relations:
\be
 \lim_{\vec{q}\rightarrow 0} P_{i\ell_0 \ldots \ell_n jj m_1\ldots m_n}(\hat{q}) \frac{\partial^{n}}{\partial q_{m_1}\cdots \partial q_{m_n}} \Bigg(\frac{1}{P_\zeta(q)} \la  \zeta_{\vec{q}} \zeta_{\vec{k}_1}\zeta_{\vec{k}_2}\ra_{\dot{\zeta}^3}' \Bigg)  = 0 \,.
\label{dotzeta3identity}
 \ee
Another way to see this is that in general $P(X,\phi)$ theories (where $X \equiv -\frac{1}{2}(\partial\phi)^2$) the coefficient of $\dot{\zeta}^3$ depends on $P_{,XXX}$, while all $\gamma\zeta\zeta$ vertices depend at most on $P_{,XX}$ but are independent of $P_{,XXX}$. Therefore the validity of the consistency relations cannot possibly rely on a cancellation between $\langle \gamma\zeta\zeta\rangle$ and $\langle \zeta\zeta\zeta\rangle$ in this case --- the scalar 3-point contribution must by itself be annihilated by all of our derivative operators.

The 3-point amplitude from $\dot{\zeta}^3$ takes the simple form~\cite{Chen:2006nt}
\be
\frac{1}{P_\zeta(q)} \la  \zeta_{\vec{q}} \zeta_{\vec{k}_1}\zeta_{\vec{k}_2}\ra_{\dot{\zeta}^3}'  = A\frac{q^2}{k_1k_2K^3}\,,
\ee
where $A$ is a constant, and $K \equiv q + k_1 + k_2$ as before. Being proportional to $q^2$, this clearly satisfies~(\ref{dotzeta3identity}) for $n=0$ and $n=1$. The first non-trivial check arises at $n=2$.

\begin{itemize} 

\item $n=2$ check: As before, to compute $q$-derivatives we work on-shell by setting $\vec{k}_2 = -\vec{k}_1 - \vec{q}$ and expand to the appropriate order in $q$. For $n=2$, this gives
\be
 \frac{\partial^{2}}{\partial q^{m_1}\partial q^{m_2}} \Bigg(\frac{1}{P_\zeta(q)} \la \zeta_{\vec{q}} \zeta_{\vec{k}_1}\zeta_{\vec{k}_2} \ra_{\dot{\zeta}^3}' \Bigg)  = \frac{A}{4k_1^5}\delta_{m_1m_2} + ({\rm higher-order~in}~q)\,.
\ee
Now, the operator $P_{i\ell_0 \ldots \ell_n j m_0 m_1\ldots m_n}$ has vanishing double-trace,
\be
P_{i\ell_0 \ldots \ell_n jj m m m_3\ldots m_n}  = 0\,,
\label{Pdoubletrace}
\ee
which can easily be seen by tracing~(\ref{Ptracecond}) over $(i,\ell_2)$ and using the symmetry under interchange of $(i,\ell_0,\ldots,\ell_n)$ and $(j,m_0,\ldots,m_n)$ sets of indices.
It follows that 
\be
 \lim_{\vec{q}\rightarrow 0} P_{i\ell_0\ell_1\ell_2 jj m_1m_2}(\hat{q}) \frac{\partial^2}{\partial q_{m_1}\partial q_{m_2}} \Bigg(\frac{1}{P_\zeta(q)} \la  \zeta_{\vec{q}} \zeta_{\vec{k}_1}\zeta_{\vec{k}_2}\ra_{\dot{\zeta}^3}' \Bigg)  = 0\,,
\ee
which verifies~(\ref{dotzeta3identity}) for $n=2$.

\item $n=3$ check: At the next order in $q$, we find
\bea
\nonumber
\frac{\partial^{2}}{\partial q^{m_1}\partial q^{m_2}\partial q^{m_3}} \Bigg(\frac{1}{P_\zeta(q)} \la \zeta_{\vec{q}} \zeta_{\vec{k}_1}\zeta_{\vec{k}_2} \ra_{\dot{\zeta}^3}' \Bigg)  &=& -\frac{A}{k_1^6} \left( \frac{5}{8}\hat{k}_1^{m_1} + \frac{9}{16} \hat{q}^{m_1}\right) \delta_{m_2m_3} + {\rm cyclic~perms.} \\
&+& \frac{9}{16} \frac{A}{k_1^6} \hat{q}^{m_1}\hat{q}^{m_2}\hat{q}^{m_3} \,.
\label{3dercheck}
\eea
When contracted with $P_{i\ell_0\ell_1\ell_2 jj m_1m_2}$, the first line on the right-hand side clearly gives vanishing contribution because of~(\ref{Pdoubletrace}).
Meanwhile, combining the trace condition~(\ref{Ptracecond}) and transversality condition~(\ref{Ptrans}), it is straightforward to show that
\be
\hat{q}^{m_1}\hat{q}^{m_2}P_{i\ell_0 \ldots \ell_n jj m_1\ldots m_n} = -P_{i\ell_0 \ldots \ell_n jj m m m_3\ldots m_n}  = 0\,.
\ee
Therefore the second line of~(\ref{3dercheck}) also gives a vanishing contribution, and thus we obtain
\be
 \lim_{\vec{q}\rightarrow 0} P_{i\ell_0\ell_1\ell_2 jj m_1m_2}(\hat{q}) \frac{\partial^2}{\partial q_{m_1}\partial q_{m_2}\partial q_{m_3}} \Bigg(\frac{1}{P_\zeta(q)} \la  \zeta_{\vec{q}} \zeta_{\vec{k}_1}\zeta_{\vec{k}_2}\ra_{\dot{\zeta}^3}' \Bigg)  = 0\,,
\ee
which verifies~(\ref{dotzeta3identity}) for $n=3$.

\end{itemize}

\noindent The argument generalizes to all $n$. Additional $q$-derivatives either generate terms proportional to $\hat{q}^{m_i}\hat{q}^{m_j}$ or $\delta_{m_im_j}$, either of which gives zero when contracted with the $P$ operator.

\section{Conclusion}
\label{conclu}

In this paper we derived an infinite number of consistency relations constraining at each $n\geq 0$ the $q^n$ behavior of cosmological
correlation functions in the soft limit $\vec{q}\rightarrow 0$. We showed how they arise as the Ward identities for an infinite set of global
symmetries, which are non-linearly realized on the perturbations. As the lowest-order identities ($n=0,1$), we recovered Maldacena's original
consistency relations for scalars and tensors~\cite{Maldacena:2002vr}, as well as the recently-discovered linear-gradient consistency relations~\cite{Creminelli:2012ed}.
The higher-order ($n\geq 2$) identities are new, and we checked as a particular example that the $n=2$ ``tensor" identity is satisfied
by known correlation functions in slow-roll inflation. Our general Ward identities hold whether the hard modes are inside or outside the horizon,
and also have implications for correlation functions with soft internal lines.

There are many directions that would be interesting to pursue:

\begin{itemize}

\item To offer further insights on the higher-order consistency relations, it would be interesting to rederive these new identities using standard
background-wave arguments. This may be technically challenging already for the $n=2$ identities, but would give an independent check
of the Ward identities.

\item It would be enlightening to apply the methods developed here to study correlation functions with multiple soft external lines. In the case of pions, it is well-known that the double-soft limit probes the non-abelian nature of the broken algebra~\cite{Weinberg:1966kf}. (See Sec.~4.1 of~\cite{ArkaniHamed:2008gz} for a recent discussion of double-soft pion theorems.)
In the present context, focusing on the scalar sector with symmetry breaking pattern~(\ref{symbreak}), correlation functions with two $\zeta$'s taken to be soft
should similarly instruct us about the underlying broken de Sitter isometries. To zeroth-order in the soft momenta, this has recently been discussed in~\cite{Senatore:2012wy}.
However, the non-abelian nature of the conformal algebra should show up at linear order in the soft momenta. This is currently in progress~\cite{austinmarko}.

\item Our derivation of the consistency relations from Ward identities crucially relied on the standard Bunch-Davies vacuum. Specifically, this assumption
was made in Sec.~\ref{LHSsec} in choosing the vacuum wavefunctional. However, the derivation straightforwardly generalizes to arbitrary
initial conditions, by substituting a modified wavefunctional, such as those of interest recently~\cite{Holman:2007na,Meerburg:2009ys,Meerburg:2009fi,Ashoorioon:2010xg,Ganc:2011dy,Chialva:2011hc,Agarwal:2012mq}, in lieu of~(\ref{Psi0}). The resulting identities would yield modified consistency relations with which to test non-standard initial conditions~\cite{Flauger:2013hra,Aravind:2013lra}.

\end{itemize}

Our derivation of consistency relations from Ward identities is very general and can be readily applied to other symmetry breaking patterns, including non-inflationary ones.
The conformal mechanism~\cite{Rubakov:2009np,Creminelli:2010ba,Hinterbichler:2011qk,Hinterbichler:2012mv,Creminelli:2012my,Hinterbichler:2012fr,Hinterbichler:2012yn}, in particular, relies the spontaneous breaking of conformal invariance on approximate flat space, with symmetry breaking pattern $SO(4,2)\rightarrow SO(4,1)$. The non-linear realization of $SO(4,2)$
implies novel consistency relations, which have been derived recently using the background-wave method~\cite{Creminelli:2012qr}. It should be straightforward to reproduce these relations from Ward identities.

{\bf Acknowledgements:} This work has greatly benefited from many discussions and initial collaboration with Walter Goldberger and Alberto Nicolis. We also thank Daniel Baumann, Lasha Berezhiani, Paolo Creminelli, Richard Holman, Austin Joyce, Eiichiro Komatsu, Juan Maldacena, Emil Mottola, Jorge Nore\~na, Koenraad Schalm, Leonardo Senatore, Marko~Simonovi\'c, Andrew Tolley, Mark Trodden, Matias Zaldarriaga, and especially Daniel Green, for helpful discussions. We warmly thank Lasha Berezhiani and Junpu Wang for fixing important typos in the original ${\cal O}(\gamma^1)$ expressions in Secs.~\ref{higherorder} and~\ref{Taylorexpand}. This work is supported in part by the DOE grant DE-FG02-92-ER40699 and NASA ATP grant NNX10AN14G (L.H.), as well as by NASA ATP grant NNX11AI95G, the Alfred P. Sloan Foundation and NSF CAREER Award PHY-1145525 (J.K.). Research at Perimeter Institute is supported by the Government of Canada through Industry Canada and by the Province of Ontario through the Ministry of Economic Development and Innovation.  This work was made possible in part through the support of a grant from the John Templeton Foundation. The opinions expressed in this publication are those of the authors and do not necessarily reflect the views of the John Templeton Foundation (K.H.). 

\appendix

\section{Conventions}
\label{convs}

In this Appendix we briefly summarize our conventions. 

\subsection{Power Spectra}

The curvature perturbation is expanded in terms of Fourier modes as
\be
\zeta(\vec{x},t) = \int \frac{{\rm d}^3k }{(2\pi)^3} \zeta(\vec{k},t) e^{-i\vec{k}\cdot\vec{x}}\,.
\ee
The scalar power spectrum $P_\zeta(k,t)$ is defined as
\be
\la \Omega| \zeta(\vec{k}) \zeta(\vec{k}')| \Omega\ra = (2\pi)^3\delta^3(\vec{k} + \vec{k}') P_\zeta(k,t)\,.
\label{fullP}
\ee
We also define the free-theory power spectrum, $P_\zeta^0(k,t)$, in terms of the 2-point function of interaction-picture
fields with respect to the free-theory vacuum state:
\be
\la 0 | \zeta_0(\vec{k}) \zeta_0(\vec{k}')| 0 \ra = (2\pi)^3\delta^3(\vec{k} + \vec{k}') P_\zeta^0(k,t)\,.
\label{freeP}
\ee
In slow-roll inflation, for instance, $P_\zeta^0 = H^2/4M_{\rm Pl}^2\epsilon k^3$.

Similarly, we expand tensor modes as
\be
\gamma_{ij}(\vec{x},t) =  \int \frac{{\rm d}^3k }{(2\pi)^3} \gamma_{ij}(\vec{k},t) e^{-i\vec{k}\cdot\vec{x}}\,.
\ee
In the helicity basis, we follow~\cite{Maldacena:2002vr} and write
\be
\gamma_{ij}(\vec{k},t) = \sum_{s = \pm} \epsilon_{ij}^s(\hat{k}) \gamma^s(\vec{k},t)\,,
\label{helbasis}
\ee
where the polarization tensors $ \epsilon_{ij}^s$ are transverse and traceless, $\hat{k}^i\epsilon_{ij}^s = \epsilon_{ii}^s = 0$. These satisfy the orthonormality condition,
\be
\epsilon_{ij}^s(\hat{k}) \epsilon_{ij}^{s'}(\hat{k}) = 2\delta_{ss'}\,,
\label{ortho}
\ee
and the completeness relation,
\be
\sum_s \epsilon^s_{ij}(\hat{k})\epsilon^{s}_{k\ell}(\hat{k})  = P_{ij k\ell}^{\rm T}(\hat{k})\,,
\label{compeps}
\ee
where the projector $P_{ijk\ell}^{\rm T}$ is defined in~\eqref{PTT}.

The tensor power spectrum is defined as
\be
\la \Omega| \gamma^s(\vec{k})  \gamma^{s'}(\vec{k}')  |\Omega\ra = (2\pi)^3\delta^3(\vec{k} + \vec{k}') P_\gamma(k,t)\delta_{ss'}\,,
\ee
or, equivalently in terms of $\gamma_{ij}$,
\be
\la \Omega| \gamma_{ij}(\vec{k})  \gamma_{ij}(\vec{k}')  |\Omega\ra = (2\pi)^3\delta^3(\vec{k} + \vec{k}') 4P_\gamma(k,t)\,.
\label{Ptens}
\ee
The free-theory tensor power spectrum is similarly given by
\be
\left \la \Omega\left\vert \gamma_{0ij}(\vec{k})  \gamma_{0ij}(\vec{k}')  \right\vert \Omega\right\ra = (2\pi)^3\delta^3(\vec{k} + \vec{k}') 4P_\gamma^0(k,t)\,.
\label{freePtens}
\ee
In slow-roll inflation, for instance, $P_\gamma^0 = H^2/M_{\rm Pl}^2k^3$.

\subsection{In-in formalism}

We briefly review the in-in formalism. Our goal is to calculate the correlator $\la\Omega |{\cal O}(t)|\Omega\ra$ for some operator ${\cal O}$,
where so far all quantities are in the Heisenberg picture. The operator ${\cal O}$ satisfies the Heisenberg equations of motion, 
$\dot{{\cal O}}=i[H(t){\cal O}(t)]$, where $H(t)$ is the full time-dependent Hamiltonian.  The state $|\Omega\ra$ is the state of the system, and does not depend on time in the Heisenberg picture. (This means that it is not an eigenstate of $H(t)$ at all times, since these eigenstates would in general depend on time).
 
As usual we split the Schr\"odinger-picture Hamiltonian into a free part and an interaction part, $H(t)=H_0(t)+V(t)$, and move
to the interaction picture with respect to this split. We have
\bea 
\nonumber
|\Omega(t)\ra_I &\equiv& U_0^\dag(t,0) U(t,0)|\Omega\ra \,; \\
{\cal O}_I(t) &\equiv&  U_0^\dag(t,0) U(t,0) {\cal O}(t) U_0(t,0)U^\dag(t,0)\,.
\eea
where $U$ and $U_0$ are the time evolution operators for the full and free Hamiltonian, respectively.
Interaction-picture operators evolve according to the free Hamiltonian: $\dot{{\cal O}}_I =i[(H_0)_I(t),{\cal O}_I(t)]$.
Interaction-picture states evolve according to a Schr\"odinger equation with $V_I(t)$ playing the role of a Hamiltonian, 
\be
i{{\rm d}\over {\rm d}t} | \Omega(t)\ra_I=V_I(t) | \Omega(t)\ra_I\,,
\ee
with general solution
\bea 
\nn
|\Omega(t)\ra_I&=&U_I(t,t_i)|\Omega(t_i)\ra_I\,;\ \ \ \\
U_I(t,t_i)&\equiv&U_0^\dag (t,0) U(t,t_i)U_0(t_i, 0)=\begin{cases} T\exp\left\{-i\int_{t_i}^t {\rm d}t' \ V_I(t')\right\} & t>t_i\,,\\ \bar{T}\exp\left\{i\int_{t}^{t_i} {\rm d}t'\ V_I(t')\right\} & t<t_i\, .\end{cases}
\eea

The expectation value is the same in any picture: $\la\Omega |{\cal O}(t)|\Omega\ra={}_I\la \Omega(t) |{\cal O}_I(t)|\Omega(t)\ra_I$.
Using the above expressions for the time evolution, we can write our correlator in terms of the state at the initial time $t_i$,
\be 
\la\Omega |{\cal O}(t)|\Omega\ra={}_I\la \Omega(t) |{\cal O}_I(t)|\Omega(t)\ra_I={}_I\la \Omega(t_i) |U_I^\dag(t,t_i) {\cal O}_I(t)U_I(t,t_i)|\Omega(t_i)\ra_I\,.
\label{corrinin}
\ee
The statement that $|\Omega\ra$ is the in-state is the statement that in the far past it was the Fock vacuum $|0\ra$ of the free field creation/annihilation operators, that is,
\be 
\lim_{t_i\rightarrow -\infty}|\Omega(t_i)\ra_I=|0\ra\,.
\ee
To see that this coincides with the usual definition of the in-state, we can write the same expression as
\be 
|\Omega\ra=\lim_{t_i\rightarrow -\infty}U^\dag(t_i,0) U_0(t_i,0)|0\ra \equiv \Omega(-\infty) |0\ra \,,
\ee
which matches the usual definition. With this, we obtain the in-in form of the correlator~(\ref{corrinin}):
\be 
\la\Omega |{\cal O}(t)|\Omega\ra = \lim_{t_i\rightarrow -\infty} \la 0|U_I^\dag(t,t_i) {\cal O}_I(t)U_I(t,t_i)|0\ra\,.
\ee

\section{Charges and Power Spectrum}
\label{chargeAppendix}

Here we justify a number of steps made in Sec. \ref{LHSsec}.
Recall from Sec. \ref{noethercharge} that
the true conserved charges are $\Delta \bar Q$, and $Q = \bar Q
+ f(t) \Delta \bar Q$ where $f(t) \equiv \int^t dt'/H(t')$.
As such, the arguments made in Sec. \ref{LHSsec},
especially (\ref{QUU1}), 
can be 
rigorously applied to them to obtain the analogs of (\ref{Qbarfull}):
\begin{eqnarray}
&& \Delta \bar Q \vert \Omega \rangle = \int {{\rm d}^3 q \over (2\pi)^3} q_i
q^2 \bar \xi_j
(-\vec q) D_{\gamma} (q) \gamma {}^{ij} (\vec q)  \vert \Omega \rangle
\nonumber \\
&& Q \vert \Omega \rangle = - \int {{\rm d}^3 q \over (2\pi)^3} q_i \bar \xi_j
(-\vec q) \left( {1\over 3} D_{\zeta} (q) \delta^{ij} \zeta (\vec q)
+ D_{\gamma} (q) \gamma {}^{ij} (\vec q) \right) \vert \Omega \rangle
\nonumber \\
&& \quad \quad + f(t) \int {{\rm d}^3 q \over (2\pi)^3} q_i
q^2 \bar \xi_j
(-\vec q) D_{\gamma} (q) \gamma {}^{ij} (\vec q)  \vert \Omega \rangle
\, .
\end{eqnarray}
On the other hand, $Q \vert \Omega \rangle
= \bar Q \vert \Omega \rangle + f(t) \Delta \bar Q \vert \Omega
\rangle$.
Thus, we are led to the conclusion that
\begin{eqnarray}
\bar Q \vert \Omega \rangle = - \int {{\rm d}^3 q \over (2\pi)^3} q_i \bar \xi_j
(-\vec q) \left( {1\over 3} D_{\zeta} (q) \delta^{ij} \zeta (\vec q)
+ D_{\gamma} (q) \gamma {}^{ij} (\vec q) \right) \vert \Omega \rangle
\, ,
\end{eqnarray}
justifying  (\ref{Qbarfull}).
No assumption has been made about $\bar Q$'s time-dependence
(or lack thereof) .

Another subtlety glossed over in Sec. \ref{LHSsec} is
that we did not distinguish the
$\zeta$ or $\gamma$ power spectrum of the full theory
from that of the free theory. The reason has to do with
the soft limit. In the $\vec q \rightarrow 0$ limit,
$\zeta (\vec q)$ and $\gamma_{ij} (\vec q)$ are time-independent.
This once again allows us to apply
the analog of~(\ref{QUU1}) at any time $t$ we wish:
\begin{eqnarray}
\zeta (\vec q) |\Omega \rangle 
= \Omega(-\infty) \zeta_0 (\vec q) | 0 \rangle
\quad , \quad
\gamma_{ij} (\vec q) |\Omega \rangle 
= \Omega(-\infty)  \gamma_0 {}_{ij} (\vec q) | 0 \rangle \,.
\end{eqnarray}
We thus have 
\begin{eqnarray}
\langle \Omega \vert \zeta (- \vec q) \zeta (\vec q) \vert \Omega
\rangle
= \langle 0 \vert \zeta_0 (- \vec q) \zeta_0 (\vec q) \vert 0
\rangle
\quad , \quad
\langle \Omega \vert \gamma_{ij} (- \vec q) \gamma_{ij} (\vec q) \vert \Omega
\rangle
= \langle 0 \vert \gamma_0 {}_{ij} (- \vec q) \gamma_0 {}_{ij} (\vec q) \vert 0
\rangle
\end{eqnarray}
in the soft limit.

\section{Connected Ward Identities}
\label{connWard}

The connected correlators $\la k_1 k_ 2\cdots k_N \ra^c$ are defined as
\be \la  k_1k_2\cdots k_N \ra^c=\sum_\pi\left(|\pi|-1\right)!\left(-1\right)^{|\pi|-1}\prod_{B\in \pi}\left\la \prod_{a\in B} k_a\right\ra\,,\ee
where $\pi$ is a partition of $1,\cdots,N$, $|\pi|$ is the number of
blocks in the partition, and $B$ represents a block of the partition.
Here we abuse the notation a bit, and use $k_1, ..., k_N$ 
as stand-in for
scalar or tensor fields at the corresponding momenta.
The full correlators $\la k_1 k_ 2\cdots  k_N\ra$ can then be written in terms of the connected correlators $\la  k_1 k_ 2\cdots  k_N\ra^c$ as
\be \la  k_1 k_ 2\cdots  k_N \ra=\sum_\pi\prod_{B\in \pi}\left\la \prod_{a\in B} k_a\right\ra^c\,.\ee

In our case we have $\la k\ra=\la k\ra^c=0$.  Thus we can write
\be \la  k_1 k_ 2\cdots  k_N \ra=\sum_{\pi'}\prod_{B\in \pi'}\left\la \prod_{a\in B} k_a\right\ra^c\,.\ee
where the sum is only over partitions $\pi'$ where each block has at least two elements.  The first few instances are
\bea 
\la k_1k_2 \ra&=&\la  k_1k_2 \ra^c\,, \\ \nn
\la k_1k_2 k_3 \ra&=&\la  k_1k_2k_3 \ra^c\,, \\ \nn
\la k_1k_2 k_3 k_4\ra&=&\la  k_1k_2k_3 k_4 \ra^c+\la  k_1k_2 \ra^c\la  k_3k_4 \ra^c+\la  k_1k_3 \ra^c\la  k_2k_4 \ra^c+\la  k_1k_4 \ra^c\la  k_2k_3 \ra^c\,, \\ \nn
\la k_1k_2 k_3 k_4 k_5\ra&=&\la  k_1k_2k_3 k_4 k_5 \ra^c+\left[ \la  k_1k_2 \ra^c\la  k_3k_4k_5 \ra^c+{\rm 9\ more\ permutations}\right] \,, \\ \nn
\la k_1k_2 k_3 k_4 k_5 k_6 \ra&=&\la  k_1k_2k_3 k_4 k_5k_6 \ra^c+\left[ \la  k_1k_2 \ra^c\la  k_3k_4k_5 k_6 \ra^c+{\rm 14\ more\ permutations}\right] \\ \nn
&&+\left[ \la  k_1k_2 k_3 \ra^c\la   k_4k_5 k_6 \ra^c+{\rm 19\ more\ permutations}\right]\,, \\ \nn
&&\vdots
\eea

Each connected correlator comes with an overall delta function, 
\be \la   k_1 k_ 2\cdots  k_N \ra^c=(2\pi)^3 \delta^3(k_1+k_ 2+\cdots+ k_N)\la   k_1 k_ 2\cdots  k_N \ra\,'.\ee
The $\la  k_1 k_ 2\cdots  k_N \ra'$ are defined on shell.
The power spectrum is $P(k)=\la k, -k\ra'$.

Our general Ward identity relation for $N\geq 1$ is in terms of the full correlators,
\bea \lim_{q\rightarrow 0}{\cal D}_q \left({1\over P(q)}\la q \, k_1\cdots k_N\ra\right)=&& (2\pi)^3\sum_{a=1}^N \left({\cal D}_a \delta^3(k_a)\right) \la k_1\cdots k_{a-1} k_{a+1}\cdots k_N\ra \nn\\
&&-\sum_{a=1}^ND_a\la k_1\cdots k_N\ra\,, \label{genrelation1}
\eea
where ${\cal D}_a$ and $D_a$ are differential operators that depend on and act only on the $a^{\rm th}$ momentum. (For example, for the dilations we have $D_a=3+k_a\cdot {\partial\over \partial k_a}$.) For $N=1$, the second line of the RHS does not contribute, the first line of the RHS gives $(2\pi)^3 {\cal D}_1 \delta^3(k_1)$ and the relation is satisfied.  

What we would like to prove is the following relation for connected correlators, for $N\geq 2$.
\be  \lim_{q\rightarrow 0}{\cal D}_q \left( {1\over P(q)}\la  q \, k_1\cdots k_N\ra^c\right) =-\sum_{a=1}^ND_a\la k_1\cdots k_N\ra^c\,. \label{genrelation2}
\ee
For $N=2$, the first line of the RHS of the relation \eqref{genrelation1} vanishes, and there is no distinction in the remaining correlators between connected and full, and so \eqref{genrelation2} is true for $N=2$.  This is the first step in a proof by induction that \eqref{genrelation2} is true for $N\geq 2$.  

For the induction step, assume that \eqref{genrelation2} is true for $N-1$.  We start by breaking up the correlator $\la  q \, k_1\cdots k_N\ra$ on the left-hand side of~\eqref{genrelation1} as follows
\bea \la  q \, k_1\cdots k_N\ra=&&\la q \, k_1\cdots k_N\ra^c+\sum_{a=1}^N\la q\, k_a\ra\la k_1\cdots k_{a-1} k_{a+1}\cdots k_N\ra.\nn\\
&&+\sum_{\bar\pi'}\left\la q \prod_{a\in \bar B} k_a\right\ra^c\prod_{B\in (\bar \pi'\backslash\bar B)}\left\la \prod_{a\in B} k_a\right\ra^c\,.\label{breakup1}\eea
Some explanation is in order: we have broken $\la  q \, k_1\cdots k_N\ra$ into its connected pieces.  The first term is the fully connected piece $\la  q \, k_1\cdots k_N\ra^c$.   The sum in the first line of the right-hand side is over all the partitions in which the soft momentum $q$ appears in a two point factor.  The rest of the partitions are in the second line of the right-hand side: here $\bar\pi'$ is the set of all non-trivial\footnote{A partition is non-trivial if it does not contain the trivial partition consisting of all the elements, which is captured already by the connected part $\la  q \, k_1\cdots k_N\ra^c$.} marked partitions of $1,2,\cdots,N$ where each block has at least two elements.   A marked partition is a partition in which one of the blocks of the partition, the one denoted by $\bar B$, is distinguished.  In this case, the special block is the one whose connected factor contains the soft momentum $q$.  We denote by $\bar\pi' \backslash\bar B$ the set of blocks of the marked partition which are not marked.  For example, with $N=3$ there are no terms in the sum over marked partitions (since in this case there are no partitions where all the blocks have at least two elements), and for $N=4$, the sum reads
\bea  &&\la k_1k_2\ra\la q\, k_3k_4\ra^c+\la k_1k_3\ra\la q\, k_2k_4\ra^c+ \la k_1k_4\ra\la q\, k_2k_3\ra^c \nn\\
& & +~\la k_3k_4\ra\la q\, k_1k_2\ra^c+\la k_2k_4\ra\la q\, k_1k_3\ra^c+ \la k_2k_3\ra\la q\, k_1k_4\ra^c\,.
\eea
Looking at the summation term in the first line of the right-hand side of~\eqref{breakup1}, using $\la q \, k_a\ra=(2\pi)^3\delta^3(q+k_a)P(q)$, and inserting into the left-hand side of~\eqref{genrelation1}, we precisely reproduce the first line of the right-hand side of~\eqref{genrelation1}.  

Thus, to complete the induction step, we need only show, under the assumption that \eqref{genrelation2} is true for $\leq N-1$, that the second line of the right-hand side of~\eqref{breakup1}, inserted into the left-hand side of~\eqref{genrelation1}, precisely reproduces the second line of the right-hand side of \eqref{genrelation1} (minus the connected part).  The second line of the right-hand side of~\eqref{breakup1}, inserted into the left-hand side of~\eqref{genrelation1} and using~\eqref{genrelation2} yields
\be \sum_{\bar\pi'}\lim_{q\rightarrow 0}{\cal D}_q \left({1 \over P(q)} \left\la q \prod_{a\in \bar B} k_a\right\ra^c\right) \prod_{B\in (\bar \pi'/\bar B)}\left\la \prod_{a\in B} k_a\right\ra^c= -\sum_{\bar\pi'}\left[\sum_{a\in \bar B}D_a\left\la  \prod_{a\in \bar B} k_a\right\ra^c\right]\prod_{B\in (\bar \pi'/\bar B)}\left\la \prod_{a\in B} k_a\right\ra^c\,.\ee
Next we break up the correlator in the second line of the right-hand side of~\eqref{genrelation1} into its connected pieces, where the sum is only over non-trivial partitions $\pi'$ where each block has at least two elements,
\be \la  k_1k_ 2\cdots k_N \ra=\la   k_1k_ 2\cdots k_N\ra^c+\sum_{\pi'}\prod_{B\in \pi'}\left\la \prod_{a\in B} k_a\right\ra^c\,.\ee
The operator $\sum_{a=1}^ND_a$ can be broken up over a given partition $\pi'$ as $\sum_{B\in \pi'}\sum_{a\in B}D_a$, and when applied to  $\sum_{\pi'}\prod_{B\in \pi'}\la \prod_{a\in B} k_a\ra^c$, we generate the sum over marked partitions,
\be
\sum_{\bar\pi'}\left[\sum_{a\in \bar B}D_a\left\la  \prod_{a\in \bar B} k_a\right\ra^c\right]\prod_{B\in (\bar \pi'/\bar B)}\left\la \prod_{a\in B} k_a\right\ra^c\,,
\ee
where the $D_a$ operator serves as the marker.  This completes  the proof.  

\section{Removing Delta Functions}
\label{deltaremove}

In this Appendix, we show how to cancel the delta function factors in~(\ref{wardalmostthere}) to obtain the primed Ward identities~(\ref{wardalmosttheren>2}). To begin with,
it is convenient to rewrite the right-hand side of~(\ref{wardalmostthere}) as
\bea
\nonumber
 & & \lim_{\vec{q}\rightarrow 0} M_{i\ell_0 \ldots \ell_n} \frac{\partial^{n}}{\partial q_{\ell_1}\cdots \partial q_{\ell_n}} \Bigg(\frac{1}{P_\gamma(q)} \la \gamma^{i\ell_0}(\vec{q}){\cal O}(\vec{k}_1,\ldots,\vec{k}_N) \ra_c + \frac{\delta^{i\ell_0}}{3P_\zeta(q)} \la  \zeta(\vec{q}) {\cal O}(\vec{k}_1,\ldots,\vec{k}_N)\ra_c  \Bigg) \\
\nonumber
& = & -  \lim_{\vec{q}\rightarrow 0} M_{i\ell_0 \ldots \ell_n} \Bigg\{ \sum_{a=1}^N \Bigg( \delta^{i\ell_0} \frac{\partial^{n}}{\partial k_{\ell_1}^a\cdots \partial k_{\ell_n}^a} + \frac{k^{i}_a}{n+1}  \frac{\partial^{n+1}}{\partial k_{\ell_0}^a \cdots \partial k_{\ell_n}^a}\Bigg) \la  {\cal O}(\vec{k}_1, \ldots, \vec{k}_a + \vec{q}, \ldots, \vec{k}_N)\ra_c  \\
\nonumber
& & -~\sum_{a=1}^M\Upsilon^{i\ell_0i_aj_a}(\hat{k}_a)  \frac{\partial^{n}}{\partial k_{\ell_1}^a\cdots \partial k_{\ell_n}^a} \la {\cal O}^\zeta(\vec{k}_1,\ldots,\vec{k}_{a-1},\vec{k}_{a+1},\ldots \vec{k}_M)\gamma_{i_aj_a}(\vec{k}_a + \vec{q}) {\cal O}^\gamma(\vec{k}_{M+1},\ldots,\vec{k}_N)\ra_c\\
\nonumber
& & -~\sum_{b=M+1}^N  \Gamma^{i\ell_0\;\;\;\;\;k_b\ell_b}_{\;\;\;\;i_bj_b}(\hat{k}_b )\frac{\partial^{n}}{\partial k_{\ell_1}^b\cdots \partial k_{\ell_n}^b}\la {\cal O}^\zeta(\vec{k}_1,\ldots,\vec{k}_M) {\cal O}^\gamma_{i_{M+1} j_{M+1},\ldots,k_b\ell_b,\ldots i_Nj_N} (\vec{k}_{M+1},\ldots,\vec{k}_b+\vec{q},\ldots,\vec{k}_N)\ra_c\Bigg\} \\
& & +~\ldots  \,,
\label{wardalmostthereapp}
\eea
where as before the ellipses stand for higher-order tensor terms. In this form, the correlators on both sides of the identity involve
the {\it same} delta function $\delta^3(\vec{P})$, where $\vec{P} \equiv \vec{q} + \vec{k}_1 + \ldots \vec{k}_N$.

First, consider the case $n = 0$. Expressing~(\ref{wardalmostthereapp}) in terms of primed correlators in this case gives
\bea
\nonumber
 & &  \lim_{\vec{q}\rightarrow 0} M_{i\ell_0} \Bigg(\frac{1}{P_\gamma(q)} \la \gamma^{i\ell_0}(\vec{q}){\cal O}(\vec{k}_1,\ldots,\vec{k}_N)\ra_c' + \frac{\delta^{i\ell_0}}{3P_\zeta(q)} \la  \zeta(\vec{q}) {\cal O}(\vec{k}_1,\ldots,\vec{k}_N)\ra_c'  \Bigg)\delta^3(\vec{P}) \\
\nonumber
&=&  -   \lim_{\vec{q}\rightarrow 0} M_{i\ell_0} \Bigg\{ \Bigg( N\delta^{i\ell_0}  +   \sum_{a=1}^N  k^{i}_a\frac{\partial}{\partial k_{\ell_0}^a}\Bigg)\la   {\cal O}(\vec{k}_1, \ldots,\vec{k}_N)\ra_c'  \\
\nonumber
& & \;\;\;\;\;\;\;\;\; -\sum_{a=1}^M\Upsilon^{i\ell_0i_aj_a}(\hat{k}_a) \la {\cal O}^\zeta(\vec{k}_1,\ldots,\vec{k}_{a-1},\vec{k}_{a+1},\ldots \vec{k}_M)\gamma_{i_aj_a}(\vec{k}_a) {\cal O}^\gamma(\vec{k}_{M+1},\ldots,\vec{k}_N)\ra_c \\
\nonumber
& & \;\;\;\;\;\;\;\;\; -\sum_{b=M+1}^N  \Gamma^{i\ell_0\;\;\;\;\;k_b\ell_b}_{\;\;\;\;i_bj_b}(\hat{k}_b )\la {\cal O}^\zeta(\vec{k}_1,\ldots,\vec{k}_M) {\cal O}^\gamma_{i_{M+1} j_{M+1},\ldots,k_b\ell_b,\ldots i_Nj_N} (\vec{k}_{M+1},\ldots,\vec{k}_N)\ra_c' \Bigg\}\,\delta^3(\vec{P})  \\
& & -  \lim_{\vec{q}\rightarrow 0}M_{i\ell_0} \la   {\cal O}(\vec{k}_1, \ldots,\vec{k}_N)\ra_c' P^{i}\frac{\partial}{\partial P^{\ell_0}}\delta^3(\vec{P}) +\ldots
\label{wardalmosttheren=1app}
\eea
Integrating the last term by parts, and using the fact that $P^i\delta^3(\vec{P}) \equiv 0$, we obtain
\bea
\nonumber
& &  \lim_{\vec{q}\rightarrow 0} M_{i\ell_0} \Bigg(\frac{1}{P_\gamma(q)} \la \gamma^{i\ell_0}(\vec{q}){\cal O}(\vec{k}_1,\ldots,\vec{k}_N)\ra_c' + \frac{\delta^{i\ell_0}}{3P_\zeta(q)} \la  \zeta(\vec{q}) {\cal O}(\vec{k}_1,\ldots,\vec{k}_N)\ra_c'  \Bigg) \\
\nonumber
&=&  -  M_{i\ell_0} \Bigg\{ \Bigg( (N-1)\delta^{i\ell_0}  +   \sum_{a=1}^N  k^{i}_a\frac{\partial}{\partial k^{\ell_0}_a}\Bigg)\la   {\cal O}(\vec{k}_1, \ldots,\vec{k}_N)\ra_c'  \\
\nonumber
& & \;\;\;\;\;\;\;\;\; -\sum_{a=1}^M\Upsilon^{i\ell_0i_aj_a}(\hat{k}_a)  \la {\cal O}^\zeta(\vec{k}_1,\ldots,\vec{k}_{a-1},\vec{k}_{a+1},\ldots \vec{k}_M)\gamma_{i_aj_a}(\vec{k}_a) {\cal O}^\gamma(\vec{k}_{M+1},\ldots,\vec{k}_N)\ra_c \\
\nonumber
& & \;\;\;\;\;\;\;\;\; -\sum_{b=M+1}^N  \Gamma^{i\ell_0\;\;\;\;\;k_b\ell_b}_{\;\;\;\;i_bj_b}(\hat{k}_b )\la {\cal O}^\zeta(\vec{k}_1,\ldots,\vec{k}_M) {\cal O}^\gamma_{i_{M+1} j_{M+1},\ldots,k_b\ell_b,\ldots i_Nj_N} (\vec{k}_{M+1},\ldots,\vec{k}_N)\ra_c' \Bigg\} +\ldots \\
& & \qquad\qquad\qquad\qquad\qquad\qquad\qquad\qquad\qquad\qquad\qquad\qquad\qquad\qquad\qquad (n = 0~{\rm identity})  \,,
\label{wardnearlyfinaln=1app}
\eea
where we have canceled the delta functions from both sides. Note that the limit $\vec{q}\rightarrow 0$ has been explicitly taken on the right-hand side,
hence the right-hand side is now independent of $\vec{q}$. This is the desired ``primed" Ward identity for $n = 0$, in agreement with~(\ref{wardalmosttheren>2}).

Next we prove that~(\ref{wardalmostthereapp}) holds in the same form for the primed correlation functions for all $n\geq 1$:
\bea
\nonumber
 & & \lim_{\vec{q}\rightarrow 0} M_{i\ell_0 \ldots \ell_n} \frac{\partial^{n}}{\partial q_{\ell_1}\cdots \partial q_{\ell_n}} \Bigg(\frac{1}{P_\gamma(q)} \la \gamma^{i\ell_0}(\vec{q}){\cal O}(\vec{k}_1,\ldots,\vec{k}_N) \ra_c' + \frac{\delta^{i\ell_0}}{3P_\zeta(q)} \la  \zeta(\vec{q}) {\cal O}(\vec{k}_1,\ldots,\vec{k}_N)\ra_c'  \Bigg) \\
\nonumber
& = & - M_{i\ell_0 \ldots \ell_n} \Bigg\{ \sum_{a=1}^N \Bigg( \delta^{i\ell_0} \frac{\partial^{n}}{\partial k_{\ell_1}^a\cdots \partial k_{\ell_n}^a} + \frac{k^{i}_a}{n+1}  \frac{\partial^{n+1}}{\partial k_{\ell_0}^a \cdots \partial k_{\ell_n}^a}\Bigg) \la  {\cal O}(\vec{k}_1, \ldots, \vec{k}_N)\ra_c'  \\
\nonumber
& & -~\sum_{a=1}^M\Upsilon^{i\ell_0i_aj_a}(\hat{k}_a)  \frac{\partial^{n}}{\partial k_{\ell_1}^a\cdots \partial k_{\ell_n}^a} \la {\cal O}^\zeta(\vec{k}_1,\ldots,\vec{k}_{a-1},\vec{k}_{a+1},\ldots \vec{k}_M)\gamma_{i_aj_a}(\vec{k}_a) {\cal O}^\gamma(\vec{k}_{M+1},\ldots,\vec{k}_N)\ra_c'\\
\nonumber
& & -~\sum_{b=M+1}^N  \Gamma^{i\ell_0\;\;\;\;\;k_b\ell_b}_{\;\;\;\;i_bj_b}(\hat{k}_b )\frac{\partial^{n}}{\partial k_{\ell_1}^b\cdots \partial k_{\ell_n}^b}\la {\cal O}^\zeta(\vec{k}_1,\ldots,\vec{k}_M) {\cal O}^\gamma_{i_{M+1} j_{M+1},\ldots,k_b\ell_b,\ldots i_Nj_N} (\vec{k}_{M+1},\ldots,\vec{k}_N)\ra_c'\Bigg\} + \ldots \\
& &  \qquad\qquad\qquad\qquad\qquad\qquad\qquad\qquad\qquad\qquad\qquad\qquad\qquad\qquad\qquad (n \geq 1~{\rm identities})\,.
\label{wardalmosttheren>2app}
\eea
We prove this by strong induction. (To simplify the notation, we will omit the momentum dependence except where necessary, and the $\vec{q}\rightarrow 0$ limit will be understood everywhere.) Starting with $n=1$,
\bea
\nonumber
& & M_{i\ell_0\ell_1} \frac{\partial}{\partial q_{\ell_1}} \Bigg(\frac{1}{P_\gamma} \la \gamma^{i\ell_0}{\cal O}\ra_c' + \frac{\delta^{i\ell_0}}{3P_\zeta} \la  \zeta {\cal O}\ra_c'  \Bigg)\delta^3(\vec{P}) +  M_{i\ell_0\ell_1} \Bigg(\frac{1}{P_\gamma} \la \gamma^{i\ell_0}{\cal O}\ra_c' + \frac{\delta^{i\ell_0}}{3P_\zeta} \la  \zeta {\cal O}\ra_c'  \Bigg)\frac{\partial}{\partial P_{\ell_1}}\delta^3(\vec{P}) \\
\nonumber
& &=  -M_{i\ell_0\ell_1}\Bigg\{ \sum_a \left(\delta^{i\ell_0} \frac{\partial}{\partial k^a_{\ell_1}}+ \frac{1}{2}k_a^{i}\frac{\partial^2}{\partial k^a_{\ell_0}\partial k^a_{\ell_1}}\right) \la {\cal O}\ra_c' -\sum_{a=1}^M\Upsilon^{i\ell_0i_aj_a}  \frac{\partial}{\partial k_{\ell_1}^a} \la {\cal O}^\zeta(\vec{k}_1,\ldots,\vec{k}_{a-1},\vec{k}_{a+1},\ldots \vec{k}_M)\gamma_{i_aj_a} {\cal O}^\gamma\ra_c'\\
\nonumber
&& ~~~~~~~~~~~~~ -\sum_{b=M+1}^N \Gamma^{i\ell_0\;\;\;\;\;k_b\ell_b}_{\;\;\;\;i_bj_b} \frac{\partial}{\partial k^b_{\ell_1}}\la {\cal O}^\zeta {\cal O}^\gamma_{i_{M+1} j_{M+1},\ldots,k_b\ell_b,\ldots i_Nj_N}\ra_c'\Bigg\} \delta^3(\vec{P}) \\
\nonumber
& & ~~- M_{i\ell_0\ell_1} \Bigg\{ \left(N\delta^{i\ell_0} + \sum_ak^{i}_a\frac{\partial}{\partial k_{\ell_0}^a}\right)  \la {\cal O}\ra_c'   -\sum_{a=1}^M\Upsilon^{i\ell_0i_aj_a}  \la {\cal O}^\zeta(\vec{k}_1,\ldots,\vec{k}_{a-1},\vec{k}_{a+1},\ldots \vec{k}_M)\gamma_{i_aj_a} {\cal O}^\gamma\ra_c'\\
\nonumber
& & ~~~~~~~~~~~~~ -\sum_{b=M+1}^N \Gamma^{i\ell_0\;\;\;\;\;k_b\ell_b}_{\;\;\;\;i_bj_b} \la {\cal O}^\zeta {\cal O}^\gamma_{i_{M+1} j_{M+1},\ldots,k_b\ell_b,\ldots i_Nj_N}\ra_c' \Bigg\}\frac{\partial}{\partial P_{\ell_1}}\delta^3(\vec{P})  \\
\nonumber
& &~~ - \frac{1}{2}M_{i\ell_0\ell_1} \la {\cal O}\ra_c' P^{i}\frac{\partial^2}{\partial P_{\ell_0}\partial P_{\ell_1}}\delta^3(\vec{P}) + \ldots \\
\nonumber
& &=  -M_{i\ell_0\ell_1}\Bigg\{ \sum_a \left(\delta^{i\ell_0} \frac{\partial}{\partial k^a_{\ell_1}}+ \frac{1}{2}k_a^{i}\frac{\partial^2}{\partial k^a_{\ell_0}\partial k^a_{\ell_1}}\right) \la {\cal O}\ra_c' -\sum_{a=1}^M\Upsilon^{i\ell_0i_aj_a}  \frac{\partial}{\partial k_{\ell_1}^a} \la {\cal O}^\zeta(\vec{k}_1,\ldots,\vec{k}_{a-1},\vec{k}_{a+1},\ldots \vec{k}_M)\gamma_{i_aj_a} {\cal O}^\gamma\ra_c'\\
\nonumber
&& ~~~~~~~~~~~~~ -\sum_{b=M+1}^N \Gamma^{i\ell_0\;\;\;\;\;k_b\ell_b}_{\;\;\;\;i_bj_b} \frac{\partial}{\partial k^b_{\ell_1}}\la {\cal O}^\zeta {\cal O}^\gamma_{i_{M+1} j_{M+1},\ldots,k_b\ell_b,\ldots i_Nj_N}\ra_c'\Bigg\} \delta^3(\vec{P}) \\
\nonumber
& & ~~- M_{i\ell_0\ell_1} \Bigg\{ \left((N-1)\delta^{i\ell_0} + \sum_ak^{i}_a\frac{\partial}{\partial k_{\ell_0}^a}\right)  \la {\cal O}\ra_c'   -\sum_{a=1}^M\Upsilon^{i\ell_0i_aj_a}  \la {\cal O}^\zeta(\vec{k}_1,\ldots,\vec{k}_{a-1},\vec{k}_{a+1},\ldots \vec{k}_M)\gamma_{i_aj_a} {\cal O}^\gamma\ra_c'\\
& & ~~~~~~~~~~~~~ -\sum_{b=M+1}^N \Gamma^{i\ell_0\;\;\;\;\;k_b\ell_b}_{\;\;\;\;i_bj_b} \la {\cal O}^\zeta {\cal O}^\gamma_{i_{M+1} j_{M+1},\ldots,k_b\ell_b,\ldots i_Nj_N}\ra_c' \Bigg\}\frac{\partial}{\partial P_{\ell_1}}\delta^3(\vec{P})  + \ldots 
\eea
where in the last step we have integrated the $\partial^2/\partial P_{\ell_0} \partial P_{\ell_1}$ once by parts using the fact that $M_{i\ell_0\ell_1}$ is symmetric in its last two indices. Collecting all $\frac{\partial}{\partial P_{\ell_1}}\delta^3(\vec{P})$ terms, we see that their coefficient is proportional to the $n=0$ primed identity~(\ref{wardnearlyfinaln=1app}) and thus vanishes. Hence we are left with
\bea
\nonumber
& &  \lim_{\vec{q}\rightarrow 0}M_{i\ell_0\ell_1} \frac{\partial}{\partial q_{\ell_1}}\Bigg(\frac{1}{P_\gamma(q)} \la \gamma^{i\ell_0}(\vec{q}){\cal O}(\vec{k}_1,\ldots,\vec{k}_N) \ra_c' + \frac{\delta^{i\ell_0}}{3P_\zeta(q)} \la  \zeta(\vec{q}) {\cal O}(\vec{k}_1,\ldots,\vec{k}_N)\ra_c'  \Bigg) \\
\nonumber
&=&  -  M_{i\ell_0\ell_1} \Bigg\{ \sum_{a=1}^N \Bigg(\delta^{i\ell_0} \frac{\partial}{\partial k^a_{\ell_1}}+ \frac{1}{2}k_a^{i}\frac{\partial^2}{\partial k^a_{\ell_0}\partial k^a_{\ell_1}}\Bigg)  \la {\cal O}(\vec{k}_1, \ldots,\vec{k}_N)\ra_c'  \\
\nonumber
& & \;\;\;\;\;\;\;\;\;\;\;\; -\sum_{a=1}^M\Upsilon^{i\ell_0i_aj_a}(\hat{k}_a)  \frac{\partial}{\partial k_{\ell_1}^a} \la {\cal O}^\zeta(\vec{k}_1,\ldots,\vec{k}_{a-1},\vec{k}_{a+1},\ldots \vec{k}_M)\gamma_{i_aj_a}(\vec{k}_a) {\cal O}^\gamma(\vec{k}_{M+1},\ldots,\vec{k}_N)\ra_c'\\
\nonumber
& & \;\;\;\;\;\;\;\;\;\;\;\; -\sum_{b=M+1}^N  \Gamma^{i\ell_0\;\;\;\;\;k_b\ell_b}_{\;\;\;\;i_bj_b}(\hat{k}_b )\frac{\partial}{\partial k^b_{\ell_1}}\la {\cal O}^\zeta(\vec{k}_1,\ldots,\vec{k}_M) {\cal O}^\gamma_{i_{M+1} j_{M+1},\ldots,k_b\ell_b,\ldots i_Nj_N} (\vec{k}_{M+1},\ldots,\vec{k}_N)\ra_c' \Bigg\} +\ldots \\
& & \qquad\qquad\qquad\qquad\qquad\qquad\qquad\qquad\qquad\qquad\qquad\qquad\qquad\qquad\qquad (n = 1~{\rm identity}) \,,
\label{wardfinaln=2}
\eea
which is the desired result for $n = 1$. 

By strong induction, we next assume that~(\ref{wardalmosttheren>2}) holds for all $m \leq n -1$ for some $n \geq 2$, and show that it must therefore hold for $m=n$.
Expressing~(\ref{wardalmostthere}) in terms of primed correlators, clearly the desired result corresponds to no derivatives acting on the delta functions.
The proof therefore boils down to showing that all terms with at least one derivative on a delta function cancel out. Let us first collect all  $\frac{\partial^j}{\partial P^j}\delta^3(\vec{P})$ terms, 
where $1\leq j \leq n-1$. Since $M_{i\ell_0 \ldots \ell_n}$ is symmetric in its last $n+1$ indices, we get
\bea
\nonumber
& & M_{i\ell_0 \ldots \ell_n} \frac{\partial^j\delta^3(\vec{P})}{\partial P_{\ell_{n-j+1}}\cdots \partial P_{\ell_n}}\Bigg\{ \left( \begin{array}{c}n \\ j \end{array} \right) \frac{\partial^{n-j}}{\partial q_{\ell_1}\cdots\partial q_{\ell_{n-j}}} \Bigg(\frac{1}{P_\gamma} \la \gamma^{i\ell_0}{\cal O}\ra_c' + \frac{\delta^{i\ell_0}}{3P_\zeta} \la  \zeta {\cal O}\ra_c'  \Bigg)\\
\nonumber
& & \;\;\;\;\;\;\;\;\;\;\;\;\;\;\; + \sum_{a} \left[\delta^{i\ell_0}\left( \begin{array}{c}n \\ j \end{array} \right)   \frac{\partial^{n-j}}{\partial k^a_{\ell_1}\cdots\partial k^a_{\ell_{n-j}}} + \frac{k_a^{i}}{n+1}\left( \begin{array}{c}n+1 \\ j\end{array} \right) \frac{\partial^{n-j+1}}{\partial k^a_{\ell_0}\cdots\partial k^a_{\ell_{n-j}}}\right] \la {\cal O}\ra_c' \\
\nonumber
& &  \;\;\;\;\;\;\;\;\;\;\;\;\;\;\;-\sum_{a=1}^M\Upsilon^{i\ell_0i_aj_a}\left( \begin{array}{c}n \\ j \end{array} \right)   \frac{\partial^{n-j}}{\partial k_{\ell_1}^a\cdots \partial k_{\ell_{n-j}}^a} \la {\cal O}^\zeta(\vec{k}_1,\ldots,\vec{k}_{a-1},\vec{k}_{a+1},\ldots \vec{k}_M)\gamma_{i_aj_a} {\cal O}^\gamma\ra_c'\\
& & \;\;\;\;\;\;\;\;\;\;\;\;\;\;\; -\sum_{b=M+1}^N  \Gamma^{i\ell_0\;\;\;\;\;k_b\ell_b}_{\;\;\;\;i_bj_b} \left( \begin{array}{c}n \\ j \end{array} \right)   \frac{\partial^{n-j}}{\partial k^b_{\ell_1}\cdots\partial k^b_{\ell_{n-j}}} \la {\cal O}^\zeta {\cal O}^\gamma_{i_{M+1} j_{M+1},\ldots,k_b\ell_b,\ldots i_Nj_N}\ra_c' \Bigg\}\,.
\eea
Using the fact that 
\be
\left( \begin{array}{c}n+1 \\ j \end{array} \right) = \left( \begin{array}{c}n \\ j \end{array} \right)\cdot \frac{n+1}{n-j+1}\,,
\ee
this reduces to
\bea
\nonumber
& & \left( \begin{array}{c}n \\ j \end{array} \right) M_{i\ell_0 \ldots \ell_n} \frac{\partial^j\delta^3(\vec{P})}{\partial P_{\ell_{n-j+1}}\cdots \partial P_{\ell_n}}\Bigg\{ \frac{\partial^{n-j}}{\partial q_{\ell_1}\cdots\partial q_{\ell_{n-j}}} \Bigg(\frac{1}{P_\gamma} \la \gamma^{i\ell_0}{\cal O}\ra_c' + \frac{\delta^{i\ell_0}}{3P_\zeta} \la  \zeta {\cal O}\ra_c'  \Bigg)\\
\nonumber
& & \;\;\;\;\;\;\;\;\;\;\;\;\;\;\;\;\;\;\;\;\;\;\;\;\;\;\;\; + \sum_{a} \left[\delta^{i\ell_0}  \frac{\partial^{n-j}}{\partial k^a_{\ell_1}\cdots\partial k^a_{\ell_{n-j}}} + \frac{k_a^{i}}{n-j+1} \frac{\partial^{n-j+1}}{\partial k^a_{\ell_0}\cdots\partial k^a_{\ell_{n-j+1}}}\right] \la {\cal O}\ra_c' \\
\nonumber
& &  \;\;\;\;\;\;\;\;\;\;\;\;\;\;\;\;\;\;\;\;\;\;\;\;\;\;\;\;-\sum_{a=1}^M\Upsilon^{i\ell_0i_aj_a}\left( \begin{array}{c}n \\ j \end{array} \right)   \frac{\partial^{n-j}}{\partial k_{\ell_1}^a\cdots \partial k_{\ell_{n-j}}^a} \la {\cal O}^\zeta(\vec{k}_1,\ldots,\vec{k}_{a-1},\vec{k}_{a+1},\ldots \vec{k}_M)\gamma_{i_aj_a} {\cal O}^\gamma\ra_c'\\
& & \;\;\;\;\;\;\;\;\;\;\;\;\;\;\;\;\;\;\;\;\;\;\;\;\;\;\;\; -\sum_{b=M+1}^N  \Gamma^{i\ell_0\;\;\;\;\;k_b\ell_b}_{\;\;\;\;i_bj_b} \frac{\partial^{n-j}}{\partial k^b_{\ell_1}\cdots\partial k^b_{\ell_{n-j}}} \la {\cal O}^\zeta {\cal O}^\gamma_{i_{M+1} j_{M+1},\ldots,k_b\ell_b,\ldots i_Nj_N}\ra_c'  \Bigg\}\,.
\eea
Thus the terms with $j$ derivatives on the delta function, with $1\leq j \leq n-1$, are proportional to the $n-j$ primed Ward identity, which holds by assumption. 
It remains to show that terms with at least $n$ derivatives on $\delta^3(\vec{P})$ also cancel out. We will see that these are in fact proportional to $n=0$ primed
identity. Collecting such terms, we obtain
\bea
\nonumber
& & M_{i\ell_0 \ldots \ell_n} \Bigg\{\frac{\partial^{n}\delta^3(\vec{P})}{\partial P_{\ell_{1}}\cdots \partial P_{\ell_n}}\Bigg[\frac{1}{P_\gamma} \la \gamma^{i\ell_0}{\cal O}\ra_c' + \frac{\delta^{i\ell_0}}{3P_\zeta} \la  \zeta {\cal O}\ra_c' 
+ \left(N\delta^{i\ell_0} + \sum_ak_a^{i}\frac{\partial}{\partial k^a_{\ell_0}}\right) \la {\cal O}\ra_c'  \\
\nonumber
& &   \;\;\;\;\;\;\;\;\;\;\;\;\;\;\;\;\;\;\;\;\;\;\;\;\;\;\;\;\;\;\;\;\;\;\;\;\;\;\; -\sum_{a=1}^M\Upsilon^{i\ell_0i_aj_a}  \la {\cal O}^\zeta(\vec{k}_1,\ldots,\vec{k}_{a-1},\vec{k}_{a+1},\ldots \vec{k}_M)\gamma_{i_aj_a} {\cal O}^\gamma\ra_c'\\
\nonumber
& &  \;\;\;\;\;\;\;\;\;\;\;\;\;\;\;\;\;\;\;\;\;\;\;\;\;\;\;\;\;\;\;\;\;\;\;\;\;\;\; -\sum_{b=M+1}^N  \Gamma^{i\ell_0\;\;\;\;\;k_b\ell_b}_{\;\;\;\;i_bj_b} \la {\cal O}^\zeta {\cal O}^\gamma_{i_{M+1} j_{M+1},\ldots,k_b\ell_b,\ldots i_Nj_N}\ra_c'\ \Bigg]  \\
& & \;\;\;\;\;\;\;\;\;\;\;\;\;\;\;\;\; + \frac{1}{n+1} P^{i} \frac{\partial^{n+1}\delta^3(\vec{P})}{\partial P_{\ell_0} \cdots \partial P_{\ell_n}}\la {\cal O}\ra_c'\Bigg\}
\eea
Integrating the last term once by parts, using the symmetry properties of $M$, we obtain
\bea
\nonumber
& & M_{i\ell_0 \ldots \ell_n} \frac{\partial^{n}\delta^3(\vec{P})}{\partial P_{\ell_{1}}\cdots \partial P_{\ell_n}}\Bigg[\frac{1}{P_\gamma} \la \gamma^{i\ell_0}{\cal O}\ra_c' + \frac{\delta^{i\ell_0}}{3P_\zeta} \la  \zeta {\cal O}\ra_c' 
+ \left((N-1)\delta^{i\ell_0} + \sum_ak_a^{i}\frac{\partial}{\partial k^a_{\ell_0}}\right) \la {\cal O}\ra_c'  \\
\nonumber
& &   \;\;\;\;\;\;\;\;\;\;\;\;\;\;\;\;\;\;\;\;\;\;\;\;\;\;\;\;\;\;\;\;\;\;\;\;\;\;\; -\sum_{a=1}^M\Upsilon^{i\ell_0i_aj_a}  \la {\cal O}^\zeta(\vec{k}_1,\ldots,\vec{k}_{a-1},\vec{k}_{a+1},\ldots \vec{k}_M)\gamma_{i_aj_a} {\cal O}^\gamma\ra_c'\\
& &  \;\;\;\;\;\;\;\;\;\;\;\;\;\;\;\;\;\;\;\;\;\;\;\;\;\;\;\;\;\;\;\;\;\;\;\;\;\;\; -\sum_{b=M+1}^N  \Gamma^{i\ell_0\;\;\;\;\;k_b\ell_b}_{\;\;\;\;i_bj_b} \la {\cal O}^\zeta {\cal O}^\gamma_{i_{M+1} j_{M+1},\ldots,k_b\ell_b,\ldots i_Nj_N}\ra_c'\ \Bigg]
\eea
As advocated, we recognize the $n=0$ identity in square brackets, hence terms with at least $n$ derivatives on the delta function cancel out. This completes the proof of~(\ref{wardalmosttheren>2app}).

\section{Component Operators}
\label{projectors}

In this Appendix, we describe a brute force method for constructing the operators used in Sec.~\ref{physward}.  We are looking for operators $P_{i\ell_0\ldots \ell_n j m_0\ldots m_n}(\hat{q)}$, composed from $\hat q_i$ and $\delta_{ij}$, which satisfy the four conditions listed in Sec.~\ref{physward}.

We start by writing down by writing down all possible terms constructed from $\hat q_i$ and $\delta_{ij}$ which have $2n+2$ indices.  First there is a one term which is just a product of $2n+2$ $q$'s, then there are terms where two of the indices are on a $\delta$ and the rest are $q$'s (there are $\left(\begin{array}{c}2n+2 \\2\end{array}\right)$ of these since we may choose any of the 2 indices for the $\delta$), then there are the terms with two $\delta$'s, and so on, until we get to the terms which are products of $n+1$ $\delta$'s.  

Now we take all these terms and write them down with an arbitrary coefficient in front of each.  Then we impose the conditions listed in Sec.~\ref{physward}.  Since the conditions are linear, this will generate a set of linear equations which the coefficients must satisfy.  There will be some subspace of solutions to this linear system.  Choosing a linearly independent basis of this subspace will yield our component operators.  

For example, in the simplest case $n=0$, we obtain the following 3 operators,
\bea && 2 P_{i m_0} P_{\ell_0 j}-P_{i \ell_0} P_{j m_0}\,, \nn\\
&& \hat q_{\ell_0} \hat q_{m_0} P_{i j}-\hat q_{\ell_0} \hat q_{j} P_{i m_0}-\hat q_{i} \hat q_{m_0} P_{\ell_0 j}+\hat q_{i} \hat q_{j} P_{\ell_0 m_0}\,, \nn\\
&& \hat q_{\ell_0} \hat q_{m_0} P_{i j}-\hat q_{\ell_0} \hat q_{j} P_{i m_0}-\hat q_{i} \hat q_{m_0} P_{\ell_0 j}-P_{i m_0} P_{\ell_0 j}+\hat q_{i} \hat q_{j} P_{\ell_0 m_0}+P_{i j} P_{\ell_0 m_0}\, ,
\eea
where $P_{ij}\equiv\delta_{ij}-\hat q_i\hat q_j$.  

For $n=1$ we obtain the follow 3 operators:

\bea &&  \half \hat q_{l_1} \hat q_{m_1} P_{i m_0} P_{l_0 j}+\half \hat q_{l_1} \hat q_{m_0} P_{i m_1} P_{l_0 j}+\half \hat q_{l_1} \hat q_{m_1} P_{i j} P_{l_0 m_0}-\half \hat q_{l_1} \hat q_{j} P_{i m_1} P_{l_0 m_0}+\half \hat q_{l_1} \hat q_{m_0} P_{i j} P_{l_0 m_1} \nn\\
&&-\half \hat q_{l_1} \hat q_{j} P_{i m_0} P_{l_0 m_1} +\half \hat q_{l_0} \hat q_{m_1} P_{i m_0} P_{l_1 j}+\half \hat q_{l_0} \hat q_{m_0} P_{i m_1} P_{l_1 j}-\half \hat q_{i} \hat q_{m_1} P_{l_0 m_0} P_{l_1 j}-\fourth P_{i m_1} P_{l_0 m_0} P_{l_1 j} \nn\\
&&-\half \hat q_{i} \hat q_{m_0} P_{l_0 m_1} P_{l_1 j}-\fourth P_{i m_0} P_{l_0 m_1} P_{l_1 j}+\half \hat q_{l_0} \hat q_{m_1} P_{i j} P_{l_1 m_0}-\half \hat q_{l_0} \hat q_{j} P_{i m_1} P_{l_1 m_0}-\half \hat q_{i} \hat q_{m_1} P_{l_0 j} P_{l_1 m_0}\nn\\
&&-\fourth P_{i m_1} P_{l_0 j} P_{l_1 m_0}+\half \hat q_{i} \hat q_{j} P_{l_0 m_1} P_{l_1 m_0}-\half P_{i j} P_{l_0 m_1} P_{l_1 m_0}+\half \hat q_{l_0} \hat q_{m_0} P_{i j} P_{l_1 m_1}-\half \hat q_{l_0} \hat q_{j} P_{i m_0} P_{l_1 m_1} \nn\\
&&-\half \hat q_{i} \hat q_{m_0} P_{l_0 j} P_{l_1 m_1}-\fourth P_{i m_0} P_{l_0 j} P_{l_1 m_1}+\half \hat q_{i} \hat q_{j} P_{l_0 m_0} P_{l_1 m_1}-\half P_{i j} P_{l_0 m_0} P_{l_1 m_1}-\half \hat q_{l_1} \hat q_{m_1} P_{i l_0} P_{j m_0} \nn\\
&&-\half \hat q_{l_0} \hat q_{m_1} P_{i l_1} P_{j m_0}+\half \hat q_{i} \hat q_{m_1} P_{l_0 l_1} P_{j m_0}+\fourth P_{i l_1} P_{l_0 m_1} P_{j m_0}+\fourth P_{i l_0} P_{l_1 m_1} P_{j m_0}-\half \hat q_{l_1} \hat q_{m_0} P_{i l_0} P_{j m_1} \nn\\
&& -\half \hat q_{l_0} \hat q_{m_0} P_{i l_1} P_{j m_1}+\half \hat q_{i} \hat q_{m_0} P_{l_0 l_1} P_{j m_1}+\fourth P_{i l_1} P_{l_0 m_0} P_{j m_1}+\fourth P_{i l_0} P_{l_1 m_0} P_{j m_1}+\half \hat q_{l_1} \hat q_{j} P_{i l_0} P_{m_0 m_1} \nn\\
&& +\half \hat q_{l_0} \hat q_{j} P_{i l_1} P_{m_0 m_1}-\half \hat q_{i} \hat q_{j} P_{l_0 l_1} P_{m_0 m_1}+\half P_{i j} P_{l_0 l_1} P_{m_0 m_1}\,,
\eea

\bea && -\half \hat q_{l_1} \hat q_{m_1} P_{i m_0} P_{l_0 j}-\half \hat q_{l_1} \hat q_{m_0} P_{i m_1} P_{l_0 j}-\half \hat q_{l_1} \hat q_{m_1} P_{i j} P_{l_0 m_0}+\half \hat q_{l_1} \hat q_{j} P_{i m_1} P_{l_0 m_0}-\half \hat q_{l_1} \hat q_{m_0} P_{i j} P_{l_0 m_1} \nn\\
&&+\half \hat q_{l_1} \hat q_{j} P_{i m_0} P_{l_0 m_1}-\half \hat q_{l_0} \hat q_{m_1} P_{i m_0} P_{l_1 j}-\half \hat q_{l_0} \hat q_{m_0} P_{i m_1} P_{l_1 j}+\half \hat q_{i} \hat q_{m_1} P_{l_0 m_0} P_{l_1 j}-\fourth P_{i m_1} P_{l_0 m_0} P_{l_1 j} \nn\\
&&+\half \hat q_{i} \hat q_{m_0} P_{l_0 m_1} P_{l_1 j}-\fourth P_{i m_0} P_{l_0 m_1} P_{l_1 j}-\half \hat q_{l_0} \hat q_{m_1} P_{i j} P_{l_1 m_0}+\half \hat q_{l_0} \hat q_{j} P_{i m_1} P_{l_1 m_0}+\half \hat q_{i} \hat q_{m_1} P_{l_0 j} P_{l_1 m_0} \nn\\
&& -\fourth P_{i m_1} P_{l_0 j} P_{l_1 m_0}-\half \hat q_{i} \hat q_{j} P_{l_0 m_1} P_{l_1 m_0}-\half P_{i j} P_{l_0 m_1} P_{l_1 m_0}-\half \hat q_{l_0} \hat q_{m_0} P_{i j} P_{l_1 m_1}+\half \hat q_{l_0} \hat q_{j} P_{i m_0} P_{l_1 m_1} \nn\\
&& +\half \hat q_{i} \hat q_{m_0} P_{l_0 j} P_{l_1 m_1}-\fourth P_{i m_0} P_{l_0 j} P_{l_1 m_1}-\half \hat q_{i} \hat q_{j} P_{l_0 m_0} P_{l_1 m_1}-\half P_{i j} P_{l_0 m_0} P_{l_1 m_1}+\half \hat q_{l_1} \hat q_{m_1} P_{i l_0} P_{j m_0} \nn\\
&&+\half \hat q_{l_0} \hat q_{m_1} P_{i l_1} P_{j m_0}-\half \hat q_{i} \hat q_{m_1} P_{l_0 l_1} P_{j m_0}+\fourth P_{i l_1} P_{l_0 m_1} P_{j m_0}+\fourth P_{i l_0} P_{l_1 m_1} P_{j m_0}+\half \hat q_{l_1} \hat q_{m_0} P_{i l_0} P_{j m_1} \nn\\
&&+\half \hat q_{l_0} \hat q_{m_0} P_{i l_1} P_{j m_1}-\half \hat q_{i} \hat q_{m_0} P_{l_0 l_1} P_{j m_1}+\fourth P_{i l_1} P_{l_0 m_0} P_{j m_1}+\fourth P_{i l_0} P_{l_1 m_0} P_{j m_1}-\half \hat q_{l_1} \hat q_{j} P_{i l_0} P_{m_0 m_1} \nn\\
&& -\half \hat q_{l_0} \hat q_{j} P_{i l_1} P_{m_0 m_1}+\half \hat q_{i} \hat q_{j} P_{l_0 l_1} P_{m_0 m_1}+\half P_{i j} P_{l_0 l_1} P_{m_0 m_1}\,,
\eea

\bea
&& \half P_{i m_1} P_{l_0 m_0} P_{l_1 j}+\half P_{i m_0} P_{l_0 m_1} P_{l_1 j}+\half P_{i m_1} P_{l_0 j} P_{l_1 m_0}-P_{i j} P_{l_0 m_1} P_{l_1 m_0}+\half P_{i m_0} P_{l_0 j} P_{l_1 m_1} \nn\\
&& -P_{i j} P_{l_0 m_0} P_{l_1 m_1}-P_{i m_1} P_{l_0 l_1} P_{j m_0}+\half P_{i l_1} P_{l_0 m_1} P_{j m_0}+\half P_{i l_0} P_{l_1 m_1} P_{j m_0}-P_{i m_0} P_{l_0 l_1} P_{j m_1} \nn\\
&& +\half P_{i l_1} P_{l_0 m_0} P_{j m_1}+\half P_{i l_0} P_{l_1 m_0} P_{j m_1}+ 2 P_{i j} P_{l_0 l_1} P_{m_0 m_1}-P_{i l_1} P_{l_0 j} P_{m_0 m_1}-P_{i l_0} P_{l_1 j} P_{m_0 m_1}\,.
\eea
At each $n$, we will find 3 operators, though their form becomes increasingly complicated as $n$ goes up.

The operators we find in this manner will not in general be projection operators, because they do not satisfy the conditions $P_AP_A=P_A\delta_{AB}$ that a set of projectors $P_A$ should satisfy.  The subspace of $M$'s satisfying the required symmetry conditions has some dimension, $d$.  If the $P$'s were projectors, the sum of the dimensions of the images of the $P$'s would equal $d$, but in our case this sum can be $>d$, so we have an overcomplete set of operators.  We can in principle enforce the conditions $P_AP_A=P_A\delta_{AB}$ to pare this down to a complete set, but since this condition is not linear in the projectors, in practice it is simpler to work with the overcomplete set, and we will not be missing anything by doing so.

\end{document}